{}%use pdflatex at arxiv

\documentclass[11pt,a4paper]{article}

\newif\ifpublic\publictrue
\newif\iffancy\fancytrue

\usepackage[a4paper,text={160mm,247mm},centering]{geometry}
\usepackage{setspace}
\usepackage{rotating}
\usepackage{amsmath,amssymb, amsfonts, geometry}

\makeatletter
\providecommand*{\shuffle}{%
  \mathbin{\mathpalette\shuffle@{}}%
}
\newcommand*{\shuffle@}[2]{%
  \sbox0{$#1\vcenter{}$}%
  \kern .15\ht0 % side bearing
  \rlap{\vrule height .25\ht0 depth 0pt width 2.5\ht0}%
  \raise.1\ht0\hbox to 2.5\ht0{%
    \vrule height 1.75\ht0 depth -.1\ht0 width .17\ht0 %
    \hfill
    \vrule height 1.75\ht0 depth -.1\ht0 width .17\ht0 %
    \hfill
    \vrule height 1.75\ht0 depth -.1\ht0 width .17\ht0 %
  }%
  \kern .15\ht0 % side bearing
}
\makeatother

\makeatletter
\g@addto@macro\bfseries{\boldmath}
\makeatother

\usepackage{xparse}

\ExplSyntaxOn
\NewDocumentCommand{\omwb}{m m}
{
 \omm\left(\begin{smallmatrix}
 \omwb_print:n {#1} \\
 \omwb_print:n {#2}
 \end{smallmatrix}\right)
}
\seq_new:N \l_omwb_list_seq
\cs_new_protected:Npn \omwb_print:n #1
{
  \seq_set_split:Nnn \l_omwb_list_seq { , } { #1 }
  \seq_use:Nn \l_omwb_list_seq { , & }
}
\ExplSyntaxOff

\usepackage[pdfencoding=auto,bookmarks=true,hyperfigures=true]{hyperref}
\usepackage{graphicx}% uncomment for latex-use
\usepackage{float}
\usepackage[nosort]{cite}
\usepackage{lmodern}
\usepackage{amsbsy}
\usepackage[utf8]{inputenc}

\usepackage[font=small,labelfont=bf]{caption}

\usepackage[usenames,dvipsnames]{xcolor}
\definecolor{dgreen}{rgb}{0,0.70,0.30}
\definecolor{gold}{rgb}{0.85,.66,0}
\definecolor{purple}{rgb}{1.0,0.3,0.6}

\numberwithin{equation}{section}

\newcommand{\eqn}[1]{eq.~\eqref{#1}}
\newcommand{\Eqn}[1]{Equation~\eqref{#1}}
\newcommand{\eqns}[2]{eqs.~\eqref{#1} and~\eqref{#2}}

\newcommand{\rcite}[1]{ref.~\cite{#1}}
\newcommand{\rcites}[1]{refs.~\cite{#1}}

\providecommand{\href}[2]{#2}

\makeatletter
\def\mr@ignsp#1 {\ifx\:#1\@empty\else #1\expandafter\mr@ignsp\fi}%
\newcommand{\multiref}[1]{\begingroup%\let\protect\string%
\xdef\mr@no@sparg{\expandafter\mr@ignsp#1 \: }%
\def\mr@comma{}%
\@for\mr@refs:=\mr@no@sparg\do{\mr@comma\def\mr@comma{,}\ref{\mr@refs}}%
\endgroup}
\renewcommand{\eqref}[1]{(\multiref{#1})}
\makeatother

\makeatletter
\newcommand{\namedref}[2]{\hyperref[#2]{#1~\ref*{#2}}}
\newcommand{\secref}{\@ifstar{\namedref{Section}}{\namedref{section}}}
\newcommand{\subsecref}{\@ifstar{\namedref{Subsection}}{\namedref{subsection}}}
\newcommand{\appref}{\@ifstar{\namedref{Appendix}}{\namedref{appendix}}}
\newcommand{\tabref}{\@ifstar{\namedref{Table}}{\namedref{table}}}
\newcommand{\figref}{\@ifstar{\namedref{Figure}}{\namedref{figure}}}
\makeatother

\providecommand{\hypersetup}[1]{}

\hypersetup{plainpages=false}
\hypersetup{pdfpagemode=UseNone}
\hypersetup{bookmarksnumbered=true}
\hypersetup{pdfstartview=FitH}
\hypersetup{colorlinks=false}
\hypersetup{citebordercolor={.5 1 .5}}
\hypersetup{urlbordercolor={.5 1 1}}
\hypersetup{linkbordercolor={1 .7 .7}}
\makeatletter
\let\@keywords\@empty
\let\@subject\@empty
\providecommand{\keywords}[1]{\gdef\@keywords{#1}}
\providecommand{\subject}[1]{\gdef\@subject{#1}}
\def\thetitle{\@title}
\def\theauthor{\@author}
\def\thesubject{\@subject}
\def\thedate{\@date}
\def\thekeywords{\@keywords}
\makeatother
\AtBeginDocument{
\hypersetup{pdftitle={\thetitle}}%
\hypersetup{pdfauthor={\theauthor}}%
\hypersetup{pdfsubject={\thesubject}}%
\hypersetup{pdfkeywords={\thekeywords}}%
}

\newif\ifnote 
\notetrue

\allowdisplaybreaks

\let\Re\relax\DeclareMathOperator{\Re}{Re}
\let\Im\relax\DeclareMathOperator{\Im}{Im}

\newcommand{\z}{\zeta}

\newcommand{\la}{\lambda}
\newcommand{\ka}{\kappa}

\newcommand{\te}{\textrm}

\newcommand{\ZC}{\mathbb C}

\newcommand{\ZN}{\mathbb N}
\newcommand{\ZZ}{\mathbb Z}

\newcommand{\CA}{\mathcal{A}}       
\newcommand{\CB}{\mathcal{B}}       
\newcommand{\CC}{\mathcal{C}}

\newcommand{\CF}{\mathcal{F}}

\def\unreg{\text{unreg}}

\DeclareMathOperator{\Gt}{\tilde{\Gamma}}
\DeclareMathOperator{\ELi}{ELi}

\DeclareMathOperator{\omm}{\omega}

\usepackage{nicefrac}

\DeclareMathOperator{\Res}{Res}
\DeclareMathOperator{\ord}{ord}
\DeclareMathOperator{\Div}{Div}
\DeclareMathOperator{\RR}{R}
\DeclareMathOperator{\DD}{D}
\DeclareMathOperator{\DE}{\DD^E}
\DeclareMathOperator{\JJ}{J}
\DeclareMathOperator{\E}{E}
\DeclareMathOperator{\Li}{Li}
\DeclareMathOperator{\Vol}{Vol}

\newcommand{\pc}[2]{\left[#1\!:#2\!:1\right]} 

\vspace*{0.2cm}
\title{\textbf{
    Functional relations for elliptic polylogarithms 
    }}
\author{Johannes Broedel$^{\,\textit{a}}$,
Andr\'e Kaderli$^{\,\textit{a},\textit{b}}$}
\date{\today}

\begin{document}
\pdfbookmark[1]{Title Page}{title} \thispagestyle{empty}
\begin{flushright}
  \verb!HU-EP-19/18!\\
  \verb!HU-Mathematik-2019-03!
\end{flushright}
\vspace*{0.4cm}
\begin{center}%
  \begingroup\LARGE\bfseries\thetitle\par\endgroup
\vspace{1.0cm}

\begingroup\large\theauthor\par\endgroup
\vspace{9mm}
\begingroup\itshape
$^{\te{a}}$Institut f\"ur Mathematik und Institut f\"ur Physik, Humboldt-Universit\"at zu Berlin\\
IRIS Adlershof, Zum Gro\ss{}en Windkanal 6, 12489 Berlin, Germany
\par\endgroup
\vspace{3mm}
\begingroup\itshape
$^{\te{b}}$Max-Planck-Institut f\"ur Gravitationsphysik, Albert-Einstein-Institut\\
Am M\"uhlenberg 1, 14476 Potsdam, Germany
\par\endgroup
\vspace{3mm}

\vspace{1.0cm}

\begingroup\ttfamily
jbroedel@physik.hu-berlin.de, kaderlia@physik.hu-berlin.de
\par\endgroup

\vspace{1.2cm}

\bigskip
%\vspace{0.7cm}

\textbf{Abstract}\vspace{5mm}

\begin{minipage}{13.4cm}
  Numerous examples of functional relations for multiple polylogarithms are
  known. For elliptic polylogarithms, however, tools for the exploration of
  functional relations are available, but only very few relations are identified. 

  Starting from an approach of Zagier and Gangl, which in turn is based on
  considerations about an elliptic version of the Bloch group, we explore
  functional relations between elliptic polylogarithms and link them to the
  relations which can be derived using the elliptic symbol formalism. The
  elliptic symbol formalism in turn allows for an alternative proof of the validity of
  the elliptic Bloch relation. 

  While the five-term identity is the prime example of a functional identity
  for multiple polylogarithms and implies many dilogarithm
  identities, the situation in the elliptic setup is more involved: there is no
  simple elliptic analogue, but rather a whole class of elliptic identities.
\end{minipage}

\vspace*{4cm}

\end{center}

\newpage

\setcounter{tocdepth}{2}
\tableofcontents

%\begin{mpostdef}
%  xu:=0.8cm;
%  ds:=3.2pt;
%  ls:=0.6pt;
%  pair vp[];
%  path c;
%  def pensize(expr s)=withpen pencircle scaled s enddef;
%  def framed(expr p)=fill bbox p withcolor white; draw bbox p pensize(0.8pt); draw p; enddef;
%  def framedd(expr p)=fill bbox p withcolor 0.94white; draw bbox p pensize(1pt); draw p; enddef;
%\end{mpostdef}
%
%\begin{mpostfig}[label=fundamentaldomain]
%  pickup pencircle scaled 1pt;
%  drawarrow (-0.15xu,0xu)--(5.2xu,0xu) withcolor black;
%  drawarrow (0xu,-0.15xu)--(0xu,2.6xu) withcolor black;
%  draw (0xu,0xu)--(3.9xu, 0xu) withcolor red;
%  draw (4.9 xu, 2 xu)--(1 xu, 2 xu) withcolor red;
%  draw (3.9 xu, 0 xu)--(4.9 xu, 2 xu) withcolor blue;
%  draw (0xu, 0xu)--(1 xu, 2 xu) withcolor blue;
%  label.bot (btex $0$ etex, (0,-.1xu));
%  label.top (btex $\tau$ etex, (0.9xu,2xu));
%  label.top (btex $\tau+1$ etex, (4.8xu,2xu));
%  label.bot(btex $1$ etex, (3.9xu,-.1xu));
%  label.lft(btex Im$(z)$ etex, (0xu,2.6xu));
%  label.top(btex Re$(z)$ etex, (5.2xu,0xu));
%\end{mpostfig}

%%%%%%%%%%%%%%%%%%%%%%%%%%%%%%%%%%%%%%%%%%%%%%%%%%%%%%%%%%%%%%%%%%%%%%%%%%%%%%
%%%%%%%%%%%%%%%%%%%%%%%%%%%%%%%%%%%%%%%%%%%%%%%%%%%%%%%%%%%%%%%%%%%%%%%%%%%%%%
%%%%%%%%%%%%%%%%%%%%%%%%%%%%%%%%%%%%%%%%%%%%%%%%%%%%%%%%%%%%%%%%%%%%%%%%%%%%%%
%%%%%%%%%%%%%%%%%%%%%%%%%%%%%%%%%%%%%%%%%%%%%%%%%%%%%%%%%%%%%%%%%%%%%%%%%%%%%%

\section{Introduction}
\label{sec:introduction}

The majority of calculations in quantum field theory, in particular when
considering quantum chromodynamics, is based on the evaluation of integrals
associated to Feynman graphs. Using Feynman parameters one can rewrite
integrations over loop variables into integrations over Feynman parameters in a
formalised manner. Along with several further advantages, the reparametrisation
allows to bring the integrals to an iterated form.

This is a rather general concept: Feynman integrals seem to be expressible in
terms of iterated integrals over a suitably chosen set of differential forms on
Riemann surfaces of various genera. The exploration of classes of these
iterated integrals and the utilisation of their algebraic properties did not
only change the way calculations are performed, but simultaneously leads to
convenient representations: once a proper class of functions is identified, one
can find functional relations and thus reduce to a basis of
integrals.  

It turns out that suitable differential forms defining classes of iterated
integrals can be identified starting from geometrical considerations: taking
the first abelian differential on the simplest genus-zero surface, the Riemann
sphere, leads to the class of multiple polylogarithms
\cite{Kummer1840,Nielsen1909,Goncharov:2001iea,Remiddi:1999ew,Gehrmann:2000zt}
while abelian differentials on a genus-one Riemann surface are the starting
point for the elliptic polylogarithms \cite{BrownLev,Enriquez:Emzv} to be
discussed in this article.  

\paragraph{Genus zero:} Multiple polylogarithms have been a very active field
of research in the last years: since their motivic version constitute a graded
Hopf algebra \cite{Goncharov:2005sla,Goncharov:1998kja,Goncharov:2001iea}, with
the shuffle product as algebra multiplication and the deconcatenation
coproduct, there are very strong tools available \cite{Duhr:2012fh} allowing in
particular to derive functional relations.  While the Duval algorithm
\cite{PIERREDUVAL1983363} delivers a basis with respect to the shuffle product,
further relations between different arguments of polylogarithms can be explored
using the coproduct, which is usually referred to as the symbol map. A
non-exhaustive list of examples, where such relations are investigated, is
\rcites{Kummer1840,Wechsung:1970,GONCHAROV1995197,wojtkowiak1996,Gangl2003,Zagier:2007knq,2015arXiv150902869S}. Examples involving the evaluation of Feynman integrals include \rcites{Tarasov:2008hw,Tarasov:2015wcd,Davydychev:2016dfi,Davydychev:2017bbl,Tarasov:2019mqy}. We are mainly interested in functional relations of the dilogarithm. Of
particular importance hereby is the so-called five-term identity
\begin{equation}
  \label{eqn:fiveterm}
   \DD\left(t\right)+\DD\left(s\right)+\DD\left(\frac{1-t}{1-ts}\right)+\DD\left(1-ts\right)+\DD\left(\frac{1-s}{1-ts}\right)=0\,,
\end{equation}
where $\DD(t)= \Im\left(\Li_2(t)-\log\left(|t|\right)\Li_1(t)\right)$ is the
Bloch-Wigner function, the single-valued version of the dilogarithm. The
five-term identity has a beautiful interpretation in terms of a volume
decomposition in hyperbolic space into (hyperbolic) tetrahedra. In addition, it
is known \cite{Kirillov:1994en,wojtkowiak1996} to create a large class of
functional equations
for the dilogarithm which are linear combinations of Bloch-Wigner functions
where the arguments are rational functions of one variable and satisfy a
particular condition, to be explained below. Similar statements are conjectured
to hold in more general situations where the arguments are allowed to be
algebraic functions or rational functions of more than one variable
\cite{Zagier:2007knq}.  
On the physics side, the idea of splitting a given volume into several polyhedra has been used to interpret and reformulate the calculation of various Feynman diagrams, see for example refs.~\cite{Davydychev:2016dfi,Davydychev:2017bbl}.
Linear combinations of values of the Bloch-Wigner
function which satisfy the mentioned condition above and which are equal modulo
finitely many applications of functional relations of the Bloch-Wigner function
are identified in the Bloch group \cite{Bloch78,Zagier91,goncharov1991}.
Similarly, higher Bloch groups have been investigated in the context of higher
order polylogarithms. 

\paragraph{Genus one:} While elliptic polylogarithms have been explored for a
long time \cite{BeilinsonLevin,LevinRacinet,BrownLev}, it is only recently that
they have been facilitated in the calculation of scattering amplitudes in
physics \cite{Bloch:2013tra, Broedel:2017siw}. However, as became apparent,
many of the structures inherent in multiple polylogarithms can be taken to
genus one easily: iterated integrals on genus one allow for a natural shuffle
multiplication and an associated coaction or symbol map \cite{Broedel:2018iwv}. 

Given the existence of the symbol map for elliptic iterated integrals, it is a
natural problem to investigate functional relations for elliptic
polylogarithms. In particular, an elliptic analogue of the Bloch group has been
considered in \rcite{ZagierGangl}, which is based on a class of
functional relations for an elliptic generalisation of the Bloch-Wigner
function, the elliptic Bloch-Wigner function $\DE$, and given by relations of
the form
\begin{equation}
  \label{eqn:ellBlochRel}
  \DE(\eta_F)=0\,
\end{equation}
where the object $\eta_F$ is parametrised by (some of the zeros and
singularities of) any non-constant elliptic function $F$ \cite{BlochRelation}.
However, a similar functional relation for the elliptic Bloch-Wigner function
and the construction of an elliptic analogue of the Bloch group has already
been discussed in \rcite{1995alg.geom..8008G}.  In contrast to the genus-zero
case, where the five-term identity suffices to represent a large class of
functional identities of the dilogarithm, a whole class of functional
identities given by \eqn{eqn:ellBlochRel} needs to be investigated in the
genus-one case \cite{ZagierGangl}. The considerations therein, however, remain
on the level of a few particular examples, e.g.\ an implicitly defined elliptic
analogue of the five-term identity. As will be described in detail below, the
answer to the question of an explicit elliptic five-term identity and the
explicit description of the other elliptic functional identities generated by
\eqn{eqn:ellBlochRel} requires substantially more technical effort than for
classical polylogarithms.

\medskip\medskip

In this article, we are going to put Zagier's and Gangl's method to work in
order to find several examples of functional identities between simple elliptic
polylogarithms.  The resulting relations are going to be contrasted with
relations derived using the elliptic symbol map. In order to compare the two
types of relations, one has to translate between different formulations of the
elliptic curve, and thus different types of (iterated) integrals, which is a
source of the complexity of the problem. Despite those difficulties we find
several relations connecting elliptic polylogarithms of rather complicated
arguments. In some cases, the relations found can be trivially accounted to
known symmetry relations for the elliptic Bloch-Wigner function. 

The translation of the elliptic Bloch-Wigner function to the torus, represented
as the complex plane modulo a two-dimensional lattice $\ZC/\Lambda$, allows a
new perspective on the elliptic Bloch relation: the condition encoded in
\eqn{eqn:ellBlochRel} above translates into rather simple relations between
iterated elliptic integrals on the torus, whose correctness is not difficult to
show.  Thus the translation combined with the elliptic symbol calculus provides
an alternative proof of the elliptic Bloch relation. 

As an aside, we are going to translate Ramakrishnan's generalisations of the
elliptic Bloch-Wigner dilogarithm \cite{Ramakrishnan86,Zagier91} as well as
Zagier's generalised single-valued elliptic polylogarithms \cite{Zagier} to the
torus formulation of the elliptic curve. These representations will be serving
as a starting point for the investigation of relations between higher
elliptic functions in a forthcoming project.   

Given the general structure of the elliptic curve, it was not to be expected
that functional relations are at the same level of simplicity as their
genus-zero cousins. On the one hand, the calculation of zeros and poles of
elliptic functions is more complicated than in the case of rational functions
on the Riemann sphere. On the other hand, the translation from the projective
formulation of the elliptic curve, where the mentioned zeros and poles may be
described in terms of rational functions, to the torus given by Abel's map is
not algebraic and highly non-trivial.

\medskip\medskip

This article is structured in the following way: in \secref{sec:review} we
present some of the well-known results for functional relations of the
Bloch-Wigner function and in particular the construction of the Bloch group and
the Bloch relation. In \secref{sec:toolbox} we review several known concepts:
we set the notation for different formulations of elliptic curves as well as
elliptic functions and review known results about the Bloch group in the
genus-one situation, which are mostly formulated on the Tate curve describing
the corresponding elliptic curve. \secref*{sec:connection} is devoted to the
translation of the above and further concepts to the torus and the projective
elliptic curve. In particular, notions of (conjecturally) single-valued
elliptic generalisations of polylogarithms defined on the Tate curve are
related to the elliptic multiple polylogarithms as holomorphic iterated
integrals on the torus, which further allows to formulate (and prove) the
elliptic Bloch relation~\eqref{eqn:ellBlochRel} on the torus and the projective
elliptic curve, respectively. 

\section{Bloch groups for polylogarithms}
\label{sec:review} The description of functional relations of polylogarithms
and in particular of the single-valued dilogarithm -- the Bloch-Wigner function
-- can be formalised using the concept of (higher) Bloch groups. These are
certain (abelian) groups $\CB_m$ which capture functional relations satisfied
by single-valued polylogarithms of order $m$.

In \subsecref{sec:review:BG} we are going to review the geometric construction
and interpretation of $\CB_2$ in terms of hyperbolic three-manifolds.
Afterwards, in \subsecref{ssec:dilogrel} we introduce the Bloch relation of the
Bloch-Wigner function, which generates functional identities such as the
five-term identity. In the subsequent section this Bloch relation will be
generalised to the elliptic curve and will be used to define the elliptic
analogue of $\CB_2$, the elliptic Bloch group, which is discussed in
\subsecref{sec:toolbox:ellBloch}. 

\subsection{The Bloch group}\label{sec:review:BG}
The functional relations of the dilogarithm $\Li_2$ often take a very simple
form when expressed in terms of the Bloch-Wigner function 
\begin{align}
  \label{eqn:BlochWigner}
\DD(t)&= \Im\big(\Li_2(t)-\log\left(|t|\right)\Li_1(t)\big)\,,
      % \nnl
      %     &= \Im\big(\Li_2(t)+\log\left(|t|\right)\log(1-t)\big)\nnl
      %     &= \Im\big(\Li_2(t)+\arg(1-t)\log|t|\big),
\end{align}
which is the single-valued version of the dilogarithm
(see~\rcite{Zagier:2007knq} for an extensive review of the Bloch-Wigner
function). The Bloch-Wigner function is continuous on the Riemann sphere and
real analytic except at the points $0$, $1$ and $\infty$, where it is defined
to vanish. 

The Bloch-Wigner function and its functional relations admit a broad variety of
mathematical interpretations and applications, ranging from  periodicities of a
cluster algebra
\cite{2001math......4151F,Nakanishi:2010uv,2011SIGMA...7..102K}, volumes in
hyperbolic space \cite{NEUMANN1985307,Thurston} and the symbol calculus
\cite{Goncharov:2010jf,Goncharov.A.B.:2009tja} to functional identities
generated by rational functions on the Riemann sphere \cite{BlochRelation}, the
latter is the main focus of our considerations.

The Bloch-Wigner function satisfies the symmetry relations 
\begin{equation}\label{sec:invitation:BlochGroup:symD}
 \DD(t)=\DD\left(1-\frac{1}{t}\right)=\DD\left(\frac{1}{1-t}\right)=-\DD\left(\frac{1}{t}\right)=-\DD\left(1-t\right)=-\DD\left(\frac{-t}{1-t}\right)
\end{equation}
and the duplication relation
\begin{equation}\label{sec:invitation:BlochGroup:dupD}
\DD(t^2)=2\DD\left(t\right)+2\DD\left(-t\right)\,,
\end{equation}
which can be easily proven using the properties of the logarithm and $\Li_2$.
In addition, there is the famous five-term identity already mentioned in the
introduction, which can be described as a consequence of the periodicity of the
$A_2$ cluster algebra \cite{2011SIGMA...7..102K}. It reads 
\begin{align}\label{sec:invitation:BlochGroup:fiveterm}
 \DD\left(t\right)+\DD\left(s\right)+\DD\left(\frac{1-t}{1-ts}\right)+\DD\left(1-ts\right)+\DD\left(\frac{1-s}{1-ts}\right)&=0\,.
\end{align}
In order for the above equation to yield a valid new relation, $t$ and $s$ are numbers
chosen such that neither of the arguments yields $0,1$ or $\infty$, i.e.\ $s,t\neq 0,1$ and $st\neq 1$. In those
special cases, however, \eqn{sec:invitation:BlochGroup:fiveterm} degenerates to
the symmetry relations in
\eqn{sec:invitation:BlochGroup:symD} above. 

Alternatively, one can interpret the five-term identity as a relation between
volumes of hyperbolic three-simplices in the so-called Poincaré half-space
model \cite{NEUMANN1985307,Thurston}. As it is this volume interpretation of
the Bloch-Wigner function which leads to an illustrative geometric construction
of the Bloch group~\footnote{The Bloch group $\mathcal{B}_2$ has originally
been introduced in \rcite{Bloch78} and has been extended in refs.\
\cite{Zagier91, goncharov1991} to higher orders.}, let us describe this
construction in a little more detail following the lines of refs.\
\cite{Zagier:2007knq,ZagierGangl}.  The volume of a complete, finite,
hyperbolic three-manifold $M$ can be triangulated and, thus, expressed as the
sum over the volumes of a finite number of three-simplices  
\begin{align}
 \label{eqn:VolM}
 \Vol(M)&=\sum_i \DD(t_i)\,,
\end{align}
each of which can be labelled by a cross ratio $t_i\in \ZC$  such that its
volume is given by $\DD(t_i)$.  Considering the geometric properties of such a
triangulation, one can show that the associated coordinates $t_i$ in
\eqn{eqn:VolM} have to satisfy the following algebraic constraint
\cite{NEUMANN1985307}:
\begin{align}\label{eqn:TriangulationCondition}
 \sum_i t_i\wedge (1-t_i)=0\in \ZC^{\ast}\wedge\ZC^{\ast}\,.
\end{align}
Correspondingly, one can in general express the volume of $M$ as  
\begin{align} \label{eqn:VolMsumDi}
  \Vol(M)&=\sum_i \DD(t_i)=\DD(\xi)\,,
\end{align}
for an element $\xi\in\CA_2\left(\ZC\right)$, where
\begin{align}\label{eqn:DefA2}
 \CA_2\left(\ZC\right)&=\Bigg\lbrace\sum_{i=1 }^n n_i(t_i)\,|\,t_i\in \ZC^{\ast}\!\setminus\!\{1\},n\in\ZN,\,n_i\in\ZZ,\,\, \sum_{i=1}^n n_i\left(t_i\wedge (1-t_i)\right)=0\Bigg\rbrace\subset\CF_{\ZC}\,,
\end{align}
$\CF_{\ZC}$ is the free abelian group\footnote{The free abelian group generated
	by a set $S$ is the group of formal finite sums $\sum_{s\in S} n_s (s)$
	with $n_s\in \ZZ$, all but finitely many equal to zero. The group
	operation is defined by $\sum_{s\in S} n_s (s)+\sum_{s\in S} m_s
	(s)=\sum_{s\in S} (n_s+m_s) (s)$ and the identity element is the empty
sum. Note that in contrast to the usual notation where square brackets are used
to denote an element of the free abelian group, we use parentheses in agreement
with the notation of divisors introduced in \subsecref{ssec:dilogrel}.}
generated by $\ZC$ and the Bloch-Wigner function is extended by linearity to
$\CF_{\ZC}$, i.e.
\begin{equation}
  \label{eqn:DlinExample}
  \DD\left(\sum_i n_i(t_i)\right)=\sum_i n_i \DD(t_i)\,.
\end{equation}
Let us briefly discuss the definition~\eqref{eqn:DefA2} of $\CA_2(\ZC)$. The
condition $t_i\notin \{0,1\}$ corresponds to the definition of
$\DD(0)=0=\DD(1)$. The fact that now, we allow in $\sum_{i=1}^n
n_i\left(t_i\wedge (1-t_i)\right)=0$ the coefficients $n_i$ to be any integer
and not only to equal $1$, as in \eqn{eqn:TriangulationCondition}, is required
to turn $\CA_2\left(\ZC\right)$ into a subgroup of $\CF_{\ZC}$ and to
(uniquely) shorten the sum~\eqref{eqn:VolMsumDi} in the case of $t_i=t_j$ for
$i\neq j$. 

The geometric interpretation of the five-term identity corresponds to a change
of triangulation: it describes two distinct triangulations of a volume defined
by five vertices and the edges being geodesics, which can either be described
by a disjoint union of two hyperbolic three-simplices or of three such
simplices. The five-term identity expresses the equality of the sum of the
volumes of the former two and the latter three simplices. The change of
triangulation and the associated applications of the five-term relation
motivate the definition of the following subgroup of $\mathcal{A}_2(\ZC)$,
which can be thought of as constituting the group of relations of the
Bloch-Wigner function with the generator\footnote{For $T\subset S$, the
subgroup $\langle t|t\in T\rangle$ of $\CF_{\ZC}$ generated by $T$ is the group
of formal finite sums $\sum_{t\in T} n_t (t)$ with $n_t\in \ZZ$, all but
finitely many equal to zero.} being the arguments occurring in the five-term
identity,
\begin{align}
 \CC_2\left(\ZC\right)&=\langle\left(t\right)+\left(s\right)+\left(\frac{1-t}{1-ts}\right)+\left(1-ts\right)+\left(\frac{1-s}{1-ts}\right)| s,t\in\ZC^{\ast}\setminus\{1\}\, , st\neq 1\rangle\, .
\end{align}
Thus, the volume of $M$ can be expressed as the value
\begin{align}
 \Vol(M)&=\DD(\xi_M)
\end{align}
for a canonical $\xi_M\in \CB_2\left(\ZC\right)$ associated to $M$ with 
\begin{align}\label{sec:invitation:BlochGroup:def}
 \CB_2\left(\ZC\right)=\frac{\CA_2\left(\ZC\right)}{\CC_2\left(\ZC\right)}
\end{align}
being the Bloch group\footnote{Higher Bloch groups $\CB_m(\ZC)$ for $m>2$ can
	be constructed recursively \cite{ZagierGangl}. In analogy to the case
	$\CC_2(\ZC)$ considered above, the subgroup $\mathcal{C}_m(\ZC)$ of the
	group of "allowable" points $\mathcal{A}_m(\ZC)$ (where allowable can
be defined recursively and corresponds to the condition $\sum_{i=1}^n
n_i\left(t_i\wedge (1-t_i)\right)=0$ in the case $m=2$) is constructed to be
the span of functional relations among polylogarithms of order $m$.}. 

Besides this geometric construction, the Bloch group $\CB_2(\ZC)$ is an
elementary algebraic structure for the description of dilogarithmic functional
relations, i.e.\ identities of finite sums such as $\sum_i n_i
\DD(t_i(s_j))=c$, for rational or algebraic functions $t_i$ of one or more
variables $s_j$ and some constant $c\in \ZC$. In the case of only one variable
$s$ and rational functions $t_i(s)\in \ZC(s)$, the element $\xi =\sum_i n_i
(t_i(s))$ evaluates under $\DD$ to a constant if and only if $\sum_i n_i
\left(t_i(s)\right)\wedge \left(1-t_i(s)\right) $ is independent of $s$
\cite{Zagier91}. For the particular condition $\sum_i n_i
\left(t_i(s)\right)\wedge \left(1-t_i(s)\right) =0$, the element $\xi$ belongs
to the Bloch group $\mathcal{B}_2(\ZC(s))$ of the field of rational functions
$\ZC(s)$. As proven in \rcite{wojtkowiak1996}, all such elements are equal to
zero in $\mathcal{B}_2(\ZC(s))$. Thus, in this case the functional equation
$\sum_i n_i \DD\left(t_i(s)\right)=0$ is indeed obtained by a finite number of
applications of the five-term identity. Similar statements are not known in the
case of algebraic functions or rational functions in more than one variable,
but they are expected to exist, see e.g.\ \cite{Zagier:2007knq}.

\subsection{Bloch's dilogarithm relations}\label{ssec:dilogrel}

Bloch describes a concept to formalise the generation of functional
identities for the Bloch-Wigner function and of its generalisation to elliptic
curves \cite{BlochRelation}. In this subsection we state his results in the
classical situation and generalise it to the elliptic case in
\secref{sec:toolbox} below.
\smallskip

In the following, we are going to make use of the concept of a
\textit{divisor}: for any meromorphic function $g$ defined on a compact Riemann
surface $X$, the divisor of $g$ is defined as 
\begin{equation}
\label{eqn:defdivisor}
\Div(g)=\sum_{p\in X}\ord_p(g)\,(p) \,,
\end{equation}
where $\ord_p(g)$ is the order of the pole (a negative integer) or the order of
the zero (a positive integer), respectively, of $g$ at $p$. If $p$ is neither a
pole nor a zero of $g$, $\ord_p(f)=0$, which renders the number of terms in the
above sum finite. In the definition above, divisors are elements of the free
abelian group generated by the Riemann surface $X$.  

Let $f: \ZC P^1\rightarrow \ZC P^1$ be a non-trivial rational function on the
Riemann sphere satisfying 
\begin{equation}\label{sec:invitation:Bloch:condf}
 f(0)=f(\infty)=1\,,
\end{equation}
which can be realized by representing $f$ as a finite product 
\begin{equation}\label{sec:invitation:Bloch:fandCoefs}
  f(t)=\prod_{i} (t-a_i)^{d_i}\,,\qquad\sum_i d_i=0\,,\qquad \prod_i a_i^{d_i}=1\,,
\end{equation}
where $a_i\in \ZC$ and $d_i\in \ZZ$. Furthermore, let us write 
\begin{align}
 1-f(t)&=b\prod_{j} (t-b_j)^{e_j}\,,
\end{align}
where $b, b_j\in \ZC$ and $e_j\in \ZZ$. The divisor of the function $f$ defined in 
\eqn{sec:invitation:Bloch:fandCoefs} reads 
\begin{equation}\label{eqn:defdivisorRiemannSphere}
 \Div(f)=\sum_i \ord_{a_i}(f)(a_i)=\sum_i d_i(a_i)\,,\qquad \Div(1-f)=\sum_j e_j(b_j)\,.
\end{equation}
In \rcite{BlochRelation} Bloch proves that for any
rational function $f$ as defined above, the Bloch-Wigner function satisfies
\begin{align}\label{sec:invitation:Bloch:BlochRel}
 \sum_{i,j}d_i e_j\DD\bigg(\frac{a_i}{b_j}\bigg)&=0\,,
\end{align}
abbreviated in terms of the element 
\begin{align}
\eta_f&=\sum_{i,j}d_i e_j\left(\frac{a_i}{b_j}\right)
\end{align}
of the free abelian group $\CF_{\ZC}$ and the Bloch-Wigner function extended by
linearity as in \eqn{eqn:DlinExample}, this so-called \textit{classical Bloch
relation} reads
\begin{align}\label{sec:invitation:Bloch:abbreviatedBlochRel}
\DD\left(\eta_f \right)&=0\,.
\end{align}
Letting the zeros $a_i$ of $f$ vary subject to the conditions in
\eqn{sec:invitation:Bloch:fandCoefs}, this dilogarithm relation of Bloch
becomes a functional relation. Choosing different rational functions satisfying
\eqn{sec:invitation:Bloch:condf} in the first place,
\eqn{sec:invitation:Bloch:abbreviatedBlochRel} yields a whole class of
functional relations for the Bloch-Wigner function parametrized by rational
functions $f$ on the Riemann sphere, which is however not independent. In fact,
it is conjecturally generated by the single example of the five-term identity
(see the discussion at the end of \subsecref{sec:review:BG}), which is
discussed in the following paragraph.

As the most fundamental example and an application of the Bloch relation, let
us discuss how to recover the five-term identity
\eqn{sec:invitation:BlochGroup:fiveterm} from
\eqn{sec:invitation:Bloch:abbreviatedBlochRel} following the lines of
\rcite{ZagierGangl}: let $a,b\in \ZC$, $a'=1-a$, $b'=1-b$ and consider the
rational function 
\begin{align}\label{sec:invitation:Bloch:5termf}
f(t)&=\frac{(t-a)(t-a')(t-bb')}{(t-b)(t-b')(t-aa')}\,.
\end{align}
It satisfies $f(0)=f(\infty)=1$ and 
\begin{align}
1-f(t)&=\frac{(bb'-aa')t^{2}}{(t-b)(t-b')(t-aa')}\,,
\end{align}
such that Bloch's relation can be applied, which yields the identity
\begin{align}
\DD\left(\frac{a}{b}\right)+\DD\left(\frac{a'}{b'}\right)+\DD\left(\frac{a}{b'}\right)+\DD\left(\frac{bb'}{aa'}\right)+\DD\left(\frac{a'}{b}\right)&=0\,,
\end{align} 
where we have used that $\DD(0)=\DD(1)=\DD(\infty)=0$ and the symmetry relations
\eqref{sec:invitation:BlochGroup:symD}. Changing variables to $t=\frac{a}{b}$,
$s=\frac{a'}{b'}$ finally leads to the five-term identity in the usual
form~\eqref{sec:invitation:BlochGroup:fiveterm}.

\section{Elliptic curves, the divisor function and Bloch's relation}
\label{sec:toolbox}

The aim of this section is twofold: after reviewing mathematical tools for the
description of elliptic curves in various formulations and a particular type of
elliptic iterated integrals in subsections \ref{sec:toolbox:ellipticFunction}
and \ref{ssec:EMP} we are going to discuss and exemplify the generalisation of
the concepts of the divisor function and the Bloch relation from the previous
section to the genus-one Riemann surfaces / elliptic curves in subsections
\ref{sec:toolbox:divisorFunction} and \ref{sec:toolbox:ellBloch}. In
particular, \subsecref{sec:toolbox:ellBloch} contains three examples of
functional relations on the elliptic curve parametrised by various rational
functions.

\subsection{Elliptic curves and functions}\label{sec:toolbox:ellipticFunction}
This subsection begins with the introduction of the torus description of elliptic
curves: being a Riemann surface of genus one, the torus is the natural geometry
underlying an elliptic curve due to its two periodicities. Along with the
discussion of the torus formulation, several properties of elliptic functions are
reviewed. 
Afterwards, two isomorphisms are discussed, where the first one relates the
torus to the projective (elliptic) curve and the second one maps the torus to
the so-called Tate curve given by the exponential map. 
These are well-known mathematical concepts, but 
in particular the map from the torus to the Tate curve is rarely mentioned in the physics
literature.  A thorough introduction which relates to the common physics
language can e.g.\ be found in \rcite{Broedel:2017kkb}, which is the basis for
the discussion in this subsection.
\medskip

\begin{figure}[H]
\begin{center}
  \includegraphics[width=6.0cm]{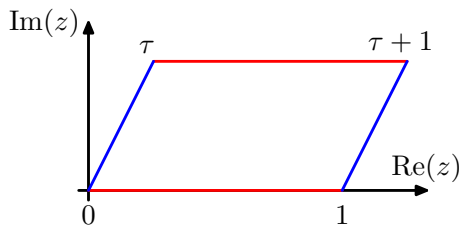}
  \caption{Fundamental domain of the torus $\ZC/\Lambda$}
  \label{fig:fundamentalParallelogram}
\end{center}
\end{figure}

A torus can be described as the quotient $\ZC/\Lambda$ of the complex plane
and a lattice 
\begin{equation}
  \label{def:lattice}
  \Lambda=\omega_1 \ZZ+\omega_2\ZZ  
\end{equation}
where the periods $\omega_1$ and $\omega_2$ are complex numbers and taken to be linearly
independent over the real numbers.
The domain $P_{\Lambda}=\{a\omega_1+b\omega_2\,|\,0\leq a,b<1\}$ is called the
\textit{fundamental parallelogram} of $\ZC/\Lambda$ which defines the torus upon
identifying the opposite sides of its closure. Due to this immediate relation,
we will simply refer to $\ZC/\Lambda$ as the torus itself. The torus is often scaled
such that $\tau=\omega_2/\omega_1$ and 1 are its periods and without loss of
generality $\tau$ is assumed to be an element of the upper half plane,
$\Im(\tau)>0$, in this case the fundamental parallelogram can be depicted as in figure \ref{fig:fundamentalParallelogram}.
A function is called \textit{elliptic} on $\ZC$ if it is $\Lambda$-periodic,
i.e.\ a function defined on $\ZC/\Lambda$, and meromorphic. However, in the
case of generalisations of multiple polylogarithms to the elliptic curve, we
sometimes also refer to multi-valued functions on the torus $\ZC/\Lambda$
(i.e.\ not necessarily $\Lambda$-periodic functions), as elliptic functions if
they are meromorphic. This is in particular the case for the elliptic multiple
polylogarithms introduced in \subsecref{ssec:EMP}. 

Two explicit examples of elliptic functions are the even
Weierstrass $\wp$ function
\begin{equation}\label{sec:toolbox_ellipticfunctions:wp}
 \wp(z)=\wp(z;\omega_1,\omega_2)=\frac{1}{z^{2}}+\sum_{(m,n)\neq (0,0)}\left(\frac{1}{(z+m\omega_1+n\omega_2)^{2}}-\frac{1}{(m\omega_1+n\omega_2)^{2}}\right)
\end{equation}
and its odd derivative $\wp'(z)$. Note that $\wp$ has a double pole at any
lattice point, whereas $\wp'$ has a triple pole at the lattice points. Closed
expressions of zeros of $\wp$ are generally complicated, while the zeros of
$\wp'$ are exactly the half periods $\omega_i/2$, for $i=1,2,3$ and
$\omega_3=\omega_2-\omega_1$.  Moreover, these elliptic functions satisfy the
differential equation 
\begin{equation}\label{sec:toolbox_ellipticfunctions:wpODE}
\wp'(z)^{2}=4(\wp(z)-e_1)(\wp(z)-e_2)(\wp(z)-e_3)=4\wp(z)^{3}-g_2\wp(z)-g_3
\end{equation}
where the three roots $e_i$ are defined by 
\begin{equation}
  \label{eqn:rootsperiods}
  e_i=\wp(\omega_i/2)
\end{equation}
and sum to zero. The Weierstrass invariants $g_2$ and $g_3$ in the above
equation can be expressed in terms of Eisenstein series
\begin{equation}
g_2=60\sum_{(m,n)\neq(0,0)}\frac{1}{(m\omega_1 +n\omega_2)^{4}}\,,\qquad g_3=140\sum_{(m,n)\neq(0,0)}\frac{1}{(m\omega_1 +n\omega_2)^{6}}
\end{equation}
and are related to the roots by 
\begin{equation}\label{sec:toolbox_ellipticfunctions:RootsAndInvariants}
 e_1+e_2+e_3=0\,,\qquad e_1e_2+e_2e_3+e_3e_1=-\frac{1}{4}g_2\,,\qquad e_1e_2e_3=\frac{1}{4}g_3\,.
\end{equation}
It turns out that the notion of ellipticity is quite restrictive: for example,
the zeros and poles of an elliptic function $F$ are subject to the
conditions\footnote{See e.g.\ the lecture notes \cite{deShalit} for the
  derivation of equation
  \eqref{sec:toolbox_ellipticfunctions:conditionsEllipticFctTorus} and the
following statements about Weierstrass functions.}
\begin{equation}\label{sec:toolbox_ellipticfunctions:conditionsEllipticFctTorus}
\sum_{z\in P_{\Lambda}}\ord_z\left(F\right)=0\,, \qquad \sum_{z\in P_{\Lambda}}\ord_z\left(F\right) z\in\Lambda\,,
\end{equation}
where the order $\ord_z(F)$ of $F$ at $z$ is the usual order of zeros and poles
of meromorphic functions, in analogy to the definition of the  order in the
context of rational functions on the Riemann sphere used in
\eqn{eqn:defdivisor}. In particular, points which are neither zeros nor
poles are of order zero. Thus, the sums over the fundamental parallelogram in
equation
\eqref{sec:toolbox_ellipticfunctions:conditionsEllipticFctTorus} are finite and
include the non-vanishing terms at zeros and poles of $F$ only. 

Moreover, an elliptic function can not have a single simple pole: using
Cauchy's residue theorem and integration along the fundamental parallelogram,
where the (reversed) parallel paths cancel pairwise due to the
$\Lambda$-periodicity, the sum of the residues has to vanish, which can not be
satisfied by a single simple pole alone.  The conditions in
\eqn{sec:toolbox_ellipticfunctions:conditionsEllipticFctTorus} follow from the
same cancellation of the integration along the fundamental parallelogram in the
(generalised) argument principle. 

Furthermore, any elliptic function is determined up to scaling by its zeros and
poles: the quotient of two elliptic functions with the same zeros and poles,
counting multiplicities, is bounded on the fundamental parallelogram
$P_{\Lambda}$ and hence, by $\Lambda$-periodicity, is a bounded entire
function, such that Liouville's theorem implies that these two elliptic
functions are proportional to each other. This fact, in turn, implies that any
elliptic function on $\ZC/\Lambda$ is a rational function in $\wp$ and $\wp'$:
those rational functions are elliptic by construction and can be combined to
have the same zeros and poles as any given elliptic function.  

Alternatively, any elliptic function can be expressed in terms of the
Weierstrass $\sigma$ function
\begin{align}\label{sec:toolbox_ellipticfunctions:sigma}
\sigma(z)&=s_C\,\exp\left(\int_{z_0}^zdz'\,\zeta(z')\right)\,,
\end{align}
where the scaling factor $s_C$ and the base point $z_0$ are
chosen\footnote{Both, $s_C$ and $z_0$, can be chosen canonically by adjusting
the integration constant $\xi(\omega_i)$ in
\eqn{sec:toolbox_ellipticfunctions:trafoZeta}.} such that $\sigma'(0)=1$.  The
logarithmic derivative (and thus the integrand in
\eqn{sec:toolbox_ellipticfunctions:sigma}) of the Weierstrass $\sigma$ function
is the Weierstrass $\zeta$ function
\begin{equation}
  \label{eqn:WeierstrassZeta}
\zeta(z)=\frac{1}{z}+\sum_{(m,n)\neq (0,0)}\left(\frac{1}{z-m\omega_1-n\omega_2}+\frac{1}{m\omega_1+n\omega_2}+\frac{z}{(m\omega_1+n\omega_2)^{2}}\right)\,,
\end{equation}
which itself is the negative odd primitive of $\wp$. 

The Weierstrass $\sigma$ function has no poles and one simple zero at the
lattice points, hence, it can not be elliptic. In fact, neither $\zeta$ nor
$\sigma$ is $\Lambda$-periodic. For the Weierstrass $\zeta$ function and a
lattice period $\omega_i$, integrating the equation $\wp(z+\omega_i)=\wp(z)$
implies that $\zeta$ changes by a $z$-independent integration constant
\begin{align}\label{sec:toolbox_ellipticfunctions:quasiperiod}
 \zeta(z+\omega_i)&=\zeta(z)+2\eta(\omega_i)\,
\end{align}
with the quasi-period $\eta(\omega_i)=\zeta(\omega_i/2)$, which follows from
the evaluation of \eqn{sec:toolbox_ellipticfunctions:quasiperiod} at
\mbox{$z=-\omega_1/2$}. In a similar manner one can determine the transformation
behaviour of the Weierstrass $\sigma$ function, which reads 
\begin{align}\label{sec:toolbox_ellipticfunctions:trafoZeta}
 \sigma(z+\omega_i)&=\exp\Big(2\eta(\omega_i)z+\xi(\omega_i)\Big)\sigma(z)\,,
\end{align}
where $\xi(\omega_i)$ is yet another integration constant (see
e.g.~\rcite{deShalit}). This shows explicitly that $\sigma$ is indeed not
elliptic. The transformation~\eqref{sec:toolbox_ellipticfunctions:trafoZeta} of
$\sigma$ and the fact that it has one simple zero at any lattice point and no
poles at all leads to the alternative representation of a given elliptic
function $F$ mentioned above: one can always choose particular representatives
$A_i$ of the zeros and poles of $F$ in $\ZC/\Lambda$ (not necessarily in the
fundamental domain) such that
\begin{equation}\label{sec:toolbox_ellipticfunctions:ZeroPolesCond}
\sum_i d_i=0\,,\qquad \sum_i d_i A_i=0\,,
\end{equation}
where $d_i=\ord_{A_i}(F)$. It is then the set of
conditions~\eqref{sec:toolbox_ellipticfunctions:ZeroPolesCond}, satisfying the
natural
constraints~\eqref{sec:toolbox_ellipticfunctions:conditionsEllipticFctTorus}
for the zeros and poles of an elliptic function, which ensures that the
combination\footnote{Note that compared to the definition of the Weierstrass
  $\sigma$ function~\eqref{sec:toolbox_ellipticfunctions:sigma}, the factors of
  $s_C$ from the product on the left-hand side of
  \eqn{sec:toolbox_ellipticfunctions:prodSigma} multiply to one and the base
point $z_0$ of the integrals in the exponential can be shifted to zero due to
the condition $\sum_i d_i=0$ in
\eqn{sec:toolbox_ellipticfunctions:ZeroPolesCond}.}
\begin{align}\label{sec:toolbox_ellipticfunctions:prodSigma}
\prod_{i}\sigma\left(z-A_i\right)^{d_i}&=\exp\left(\sum_i d_i \int_0^{z-A_i}dz'\zeta(z')\right)
\end{align}
is elliptic. Indeed, under a lattice displacement the exponential
proportionality factor in \eqn{sec:toolbox_ellipticfunctions:trafoZeta} from
the transformation of the individual factors $\sigma(z-A_i)^{d_i}$ in
\eqn{sec:toolbox_ellipticfunctions:prodSigma} form a product with an exponent
which is a linear combination of the left-hand sides of the two conditions
\eqref{sec:toolbox_ellipticfunctions:ZeroPolesCond}, such that the overall
proportionality constant evaluates to one. Since $\sigma$ has only one simple
zero at the lattice points and no pole, the above product has exactly the same
zeros and poles including multiplicities as the function $F$. Correspondingly,
any elliptic function $F$ can be written as
\begin{equation}\label{sec:toolbox_ellipticfunctions:FandSigma}
  F(z)=s_A\prod_{i}\sigma\left(z-A_i\right)^{d_i}=s_A\exp\left(\sum_i d_i \int_0^{z-A_i}dz'\zeta(z')\right)
\end{equation}
for some scaling factor $s_A\in\ZC$. The behaviour of the zeros and poles of an
elliptic function can be conveniently captured in terms of divisors, which are
introduced in \subsecref{sec:toolbox:divisorFunction}.
\smallskip

The fact that all elliptic functions can be expressed as rational functions of
$\wp$ and $\wp'$ facilitates their description in terms of rational functions
on a complex projective algebraic curve. The Weierstrass $\wp$ function induces an
isomorphism between $\ZC/\Lambda$ and the complex projective algebraic curve
\begin{align}\label{sec:toolbox_ellipticfunctions:ProjectiveCurve}
E\left(\ZC\right)&=\{\pc{x}{y}|\,y^{2}=4x^{3}-g_2\left(\Lambda \right) x-g_3\left(\Lambda \right)\}\cup \{\left[0\!:1\!:0\right]\}\,,
\end{align}
where $\left[0\!:1\!:0\right]$ is denoted by infinity $\infty$. Note that the
cubic equation in $x$ and $y$ of the curve in definition
\eqref{sec:toolbox_ellipticfunctions:ProjectiveCurve} is of the same form as
the differential equation~\eqref{sec:toolbox_ellipticfunctions:wpODE} for
$\wp$: this representation of the constraint equation on the projective
formulation of the elliptic curve is called the \textit{Weierstrass form} or
\textit{Weierstrass equation}. Furthermore, the projective algebraic curve $E(\ZC)$ is
often called the projective formulation of the elliptic curve or the projective elliptic curve. 

The isomorphism of Riemann surfaces is given by
\begin{equation}\label{sec:toolbox_ellipticfunctions:xiLE}
\xi_{\Lambda,E}: \ZC/\Lambda\rightarrow E\left(\ZC\right)\,, \qquad 0\neq z\mapsto \pc{\wp(z)}{\wp(z)'}\,,\qquad 0\mapsto \left[0\!:1\!:0\right]=\infty\,,
\end{equation}
see e.g.~\rcite{Broedel:2017kkb} for more details. The addition on
$E\left(\ZC\right)$ is provided by the so-called chord-tangent construction with the
additive unity being $\infty$. It has a nice geometric interpretation, which is
described in \appref{app:chord-tangent}. 

The inverse of the isomorphism $\xi_{\Lambda,E}$ is called Abel's map and can
be determined from the differential
equation~\eqref{sec:toolbox_ellipticfunctions:wpODE}. Given a point
$P=\pc{x_P}{y_P}$ with $ y_P\neq 0$, one finds
\begin{align}
  \label{sec:toolbox_ellipticfunctions:Abelsmap}
  z&=\pm \int_{\infty}^{x_P}\frac{dx}{y}\mod \Lambda\,,
\end{align}
where the correct sign is determined by the requirement that $\wp'(z)=y_P$, and
$\xi^{-1}_{\Lambda,E}(e_i)=\omega_i/2$ for $P=\pc{e_i}{0}$. Upon identifying
\begin{align}
  \label{sec:toolbox_ellipticfunctions:xy}
  x=\wp(z)\,,\qquad y=\wp'(z)
\end{align}
as well as using the fact that these two functions generate any elliptic
function on the torus in terms of rational functions, it follows that the
elliptic functions can be described as the rational functions in $x$ and $y$
on the projective elliptic curve $E(\ZC)$.

The above choice of signs in Abel's map
\eqref{sec:toolbox_ellipticfunctions:Abelsmap} is not the only issue that needs
some care if a translation from a given projective elliptic curve $E(\ZC)$ with
elliptic invariants $g_2$ and $g_3$ to the torus has to be implemented
explicitly. 

A first ambiguity has to be addressed by making a choice for the periods
$\omega_1$ and $\omega_2$ associated to the elliptic curve with Weierstrass
equation $y^{2}=4x^{3}-g_2x-g_3=4(x-e_1)(x-e_2)(x-e_3)$. The roots $e_i$ are
defined by $g_2$ and $g_3$ up to relabelling according to
\eqn{sec:toolbox_ellipticfunctions:RootsAndInvariants}. Simultaneously, Abel's
map together with \eqn{eqn:rootsperiods} implies
\begin{equation}\label{sec:toolbox_ellipticfunctions:HalfPeriods}
  \frac{\omega_1}{2}=\frac{\omega_2}{2}-\frac{\omega_3}{2}=\bigg(\int_{e_3}^{e_2}\frac{dx}{y}\bigg)\!\!\!\mod \Lambda\,,
\end{equation}
where the $\omega_i$, or the fundamental parallelogram, respectively, are
chosen such that the periods are given by the integrals \cite{NIST}
\begin{equation}\label{sec:toolbox_ellipticfunctions:Periods}
 \omega_1=2\int_{e_3}^{e_2}\frac{dx}{y}\,, \qquad \omega_2=2\int_{e_3}^{e_1}\frac{dx}{y}\,,\qquad \omega_3=\omega_2-\omega_1=2\int_{e_2}^{e_1}\frac{dx}{y}\,.
\end{equation}
Any other choice of labelling the roots will yield an integer linear
combination of the periods defined in
\eqn{sec:toolbox_ellipticfunctions:Periods} above, i.e.\
\begin{align}
 2\int_{e_i}^{e_j}&=m_{ij}\omega_1+n_{ij}\omega_2
\end{align}
with $m_{ij}, n_{ij}\in\ZZ$. Hence, the choice of periods corresponds to
choosing different basis vectors for spanning the lattice $\Lambda$.
Correspondingly, the six possible labellings of the roots define six pairs of
periods $(\omega_1,\omega_2)$, whereas the associated different tori are
isomorphic to a particular elliptic curve. 

The second, but related issue is that the complex plane may always be rescaled
by $\omega_1$. Hence, only the ratio $\tau=\omega_2/\omega_1$ matters when
dealing with the $\Lambda$-periodicity, i.e.\ the geometry of the torus.
Therefore, a torus is usually only defined by the modular parameter $\tau$ with
positive imaginary part $\Im(\tau)>0$ while the second period is chosen to be
one. Under scaling $\omega_1$, the
Weierstrass $\wp$ function rescales as 
\begin{align}
  \wp(z;1,\tau)&=\omega_1^{2}\,\wp(\omega_1z;\omega_1,\omega_2)\,.
\end{align}
Choosing $\tau$ in the upper half-plane means that three possible labellings of
the roots $e_i$ are disregarded. The remaining three period ratios obtained
from the different labellings of the roots $e_i$ are related to the $\tau$ in
the upper half plane by modular transformations. 

In summary, the Weierstrass invariants of an elliptic curve completely
define the torus up to modular transformations. Conversely, given two tori with
period ratios $\tau$ and $\tau'$ related by a modular transformation, the
Weierstrass equations of the projective elliptic curves obtained by
$\xi_{\tau\ZZ+\ZZ,E}$ and $\xi_{\tau'\ZZ+\ZZ,E'}$, respectively, are related by
a coordinate transformation of the form 
\begin{equation}
  \label{eqn:coordinatetransformation}
  x\mapsto a^{2} x+b\,,\quad y\mapsto
  a^{3}y+ca^{2} x+d\,,\quad\text{with } a,b,c,d\in\ZC\,,\quad a\neq0\,.  
\end{equation}
Two elliptic curves are called isomorphic if they are related in this way. 
For example, the transformation $x\mapsto a^{2}x$, $y\mapsto a^{3}y$ only
changes the roots by the constant rescaling $e_i\mapsto a^{-2}e_i$, which is an
isomorphism on the complex plane (with respect to addition). Thus, the period
ratio $\tau$ modulo modular transformations uniquely defines the isomorphism
class of elliptic curves defined by $\xi_{\tau\ZZ+\ZZ,E}$ and vice versa.

Note that the computer algebra system \texttt{Mathematica}\footnote{See e.g.\
\rcite{2015arXiv151007818B} for a guideline of the use and the choices of
\texttt{Mathematica}'s built-in conversions from the projective elliptic curve to the torus,
which is based on the conventions of \cite{NIST}.} offers built-in functions
for translating from the projective formulation of elliptic curves to the torus.
In those functions, however, the ambiguities described above are chosen
implicitly. For example, the determination of the half periods $\omega_1/2$,
$\omega_2/2$ with the \texttt{Mathematica} function
\texttt{WeierstrassHalfPeriods[\{g2,g3\}]} relies on a choice of the labellings
of the roots which is selected depending on the signs of the Weierstrass
invariants and the modular discriminant $\Delta=g_2^3-27 g_3^2$.
\smallskip

In order to define a third formulation of the elliptic curve, let us
consider a torus defined by the modular parameter $\tau$ and define $q=e^{2\pi
i\tau}$. The exponential map induces another isomorphism 
\begin{equation}\label{sec:toolbox_ellipticfunctions:toTatecurve}
  \xi_{\tau,q}:\ZC/\left(\tau\ZZ+\ZZ\right)\rightarrow \ZC^{\ast}/q^{\ZZ}\,,\qquad z\mapsto e^{2\pi i z}\,.
\end{equation}
where the codomain $\ZC^{\ast}/q^{\ZZ}$ is called Tate curve~\footnote{See
\rcite{Katz73}, appendix A.1.2, or \rcite{Lur09}, section 4.3, for a more
recent introduction to the Tate curve.} and is endowed with the multiplicative
group structure inherited by the exponential map from addition on the torus.
For example, the representatives $z_1+n_1+ m_1\tau$ and $z_2+n_2+ m_2\tau$ of
$z_1$ and $z_2$ modulo lattice displacements in $\ZC/\Lambda$, where $n_i,
m_i\in \ZZ$, are mapped to the elements
\begin{equation}\label{sec:toolbox_ellipticfunctions:mapToTateCurve}
\xi_{\tau,q}(z_i+n_i+ m_i\tau)=e^{2\pi i(z_i+n_i+ m_i\tau) }=e^{2\pi
iz_i}q^{m_i}, \end{equation} which are representatives of $t_1=e^{2\pi i z_1}$
and $t_2=e^{2\pi i z_2}$, respectively, modulo integer powers of $q$.
Similarly, the sum $z_1+z_2$ modulo lattice displacements is mapped to the
product $t_1 t_2$ modulo integer powers of $q$ on the Tate curve.

The description of elliptic functions on the Tate curve offers a connection to
rational functions on the Riemann sphere $\ZC P^1$ and, in particular, admits a
convenient tool to take the classical limit $q\rightarrow 0$. In order to
reveal this connection to functions on the Riemann sphere, let $f:\ZC
P^1\rightarrow \ZC P^1$ be a non-trivial rational function on the Riemann
sphere satisfying the condition 
\begin{equation}\label{sec:toolbox_ellipticfunctions:condf}
f(0)=f(\infty)=1\,,
\end{equation}
which will be justified in a moment. Note that this class of functions was
already discussed in the context of the classical Bloch relation in
\subsecref{ssec:dilogrel}; the two approaches will be related below. For now,
recall from the discussion of the classical Bloch relation that this ensures
that $f$ is of the form~\eqref{sec:invitation:Bloch:fandCoefs}, i.e.\
\begin{align}
f(t)&=\prod_{i} (t-a_i)^{d_i}\,,
\end{align}
with 
\begin{equation}\label{sec:toolbox_ellipticfunctions:conditionsRationalFctRiemannSphere}
\sum_i d_i=0\,,\qquad \prod_i a_i^{d_i}=1\,.
\end{equation}
Averaging $f$ multiplicatively as follows over the Tate curve yields a function
\begin{align}\label{sec:toolbox_ellipticfunctions:FTateCurve}
F(t)&=\prod_{l\in\ZZ}f(t q^l)\,,
\end{align}
which obeys the transformed $\Lambda$-periodicity condition $F(tq)=F(t)$, cf.\
\eqn{sec:toolbox_ellipticfunctions:mapToTateCurve} for the transformation
behaviour of lattice displacements under the isomorphism $\xi_{\tau,q}$, and,
can therefore be called elliptic on the Tate curve. A discussion of the
properties of such elliptic functions on the Tate curve can be found in
\rcite{BlochRelation}.

The so far unexplained condition~\eqref{sec:toolbox_ellipticfunctions:condf}
can be justified as follows: on the one hand it ensures that in the limit
$q\rightarrow 0$ we recover $f(t)$, on the other hand it implies the
condition~\eqref{sec:toolbox_ellipticfunctions:ZeroPolesCond} on the zeros and
poles $a_i$ of the elliptic generalisation $F$ of $f$ from
\eqn{sec:toolbox_ellipticfunctions:FTateCurve} after the application of the
isomorphism $\xi_{\tau,q}^{-1}$ and the identification $a_i=e^{2\pi i
  A_i}$.~\footnote{We generally denote an elliptic function by a capital Latin
  letter $F$ (while functions on the Riemann sphere are denoted by small Latin
  letters) and its zeros and poles on the Tate curve by small letters $a_i$,
  while their images on the torus are denoted by the corresponding capital
  letters $A_i$. However, for a point on the torus, which corresponds to a
point $t$ on the Tate curve, we usually write $z_t$. The same applies for a
given point $P$ on the elliptic curve and its image $z_P$ on the torus.} As we
will see in \subsecref{sec:toolbox:divisorFunction}, these two conditions
(modulo lattice displacements) are not only necessary, but also sufficient to
be the zeros and poles of some elliptic function. Therefore, we can summarize
that the function $F$ is the elliptic generalisation of $f$ on the Tate curve
and all elliptic functions on the Tate curve can be obtained by this method up
to scaling.

\subsection{Elliptic multiple polylogarithms}\label{ssec:EMP}
There are several descriptions of elliptic generalisations of multiple
polylogarithms, so-called elliptic multiple polylogarithms. Based on the
fact that there is no elliptic function on the torus with just one simple pole,
such generalisations are either not meromorphic or not $\Lambda$-periodic.

However, in Feynman integral calculations one usually chooses to work with
meromorphic rather than single-valued functions. Motivated by this physical
reason, we focus on the holomorphic iterated integrals $\Gt$ on the torus
described in \rcite{Broedel:2017kkb} and relate some other notions of
(single-valued but non-holomorphic) elliptic multiple polylogarithms to these
iterated integrals in \subsecref{sec:connection:TatetoTorus}. In analogy with
the multi-valuedness of the logarithm function, we still refer to these
holomorphic iterated integrals as elliptic multiple polylogarithms defined on
the torus.
\medskip

Consider a torus with periods $1$ and $\tau$, where $\Im(\tau)>0$ as described
in \subsecref{sec:toolbox:ellipticFunction} above, and denote
\begin{equation}\label{sec:toolbox_eMPL:tqw}
  t=e^{2\pi i z}\,,\qquad q=e^{2\pi i \tau}\quad\text{and}\quad  w=e^{2\pi i \alpha}\,. 
\end{equation}
The holomorphic functions $g^{(n)}(z,\tau)$, which satisfy
$g^{(n)}(z,\tau)=g^{(n)}(z+1,\tau)$, constitute the integration kernels of the holomorphic
iterated integrals $\Gt$ described in \rcite{Broedel:2017kkb}. They are
generated by the Eisenstein-Kronecker series \cite{Kronecker,ZagierF}
\begin{equation}\label{sec:toolbox_eMPL:GeneratingEKSeries}
 F(z,\alpha,\tau)=\frac{\theta_1'(0,\tau)\theta_1(z+\alpha,\tau)}{\theta_1(z,\tau)\theta_1(\alpha,\tau)}=\frac{1}{\alpha}\sum_{n\geq 0}g^{(n)}(z,\tau)\alpha^n\,,
\end{equation}
where $\theta_1(z,\tau)$ is the odd Jacobi $\theta$ function. The corresponding
iterated integrals are defined via
\begin{equation}\label{sec:toolbox_eMPL:Gammas}
\Gt(\begin{smallmatrix}n_1&\dots&n_k\\z_1&\dots&z_k\end{smallmatrix}; z,\tau)=\int_0^zdz'\text{ }g^{(n_1)}(z'-z_1,\tau) \Gt(\begin{smallmatrix}n_2&\dots&n_k\\z_2&\dots&z_k\end{smallmatrix}; z',\tau)\,, \qquad \Gt(; z,\tau)=1\,.
\end{equation}
The integration kernels $g^{(n)}(z,\tau)$ are not as abstract as they might
seem at first glance: they are closely related to the doubly-periodic kernels
introduced and used in \rcite{BrownLev}. Furthermore, the kernel
$g^{(1)}(z,\tau)$ can be expressed as 
\begin{align}\label{sec:toolbox_eMPL:g1Zeta}
 g^{(1)}(z,\tau)&=\zeta(z)-2\eta_1 z\,,
\end{align}
where $\zeta(z)$ is the Weierstrass $\zeta$ function introduced in
\eqn{eqn:WeierstrassZeta} and $\eta_1=\zeta(1/2)$ is a quasi-period of the
elliptic curve (cf.~\eqn{sec:toolbox_ellipticfunctions:quasiperiod}). For
$n>1$, the integration kernels can be expressed as polynomials of degree $n$ in
$g^{(1)}(z,\tau)$ and the coefficients depend polynomially on $\wp(z)$ and
$\wp'(z)$, where the first two examples are
\begin{equation}
g^{(2)}(z)=\frac{1}{2}\left(g^{(1)}(z)\right)^2-\frac{1}{2}\wp(z)\,,\qquad g^{(3)}(z)=\frac{1}{6}\left(g^{(1)}(z)\right)^3-\frac{1}{2}\wp(z)g^{(1)}(z)-\frac{1}{6}\wp'(z)\,.
\end{equation}

More suitable for numerical evaluation is the description of the integration
kernels $g^{n}(z,\tau)$ by their $q$-expansions, which are stated in the
\appref{app:qExpansion}.  Furthermore, since the Eisenstein-Kronecker series
satisfies the mixed heat equation $2\pi i \frac{\partial}{\partial
\tau}F(z,\alpha,\tau)=\frac{\partial^{2}}{\partial z\partial
\alpha}F(z,\alpha,\tau)$ \cite{BrownLev}, the integration kernels solve the
partial differential equation
\begin{align}\label{sec:toolbox_eMPL:gnHeatEq}
 2\pi i \frac{\partial}{\partial \tau} g^{(n)}(z,\tau)&=n\frac{\partial}{\partial z} g^{(n+1)}(z,\tau)\,.
\end{align}
At this point, some facts about the regularisation of those iterated integrals
need to be mentioned. Considering the $q$-expansions
\eqref{sec:toolbox_eMPL:qExpansiong0}-\eqref{sec:toolbox_eMPL:qExpansiong2n1}
it is obvious that only the kernel $g^{(1)}$ has a singularity at $z=0$. This
singularity is a simple pole, which renders the iterated integrals
$\Gt(\begin{smallmatrix}n_1&\dots&n_k\\z_1&\dots&z_k\end{smallmatrix};
z,\tau)$ with $z_k=0, n_k=1$ singular. 

Employing the shuffle product of iterated integrals, any singular integral can
be rewritten such that the only singular terms are of the form
$\Gt(\underbrace{\begin{smallmatrix}1 &\dots &1\\ 0&\dots&
0\end{smallmatrix}}_{n}; z,\tau)$. Those singular terms can then be regularised
in a way that preserves the shuffle algebra. Following the prescription
described in \rcite{Broedel:2018iwv} the logarithmic singularity at the lower integration boundary of the
integral $\Gt(\begin{smallmatrix}1\\0 \end{smallmatrix}; z,\tau)$ for $z\neq 0$ can be subtracted by defining its regularised value as follows:
\begin{align}\label{sec:toolbox_eMPL:qExpG1}
\Gt_{\text{reg}}(\begin{smallmatrix}1\\0 \end{smallmatrix}; z,\tau)&=\lim_{\epsilon\to 0}\int_{\epsilon}^z dz'\text{ }g^{(1)}(z',\tau)+\log(1-e^{2\pi i \epsilon})\nonumber\\
&= \log(1-e^{2\pi i z})-\pi i z+4\pi \sum_{k,l>0}\frac{1}{2\pi k}\left(1-\cos(2\pi k z)\right)q^{kl}\,.
\end{align}
Note that while the original integral $\Gt(\begin{smallmatrix}1\\0
\end{smallmatrix}; z,\tau)$ vanishes at $z=0$ and is divergent at any other
value of $z$, the regularised version
$\Gt_{\text{reg}}(\begin{smallmatrix}1\\0 \end{smallmatrix};
z,\tau)$ is finite at any $z\neq 0$, but has a logarithmic divergence at $z=0$.
The prescription can be easily generalized to iterated integrals with multiple
successive divergent entries: with the generalisation
\begin{align}
\Gt_{\text{reg}}(\underbrace{\begin{smallmatrix}1 &\dots &1\\ 0&\dots& 0\end{smallmatrix}}_{n}; z,\tau)&=\frac{1}{n!}\left(\Gt_{\text{reg}}(\begin{smallmatrix}1\\0 \end{smallmatrix}; z,\tau) \right)^n\,.
\end{align}
From here on -- unless stated otherwise -- we denote by
$\Gt(\underbrace{\begin{smallmatrix}1 &\dots &1\\ 0&\dots&
0\end{smallmatrix}}_{n}; z,\tau)$ its regularised value and refer to the
unregularised version as follows
\begin{equation}
  \label{eqn:unregGamma}
  \Gt_\unreg(\underbrace{\begin{smallmatrix}1 &\dots &1\\
  0&\dots& 0\end{smallmatrix}}_{n}; z,\tau)=\int_0^z dz'
  g^{(1)}(z',\tau)\Gt_\unreg(\underbrace{\begin{smallmatrix}1
  &\dots &1\\ 0&\dots& 0\end{smallmatrix}}_{n-1}; z',\tau)\,.
\end{equation}

In \subsecref{sec:connection:TatetoTorus} below, a particular class of the
iterated integrals $\Gt$ is discussed in detail, which is the one given by the
regularised elliptic polylogarithms of the form
\begin{align}\label{sec:toolbox.ellPolylog}
\Gt(\underbrace{\begin{smallmatrix}0 &\dots &0&m\\ 0&\dots &0&0 \end{smallmatrix}}_{n}; z,\tau)\,,
\end{align}
where $n,m\geq 1$. The numerical evaluation of this class of functions is
particularly simple, since their $q$-expansions can directly by given by n-fold
integration of the $q$-expansions \eqref{sec:toolbox_eMPL:qExpansiong2n} and
\eqref{sec:toolbox_eMPL:qExpansiong2n1} of the integration kernels
$g^{(m)}(z,\tau)$ for $m>1$ and the $q$-expansion \eqref{sec:toolbox_eMPL:qExpG1}
of $\Gt_{\text{reg}}(\begin{smallmatrix}1\\0 \end{smallmatrix}; z,\tau)$. The
results are given in equations \eqref{app:qExp:G12n}-\eqref{app:qExp:G2m12n1}.

The values of the (regularised) iterated integrals at $z=1$ are particularly
interesting since they can be used to define a class of elliptic multiple zeta
values \cite{Broedel:2014vla, Matthes:Thesis}. Ordinary zeta values $\zeta_m$,
for $m>1$, are defined as the values of the corresponding polylogarithms
evaluated at one
\begin{align}
\zeta_m&=\Li_m(1)\,.
\end{align}
Analogously, we consider the elliptic zeta values defined by evaluation at one
of the above class of elliptic polylogarithms
\begin{align}\label{sec:toolbox:ezv}
\omega_n(m;\tau)&=\Gt(\underbrace{\begin{smallmatrix}0 &\dots &0&m\\ 0&\dots &0&0 \end{smallmatrix}}_{n}; 1,\tau)\,.
\end{align}
Note that this class of elliptic zeta values agrees with the definition of
elliptic multiple zeta values in \rcite{Broedel:2014vla}. Furthermore, the even
zeta values are related to the elliptic zeta values according to 
\begin{align}
\omega_{1}(2m;\tau)&=-2\zeta_{2m},
\end{align}
which can be seen from the $q$-expansion \eqref{app:qExp:G2m2n1}

\subsection{The divisor function}
\label{sec:toolbox:divisorFunction}
The last paragraph in \subsecref{sec:toolbox:ellipticFunction} was devoted to
illuminating the relation between an elliptic function $F$ on the Tate curve
and the corresponding rational function $f$ on the Riemann sphere. As mentioned
at that point, this is closely connected to the formulation of the classical
Bloch relation in terms of divisors of such rational functions $f$ discussed in
\subsecref{ssec:dilogrel}. The combination of these two considerations leads to
the formulation of the elliptic Bloch relation using the concept of divisors of
elliptic functions.
\smallskip

The group of divisors $\Div(\ZC/\Lambda)$ of the torus $\ZC/\Lambda$ is the
free abelian group $\CF_{\ZC/\Lambda}$ generated by the points on the torus
$\ZC/\Lambda$ and similarly for the projective elliptic curve as well as the Tate curve,
which are related via the isomorphisms introduced in
\eqns{sec:toolbox_ellipticfunctions:xiLE}{sec:toolbox_ellipticfunctions:toTatecurve}
above. Hence, a generic divisor is a finite sum of the form 
\begin{align}
\sum_i n_i (z_i)\in \Div(\ZC/\Lambda)\,,\quad\sum_i n_i
(P_i)\in \Div(E(\ZC))\quad\text{or}\quad\sum_i n_i (t_i)\in \Div(\ZC^{\ast}/q^{\ZZ})\,,
\end{align}
respectively, with $n_i\in\ZZ$, $z_i\in\ZC/\Lambda$,
$\xi_{\Lambda,E}(z_i)=P_i\in E(\ZC)$ and $\xi_{q,E}(z_i)=t_i\in
\ZC^{\ast}/q^{\ZZ}$. Analogously to the case of rational functions on the
Riemann sphere, cf.\ \eqn{eqn:defdivisorRiemannSphere}, and according to the
general definition~\eqref{eqn:defdivisor}, the divisor of an elliptic function
$F$ captures the structure of the zeros and poles of $F$ and is defined by
\begin{align}\label{sec:toolbox_divisors:DivF}
  \Div(F)&=\sum_{z\in P_{\Lambda}}\ord_{z}(F)(z)\in \Div(\ZC/\Lambda)
\end{align}
where the sum runs over all points in the fundamental domain $P_\Lambda$ of $\ZC/\Lambda$.

According to the identification of elliptic functions on
the torus with rational functions on the projective elliptic curve and elliptic
functions on the Tate curve alluded to above, the divisor
\eqref{sec:toolbox_divisors:DivF} of an elliptic function $F$ can be translated
by the usual isomorphisms to the projective formulation and the Tate curve via 
\begin{equation}
  \Div(F)=\sum_{P\in E(\ZC)}\ord_{P}(F)(P)\in \Div(E(\ZC))
\end{equation}
and
\begin{equation}\label{sec:toolbox_divisors:DivFTateCurve}
  \Div(F)=\sum_{t\in \ZC^{\ast}/q^{\ZZ}: |q|<|t|\leq 1}\ord_{t}(F)(t)\in \Div(\ZC^{\ast}/q^{\ZZ})\,,
\end{equation}
where the orders of the rational function $F(x,y)$ and the elliptic function on
the Tate curve $F(t)$ are defined by the order of the elliptic function $F(z)$
on the torus at the corresponding points.
For two divisors $D=\sum_i d_i (P_i)$
and $E=\sum_j e_j (Q_j)$, a new divisor
\begin{align}
D^{-}=\sum_i d_i (-P_i)
\end{align}
and the binary product
\begin{align}
D\ast E&=\sum_{i,j}d_i e_j (P_i+Q_j)
\end{align}
can be defined, such that a divisor of the form
\begin{align}\label{sec:toolbox_divisors:etaF}
\eta^{\kappa}_F&=\Div(F)\ast\Div(\kappa-F)^{-}\,,
\end{align}
can be associated to any rational function $F$, and similarly on the torus and
the Tate curve, respectively. In the above definition, $\kappa\in\ZC$ is a
scaling parameter, which needs to equal one in the classical situation
described in \subsecref{ssec:dilogrel}.
The divisor $\eta_F^{\kappa}$ associated to an elliptic function $F$ plays an
important role in later sections of this article and in particular in the
formulation of the elliptic Bloch relation in \subsecref{sec:toolbox:ellBloch}.

The fact that two elliptic functions are equal up to scaling if they have the
same zeros and poles (counted with multiplicities) translates to the condition
of having the same divisors. On the other hand, a divisor $D$ is said to be
\textit{principal}, if there exists an elliptic function $F$ such that
$D=\Div(F)$. Now, we can properly rephrase the last two sentences in
\subsecref{sec:toolbox:ellipticFunction}: It turns out that a divisor $D$ is
principal if and only if it is of the form $D=\sum_i d_i (A_i)$ where
\begin{equation}\label{sec:toolbox_ellipticfunctions:conditionsPrincipalDivisor}
\sum_i d_i =0\,,\qquad \sum_i d_i A_i\in\Lambda\,.
\end{equation}
A proof of this equivalence can be outlined as follows (cf.\ \rcite{deShalit}):
the necessary implication follows from the conditions
\eqref{sec:toolbox_ellipticfunctions:conditionsEllipticFctTorus} on the zeros
and poles of an elliptic function. In order to prove sufficiency, first note that any
divisor $D=\sum_i d_i (A_i)$ satisfying equations
\eqref{sec:toolbox_ellipticfunctions:conditionsPrincipalDivisor} can be written
as a linear combination of divisors of the form $(A_1)+(A_2)-(0)-(A_1+A_2)$.
Now, consider elliptic functions of the form
\begin{align}\label{sec:toolbox_ellipticfunctions:Flambda}
F_{\lambda}(z)&= (1-\lambda)\frac{\wp'(z)-\wp'(-A_1-A_2)}{\wp(z)-\wp(-A_1-A_2)}+\lambda\,.
\end{align}
In \rcite{deShalit} it is shown that one can always find a complex parameter
$\lambda$ such that the divisor associated to the above function reads:
\begin{align}\label{sec:toolbox_ellipticfunctions:spanningFlambda}
\Div(F_{\lambda})&= (A_1)+(A_2)-(0)-(A_1+A_2)
\end{align}
and $D$ can indeed be written as a divisor of an elliptic function
$D=\sum_je_j\Div(F_{\lambda_j})=\Div\left(\prod_{j}F_{\lambda_j}^{e_j}
\right)$, since the divisor function satisfies $\Div(F_1
F_2)=\Div(F_1)+\Div(F_2)$ for two elliptic functions $F_1$ and $F_2$.
Alternatively, the elliptic function $F$ such that $\Div(F)=D$ can be
constructed by means of the Weierstrass $\sigma$ function as in
\eqn{sec:toolbox_ellipticfunctions:FandSigma}.

\subsection{The elliptic Bloch relation}
\label{sec:toolbox:ellBloch}
After having introduced the mathematical background for elliptic curves and
elliptic functions in the previous subsections, the elliptic version of Bloch's
dilogarithm identity~\eqref{sec:invitation:Bloch:BlochRel} can be discussed: in
order to do so, an elliptic generalisation of the Bloch-Wigner function $\DD$
defined in \eqn{eqn:BlochWigner} is required. Since the Bloch-Wigner function
satisfies $\DD(0)=\DD(\infty)=0$, the elliptic generalisation on the Tate curve
in terms of an infinite product as in
\eqn{sec:toolbox_ellipticfunctions:FTateCurve} is not applicable. However, an
additive average over the Tate curve yields
\begin{align}\label{sec:toolbox_Bloch:DE}
  \DE(t,q)&=\sum_{l\in\ZZ}\DD(tq^l)\,.
\end{align}
This function $\DE$ is referred to as the \textit{elliptic Bloch-Wigner function}
\cite{BlochRelation}. It inherits some symmetry properties from the
classical Bloch-Wigner function $\DD$, in particular the inversion relation
\begin{align}\label{sec:toolbox_ellBloch:inversion}
\DE(t^{-1},q)&=-\DE(t,q)
\end{align}
and the duplication relation 
\begin{align}\label{sec:toolbox_ellBloch:duplication}
\DE(t^2,q)&=2\left(\DE(t,q)+\DE(t\sqrt{q},q)+\DE(-t,q)+\DE(-t\sqrt{q},q)\right)
\end{align}
from \eqns{sec:invitation:BlochGroup:symD}{sec:invitation:BlochGroup:dupD},
respectively. The elliptic version of Bloch's dilogarithm identity
\eqref{sec:invitation:Bloch:BlochRel} is that for any elliptic function $F$,
any $\kappa\in\ZC$ and the divisors $\Div\left(F\right)=\sum_i d_i (a_i)$ and
$\Div\left(\kappa-F\right)=\sum_j e_j (b_j)$ expressed on the Tate curve, the
following identity holds
\begin{align}
\sum_{i,j}d_i e_j\DE\left(\frac{a_i}{b_j},q\right)&=0\,,
\end{align}
which takes the form
\begin{align}\label{sec:toolbox_ellBloch:ellBlochRelation}
\DE\left(\eta_F^{\kappa},q\right)&=0\,,
\end{align}
when expressed in terms of $\eta^{\kappa}_F$, the divisor defined in
\eqn{sec:toolbox_divisors:etaF}. Here again $\DE$ is extended by linearity to
the group of divisors. The above identity is referred to as the
\textit{elliptic Bloch relation}.  

Bloch proves this statement starting from the classical case with a rational
function $f$ satisfying $f(0)=f(\infty)=1$, approximating its elliptic
generalisation $F$ on the Tate curve, constructed according to
\eqn{sec:toolbox_ellipticfunctions:FTateCurve}, by $F_N=\prod_{|l|\leq
N}f(tq^l)$ and an error estimation as $N\rightarrow\infty$
\cite{BlochRelation}. Note that in contrast to the classical case, cf. the
second equation in~\eqref{eqn:defdivisorRiemannSphere}, the constant $\kappa$
defined in \eqn{sec:toolbox_divisors:etaF} does not need to equal one. But
since any scaling of the elliptic function $F$ is allowed, this condition is
redundant and the elliptic Bloch relation can be stated without loss of
generality with $\kappa=1$. In the classical limit $q\rightarrow 0$ the Tate
curve $\ZC^{\ast}/q^{\ZZ}$ degenerates to $\ZC^{\ast}$, and, simultaneously,
the elliptic Bloch-Wigner function degenerates to its classical version, the
Bloch-Wigner function.  Finally, for an elliptic function on the Tate curve of
the form
\begin{equation}
F=\lim_{N\rightarrow\infty}F_N 
\end{equation}
with $F_N$ and $f$ (scaled) as before, the elliptic Bloch relation degenerates to the
classical Bloch relation, cf.\ \eqref{sec:invitation:Bloch:BlochRel},
\begin{equation}\label{eqn:BlochLimit}
\DE\Big(\Div(F)\ast\Div(1-F)^{-},q\Big)\rightarrow \sum_{i,j}\ord_{a_i}(f)\ord_{b_j}(1-f)\DD\left(\frac{a_i}{b_j}\right)\,.
\end{equation}
However, the above limit, i.e.\ the transition from elliptic to classical
and vice-versa, is subtle, as it can be seen by Bloch's careful proof of the
elliptic Bloch relation in terms of the classical Bloch relation. Another
hint for this subtlety is the following: If on the left-hand side in the limit
\eqref{eqn:BlochLimit} instead of $1-F$, the difference $\kappa-F$ for
$\kappa\neq 1$ is chosen, the left-hand side still vanishes identically
according to \eqn{sec:toolbox_ellBloch:ellBlochRelation}. But the right-hand
side of \eqref{eqn:BlochLimit} in general only vanishes for $1-f$, but not for
$\kappa-f$. Therefore, in such a case the elliptic Bloch relation does not
degenerate to its classical analogue.
\medskip

Analogously to the classical Bloch group $\mathcal{B}_2(\ZC)$, Zagier and Gangl
define the group of functional relations $\mathcal{C}_2(E)$ in the construction
of the elliptic Bloch group
$\mathcal{B}_2(E)=\mathcal{A}_2(E)/\mathcal{C}_2(E)$ as a subgroup of the group
$\mathcal{A}_2(E)$ of "allowable" elements in the free abelian group generated
by points on the elliptic curve $E$, the precise meaning of allowable is
reviewed in \cite{ZagierGangl}. It is generated by the elliptic Bloch
relation~\eqref{sec:toolbox_ellBloch:ellBlochRelation}, the inversion
relation~\eqref{sec:toolbox_ellBloch:inversion} and the duplication
relation~\eqref{sec:toolbox_ellBloch:duplication}, which are expected to form a
full set of relations for the elliptic Bloch-Wigner function on the points in
$\mathcal{A}_2(E)$ \cite{ZagierGangl}. Thus, in contrast to the classical case
discussed in \subsecref{sec:review:BG}, where the five-term identity is
sufficient to generate the subgroup of functional relations for the dilogarithm
on the points in $\mathcal{A}_2(\ZC)$, the elliptic analogue may require a
larger class of functional relations generated by the elliptic Bloch relation.
Using the construction of elliptic functions on the Tate curve described in
\subsecref{sec:toolbox:ellipticFunction}, the class of functional relations of
the elliptic Bloch-Wigner function is parametrised by rational functions on the
Riemann sphere and the complex number $\kappa$. We refer to this procedure to
generate functional identities for the elliptic Bloch-Wigner function as Zagier
and Gangl's method. In the following three subsections we discuss some examples
and show explicit calculations which use the above concepts and in particular
the elliptic Bloch relation.

\subsubsection{First example: a divisor on $y^2=4x^3-4x+1$}
Let us consider the following example\footnote{This is the example
$E_{37}:y^2-y=x^3-x$ in \rcite{ZagierGangl}. However, we directly work in the
Weierstrass form, which can be obtained from the original example by the
coordinate transformation $y\mapsto \frac{y+1}{2}$.} of \rcite{ZagierGangl} to
approve the elliptic Bloch relation. Take the elliptic curve with Weierstrass
equation $ y^{2}=4x^{3}-4 x+1$, i.e.\ $g_2=4$ and $g_3=-1$, and the rational
function \mbox{$F(x,y)=\frac{y+1}{2}$} on $E(\ZC)$. The three zeros of $F$ are
$P=\pc{0}{-1}$, $P_1=\pc{1}{-1}$ and \mbox{$P_2=\pc{-1}{-1}$} and since
$F(\wp(z),\wp'(z))=\frac{\wp'(z)+1}{2}$, the (pull-back of the) rational
function $F$ as an elliptic function on the torus has a triple pole at the
lattice points, such that on the elliptic curve $\ord_{\infty}(F)=-3$. Using
the group addition on the elliptic curve described in
\appref{app:chord-tangent}, one obtains $P_1=2P$ and $P_2=-3P$ and more generally
\begin{alignat}{7}
-3P&=\pc{-1}{-1}\,,{}&{}-2P&=\pc{1}{1},{}&{} -P&=\pc{0}{1}\,, \nonumber\\
P&=\pc{0}{-1},{}&{}2P&=\pc{1}{-1}\,,{}&{} 3P&=\pc{-1}{1}\,, \nonumber\\
4P&=\pc{2}{5}\,,{}&{} 5P&=\pc{\frac{1}{4}}{\frac{1}{4}}\,,{}&{} 6P&=\pc{6}{-29}\,.
\end{alignat}
Therefore, the divisor of $F$ on the projective elliptic curve is 
\begin{align}
\Div(F)&=(P)+(2P)+(-3P)-3(\infty)\,.
\end{align}
Similarly, the divisor of $1-F$ is given by
\begin{align}
\Div(1-F)&=(-P)+(-2P)+(3P)-3(\infty)\,,
\end{align}
such that the associated divisor $\eta_F^{1}$ of $F(x,y)=\frac{y+1}{2}$ defined
in \eqn{sec:toolbox_divisors:etaF} is
\begin{align}\label{sec:toolbox_divisors:ex}
\eta^{1}_F&=(-6P)-6(-3P)+2(-2P)+2(-P)+9(\infty)-6(P)-5(2P)+2(3P)+(4P)\,.
\end{align}
The roots of the elliptic curve \mbox{$y^2=4x^3-4x+1=4(x-e_1)(x-e_2)(x-e_3)$}
are\footnote{We have chosen to display the first ten digits of
numbers only. Of course, all calculations have been performed with much higher
precision.}
\begin{equation}
e_1=0.8375654352\,,\qquad e_2=0.2695944364\,,\qquad e_3=-1.1071598716\,,
\end{equation}
such that according to equations~\eqref{sec:toolbox_ellipticfunctions:Periods}
the periods of the corresponding tori are given by
\begin{equation}
\omega_1=2.9934586462  \,,\qquad \omega_3=2.4513893819i\,,\qquad \omega_2=\omega_1+\omega_3
\end{equation}
with the period ratio
\begin{equation}
\tau=\frac{\omega_2}{\omega_1}=1 + 0.8189153991i\,.
\end{equation}
The point $z$ on the torus with lattice $\Lambda=\omega_1\ZZ+\omega_2\ZZ$ which
corresponds to $P=\pc{0}{-1}$ is determined by Abel's map
\eqref{sec:toolbox_ellipticfunctions:Abelsmap}
\begin{align}
\tilde{z}_P&= \int_0^{\infty}\frac{dx}{\sqrt{4(x-e_1)(x-e_2)(x-e_3)}}\nonumber\\
&=2.0638659408 + 1.2256947056i\,.
\end{align}
The rescaled point corresponding to $P$ in the fundamental parallelogram of the
torus defined by $\tau$ is
\begin{equation}
  z_P=\frac{\tilde{z}_P}{\omega_1}\!\!\!\mod (\ZZ+\tau\ZZ)=0.6894586481 + 0.4094577022 i
\end{equation}
which maps to 
\begin{equation}
t_P=e^{2\pi i z_P}=-0.0283399159 - 0.0708731874 i
\end{equation}
on the Tate curve, while the parameter $q$ takes the value
\begin{equation}
q=e^{2\pi i \tau}=0.0058261597\,.
\end{equation}
For practical purposes, let us define the following approximation of $\DE(t,q)$:
\begin{align}
  \label{eqn:approx}
  \DD_k^{\E}(t,q)&=\sum_{l=-k}^k\DD(tq^l),
\end{align}
which allows to control the accuracy of convergence depending on the number of
terms $2k+1$. According to the elliptic Bloch relation, our example $\eta_F^1$
from \eqn{sec:toolbox_divisors:ex} is expected to satisfy
\begin{align}\label{Ell:EBW:BlochRelExample}
  -8\DE\left(t_P,q\right)-7\DE\left(t_P^{2},q\right)+8\DE\left(t_P^{3},q\right)+\DE\left(t_P^{4},q\right)-\DE\left(t_P^{6},q\right)&=0\,,
\end{align}
where we already used the inversion relation
\eqref{sec:toolbox_ellBloch:inversion} to simplify the evaluation of the
divisor $\eta_F^1$. 

Using the approximation~\eqref{eqn:approx} for numerical evaluation of the
above equation, we find agreement up to $10^{-7}$ already for $k=10$.
For other permutations of labelling the roots $e_i$ the elliptic Bloch relation holds as
well, as can be tested numerically.

\subsubsection{Second example: lines on the projective elliptic curve}
As a second example, consider a line on the projective elliptic curve, i.e.\ a rational
function of the form
\begin{align}\label{Ell:EBW:Labc}
 L_{a,b,c}(x,y)&=ax+by+c
\end{align}
with $a$ or $b$ not equal to zero, $x$ and $y$ satisfying $y^{2}=4x^3-g_2
x-g_3$. The poles of $L_{a,b,c}$ are located at $\infty$ with multiplicities
$2$ if $b=0$ and $3$ otherwise (They correspond to the double and triple pole
of $x=\wp(z)$ and $y=\wp'(z)$, respectively. See the discussion around
\eqn{sec:toolbox_divisors:DivFTateCurve}.). The zeros of $L_{a,b,c}$ can be
determined explicitly as algebraic functions of the coefficients $a$, $b$ and
$c$, they satisfy the cubic equation 
\begin{equation}\label{372}
\left(\frac{a}{b}x+\frac{c}{b}\right)^2=4x^3-g_2x-g_3\,,\qquad y=-\frac{a}{b}x-\frac{c}{b}\,.
\end{equation}
Similarly, the zeros of $1-L_{a,b,c}$ satisfy
\begin{equation}\label{373}
\left(\frac{a}{b}x+\frac{c-1}{b}\right)^2=4x^3-g_2x-g_3\,,\qquad y=-\frac{a}{b}x-\frac{c-1}{b}\,.
\end{equation}
These cubic equations can be solved by radicals, such that
$\Div\left(L_{a,b,c}\right)$ and $\Div\left(1-L_{a,b,c}\right)$ depend
algebraically on the coefficients $a$, $b$ and $c$. Furthermore, since the
group addition on the projective formulation $\E(\ZC)$ of the elliptic curve
also only involves algebraic operations, as can be seen from the explicit
equations in the \appref{app:chord-tangent}, the resulting divisor
$\eta_{L_{a,b,c}}^1$ expressed on the projective elliptic curve is algebraic in $a$, $b$
and $c$. However, applying Abel's map and translating the solutions to the Tate
curve, where the elliptic Bloch-Wigner
relation~\eqref{sec:toolbox_ellBloch:ellBlochRelation} is defined (so far),
generally turns the zeros into integral expressions of the variables $a$, $b$
and $c$. Thus, in order to obtain a functional relation with algebraic
arguments, the elliptic Bloch-Wigner relation has to be expressed on the torus
and ultimately on the projective elliptic curve, which is done in
\secref{sec:connection}.

Alternatively, instead of solving for the zeros starting from a particular
choice of parameters $a,b$ and $c$, one could as well choose three zeros
directly and obtain another three from \eqns{372}{373}. However, since there
are only three free parameters (the lines $L_{a,b,c}$ and $1-L_{a,b,c}$ have
the same slope), i.e.\ three roots which determine the remaining roots in terms
of at least one non-linear equation, this still involves some non-trivial
algebraic dependencies of the arguments in the final functional relation
induced by the elliptic Bloch-Wigner relation on the Tate curve. 

Let us illustrate the above argumentation by an example: if e.g.\ the zeros
$P_1$, $P_2$ of $L_{a,b,c}$ and one zero $Q_1$ of $1-L_{a,b,c}$ are called
$\pc{x_1}{y_1},\pc{x_2}{y_2}$ and $\pc{x_3}{y_3}$, respectively, the third zero of
$L_{a,b,c}$ is $P_3=-P_1-P_2$ according to the definition of addition in
\appref{app:chord-tangent} and thus algebraic in $x_i,y_i$, $i=1,2$. 
This ensures that the $x$-coordinates of the divisor
of $L_{a,b,c}$ are very simple. One of the two remaining zeros, $Q_2$ and $Q_3$,
of $1-L_{a,b,c}$ is defined by $Q_3=-Q_1-Q_2$. But the last zero, $Q_2$, is still
determined by a quadratic equation in terms of $Q_1$, such that mapping the
divisor to the Tate curve yields again a non-algebraic functional relation. 

\subsubsection{Third example: the five-term identity}
As a last example, consider $f$ given by \eqn{sec:invitation:Bloch:5termf}
which is the rational function generating the classical five-term identity when
inserted in the classical Bloch relation~\eqref{sec:invitation:Bloch:BlochRel}.
It satisfies $f(0)=f(\infty)=1$, such that its elliptic generalisation on the
Tate curve $F(t)=\prod_{l\in\ZZ}f(tq^l)$
(cf.~\eqn{sec:toolbox_ellipticfunctions:FTateCurve}) with the following
associated divisor can be formed
\begin{align}
\Div(F)&=(a)+(a')+(bb')-(b)-(b')-(aa')\,,
\end{align}
where all variables have been defined after \eqn{sec:invitation:Bloch:5termf}.
Since the elliptic Bloch relation $\DE(\eta^1_F,q)=0$ degenerates to the
classical one $\DD(\eta_F)=0$ for $q\rightarrow 0$, it can be expected that
the elliptic Bloch relation evaluated for $F$ generates an elliptic analogue of
the five-term identity \cite{ZagierGangl}: it is an elliptic
dilogarithm identity generated by the same rational function $f$ on the Riemann
sphere, which implies the five-term identity.
In order to write it down explicitly, the zeros and poles of $1-F$ need to be
known. While the poles are the same as the ones of $F$, finding the zeros of
$1-F$ is a major obstacle, which was already encountered in the example with
the lines in the previous paragraph.  While the cubic equation for the zeros of
$1-F$ in the line example could be solved by radicals, the current situation
involves a quintic equation in the $x$-coordinate of the projective elliptic
curve, which cannot be solved in general.  Correspondingly, an elliptic
analogue of the five-term identity can in general not be written down
explicitly in terms of algebraic arguments.  

The quintic equation is obtained as follows: following the argumentation at the
end of \subsecref{sec:toolbox:divisorFunction}, there exist
$\lambda_a,\lambda_b\in\ZC$, such that
\begin{align}
\Div\left(\frac{F_{\lambda_a}}{F_{\lambda_b}}\right)&=\Div(F)\,,
\end{align}
where $F_{\lambda}$ is a rational function on the projective elliptic curve of the form
\eqref{sec:toolbox_ellipticfunctions:Flambda} and the divisor of $F$ is
expressed on the projective elliptic curve via the usual isomorphisms 
\begin{equation}
  \label{eqn:}
   a\in\ZC^{\ast}/q^{\ZZ}\mapsto \xi_{\tau,q}^{-1}(a)=A\in\ZC/\left(\tau\ZZ+\ZZ\right)\mapsto \xi_{\tau\ZZ+\ZZ,E}(A)=P_a=\pc{x_a}{y_a}\in E(\ZC)\,.
\end{equation}
Performing the translation, $F$ can be expressed on the projective elliptic curve as the
rational function
\begin{equation}\label{sec:toolbox_ellipticfunctions:Ffiveterm}
F(x,y)=\frac{1}{\kappa}\frac{F_{\lambda_a}(x,y)}{F_{\lambda_b}(x,y)}=\frac{(1-\la_a)(y-y_{\frac{1}{aa'}})(x-x_{\frac{1}{bb'}})+\la_a(x-x_{\frac{1}{bb'}})(x-x_{\frac{1}{aa'}})}{\ka(1-\la_b)(y-y_{\frac{1}{bb'}})(x-x_{\frac{1}{aa'}})+\ka\la_b(x-x_{\frac{1}{bb'}})(x-x_{\frac{1}{aa'}})}
\end{equation}
for some scaling factor $\ka\in\ZC$, $x_a=\wp(A)$ and $y_a=\wp'(A)$. The poles
of $1-F$ are the same as the ones of $F$, i.e.\ $P_a$, $P_{a'}$ and $P_{bb'}$.
The zeros of $1-F$ are determined by $\kappa F_{\la_b}-F_{\la_a}=0$, which translates by
the Weierstrass equation to the quintic equation
mentioned above. Since $a$ and $b$ are variables as well as based on the fact
that $\la_a$ and $\la_b$ depend non-trivially on $a$ and $b$ the resulting
quintic equation is not solvable by radicals in general. Even though the elliptic analogue
of the five-term identity generated by the elliptic Bloch relation can not be
written down explicitly, it may however be described implicitly as
above.
\smallskip

In summary, the elliptic Bloch relation
\eqref{sec:toolbox_ellBloch:ellBlochRelation} generates many (conjecturally
all) functional relations of the elliptic
Bloch-Wigner function. But for most of these relations, the relevant divisor
$\eta_F^1$ can not be expressed as a linear combination of variables depending
algebraically on each other. The most notable exceptions are the divisors
$\eta_{L_{a,b,c}}^1$ generated by lines expressed on the projective elliptic curve. It
is this situation, it would still be possible to explicitly write down
functional relations. However, they are by no means nice and elucidating and we
will thus refrain from doing so.  Instead, all relations are going to be
formulated on the torus in order to be contrasted with relations between
elliptic polylogarithms on the torus introduced in \subsecref{ssec:EMP}.

\section{Elliptic multiple polylogarithms: connecting two languages}
\label{sec:connection}
The aim of this section is to translate the elliptic Bloch relation
\eqref{sec:toolbox_ellBloch:ellBlochRelation} from the Tate curve to the torus
and to the projective elliptic curve, respectively. From the previous section, it is
known how elliptic functions and their divisors can be translated between the
three descriptions of an elliptic curve. Hence, we are left with the
translation of the elliptic Bloch-Wigner function $\DE$, defined in
\eqn{sec:toolbox_Bloch:DE}, to the iterated integrals $\Gt$ on the torus, which
will be performed in \subsecref{sec:connection:TatetoTorus}.  Moreover, a
further translation will allow to express the Bloch-Wigner function in the
projective formulation of the elliptic curve. 

In \subsecref{sec:connection:HigherEllPolylogs} we show how these translations
can be generalised to two families of elliptic polylogarithms of higher weight,
both of which include the elliptic Bloch-Wigner function. Finally, in
\subsecref{sec:BlochRelationTorus} we combine
our results and write down the elliptic Bloch relation on the torus and on the
projective elliptic curve explicitly. Moreover, we discover some holomorphic functional
relations on the torus which imply the elliptic Bloch relation and thereby give
an interpretation of the elliptic Bloch relation in terms of the elliptic
symbol calculus. 

\subsection{The elliptic dilogarithm: from the Tate curve to the torus}
\label{sec:connection:TatetoTorus}
We begin with establishing a connection between the iterated integrals
$\Gt$ defined in \eqn{sec:toolbox_eMPL:Gammas} above and the sum
\begin{align}\label{Ell:EBW:Enm}
\E_{n,m}(t,s,q)&=-\left(\ELi_{n,m}(t,s,q)-(-1)^{n+m}\ELi_{n,m}(t^{-1},s^{-1},q)\right),
\end{align}
where the objects 
\begin{align}
\ELi_{n,m}(t,s,q)&= \sum_{k,l>0}\frac{t^k}{k^n}\frac{s^l}{l^m}q^{kl}
\end{align}
have been introduced and described in \rcite{Adams:2016xah}. 

In the end, it will turn out that the value $\E_{n,-m}(t,1,q)=\E_{n,-m}(e^{2
    \pi i z},1,q)$ of $\E_{n,-m}$ defined on the Tate curve is, up to
    polynomials in $z$, equal to the $n$-fold iterated integral of the
    integration kernel $g^{(m+1)}(z,\tau)$, i.e.\
    $\Gt(\underbrace{\begin{smallmatrix}0&\dots&m+1\\
0&\dots&0\end{smallmatrix}}_{n}; z,\tau)$, which is an iterated integral
    defined on the torus.
    
In order to show this, the case $m=0$ is discussed first, for which the definition
\begin{align}\label{Ell:EBW:BarEnDef}
\E_n(t,s,q)&=
-\left(\frac{1}{2}\Li_n(t)-(-1)^n\frac{1}{2}\Li_n(t^{-1})\right)+\E_{n,0}(t,s,q)
\end{align}
turns out to be useful. In terms of the variables $t$, $q$ and $w$ defined in
\eqref{sec:toolbox_eMPL:tqw}, the Eisenstein-Kronecker series
\eqref{sec:toolbox_eMPL:GeneratingEKSeries} can be rewritten as \cite{Weil76}
\begin{align}\label{Ell:EBW:EisensteinKroneckerXWQ}
F(t,w,q)&
=-2\pi i\left(\frac{t}{1-t}+\frac{1}{1-w}+\sum_{k,l>0}(t^k w^l-t^{-k}w^{-l})q^{kl}\right),
\end{align}
such that from the limit
$g^{(1)}(z,\tau)=\lim_{\alpha\rightarrow 0} \left(F(z,\alpha,\tau) -
\frac{g^{(0)}(z,\tau)}{ \alpha}\right)$ a straightforward calculation
implies\footnote{This calculation has been pointed out in ref.\
\cite{Panzer17} and motivated to consider the generalisations for
$E_{n,-m}(t,1,q)$ with $n,m>0$ described in the following parts of this subsection. Similar considerations can be found in \rcite{Levin}.
}
\begin{align}\label{Ell:EBW:BarE0g1}
 \E_0(t,1, q)&=\frac{1}{2\pi i}g^{(1)}(z,\tau)\,.
\end{align}
The iterated integrals $\Gt$ on the torus may be recovered using the partial
differential equation
\begin{align}\label{sec:connecting_eMPL:deqEn(z)}
 \frac{\partial}{\partial z}\E_n(z,1,\tau)&= 2\pi i \E_{n-1}(z,1,\tau)\,,
\end{align}
where the function $\E_n$ is pulled-back to the torus by the exponential map.
This leads to the following integral representation of $\E_1(t,1,q)$
\begin{align}
 \E_1(t,1,q)&=\lim_{\epsilon\to 0}\int_{\epsilon}^{z}dz'\frac{\partial}{\partial z'}\E_1(z',1,\tau)+\E_1(e^{2\pi i \epsilon},1,q)\nonumber\\
&=\Gt(\begin{smallmatrix}1\\0\end{smallmatrix}; z,\tau)-2\ELi_{1,0}(1,1,q)+\frac{\pi i}{2}\,,
\end{align}
where $\Gt(\begin{smallmatrix}1\\0\end{smallmatrix}; z,\tau)$ is the
  regularised integral (see \subsecref{ssec:EMP}). Note that the logarithmic singularity of
$\Gt_\unreg(\begin{smallmatrix}1\\ 0\end{smallmatrix};
z,\tau)=\int_0^z dz' g^{(1)}(z',\tau)$ cancels the singular contribution
$\Li_1(1)$ of $\E_{1}(1,1,q)$, leaving only a phase shift $\frac{\pi i}{2}$
caused by the different directions of the paths approaching the singularity of
$\Li_1(1)$. For $n>1$, there is no singularity at all if the regularised
iterated integrals are used, since for $n>1$
\begin{align}\label{sec:connecting_eMPL:En1}
 \E_n(1,1,q)&=
 -\left(\frac{1}{2}(1-(-1)^n)\Li_n(1)+(1-(-1)^n)\ELi_{n,0}(1,1,q)\right)\nonumber\\
 &=\begin{cases}
 0&n\text{ even}\\ -\zeta_n-2\ELi_{n,0}(1,1,q)&n\text{ odd}
 \end{cases}
\end{align}
is finite as well. This can be seen by considering \eqn{Ell:EBW:Enm} for $s,t=1$: 
\begin{equation}
\ELi_{n,0}(1,1,q)= \sum_{k>0}\frac{q^k}{(1-q^k)k^n}=-2i\sum_{k>0}\frac{e^{k\pi i \tau}}{\sin\left(k\pi  \tau\right)k^n}\,.
\end{equation}
Fortunately, the calculation of the above series can be circumvented by
considering the integral representation of $\E_2$ on the torus: taking into
account that $\E_2(1,1, q)=0$, a representation of $\E_2(t,1, q)$ can be
obtained by the following calculation
\begin{align}
\E_2(t,1, q)&=
2\pi i \int_0^{z}dz'\text{ }\E_1(z',1,\tau)\nonumber\\
&=2\pi i \Gt(\begin{smallmatrix}0&1\\0&0\end{smallmatrix}; z,\tau)+2\pi i \left(\frac{\pi i}{2}-2\ELi_{1,0}(1,1,q)\right)z\,.
\end{align}
Evaluation at $z=1$ of \eqn{Ell:EBW:E2Gamma} together with
\eqn{sec:connecting_eMPL:En1} yields the value of $\ELi_{1,0}(1,1,q)$ in terms
of the regularised iterated integrals
\begin{align}\label{sec:connecting_eMPL:ELi1}
2  \ELi_{1,0}(1,1,q)&=\omega_2(1;\tau)+\frac{\pi i}{2}\,,
\end{align}
such that 
\begin{align}\label{Ell:EBW:E2Gamma}
\E_2(t,1, q)&=
2\pi i \int_0^{z}dz'\text{ }\E_1(z',1,\tau)\nonumber\\
&=2\pi i \left(\Gt(\begin{smallmatrix}0&1\\0&0\end{smallmatrix}; z,\tau)-\omega_2(1;\tau)z\right)\,.
\end{align}
Turning back to the functions $\E_{n}(t,1,q)$, one finds recursively that for
$n\geq 1$
\begin{align}\label{sec:connecting_eMPL:EnandGammaPn}
\E_n(t,1,q)
&=(2\pi i)^{n-1} \Gt(\underbrace{\begin{smallmatrix}0 &\dots &0&1\\ 0&\dots&0& 0\end{smallmatrix}}_{n}; z,\tau)+P_n(z,q)\,,
\end{align}
where $P_n(z,q)$ is the polynomial of degree $n-1$ in $z$
\begin{align}
 P_n(z,q)&=-(2\pi i)^{n-1}\omega_2(1;\tau)\frac{z^{n-1}}{(n-1)!}+\sum_{j=2}^{n}(2\pi i)^{n-j} \E_{j}(1,1,q)\frac{z^{n-j}}{(n-j)!}.
\end{align}
In \eqn{sec:connecting_eMPL:EnandGamma}, the whole $z$ dependence of
$\E_n(t,1,q)$ is expressed solely in terms of (polynomials of) the regularised
iterated integrals $\Gt$ with at most weight one, since
$z=\Gt(\begin{smallmatrix}0\\0\end{smallmatrix}; z,\tau)$. The integration
constants $\E_{j}(1,1,q)$, given in \eqref{sec:connecting_eMPL:En1} and
appearing in the polynomial $P_n(z,q)$, can be expressed as a linear
combination of elliptic zeta values. The result can be obtained recursively by
evaluation of \eqn{sec:connecting_eMPL:EnandGamma} at one. The full calculation
is shown in the \appref{app:eMZV} and results in the explicit expression for
$n>1$
\begin{align}\label{sec:connecting_eMPL:explicitEn1}
\E_n(1,1,q)&=
-\left(\frac{1}{2}(1-(-1)^n)\Li_n(1)+(1-(-1)^n)\ELi_{n,0}(1,1,q)\right)\nonumber\\
&=\begin{cases}
(2\pi i)^{n-1}\sum_{k=0}^{\frac{(n-1)}{2}}d_{2k+1}\omega_{n+1-2k}(1;\tau)&n\text{ odd}\\
0&n\text{ even,}
\end{cases}
\end{align}
cf.\ \eqn{app:intConst:EnExplicit}, where $d_k$ is the sequence defined by 
\begin{equation}
d_k=\begin{cases}
-1&k=1\\
0&k\text{ even}\\
-\frac{d_1}{k!}-\frac{d_{3}}{(k-2)!}-\dots-\frac{d_{k-2}}{3!}&k\text{ odd,}
\end{cases}
\end{equation}
such that e.g.\
\begin{equation}
d_1=-1\,,\qquad d_3=\frac{1}{3!}\,,\qquad d_5=\frac{1}{5!}-\frac{1}{3!3!}\,,\qquad d_7=\frac{1}{7!}-\frac{1}{5!3!}-\frac{1}{3!5!}+\frac{1}{3!3!3!}\,.
\end{equation}
Therefore, the polynomial $P_n(z,q)$ can be rewritten in terms of elliptic zeta
values as
\begin{align}
P_n(z,q)&=(2\pi i)^{n-1}\sum_{j=0}^{\lfloor\frac{n-1}{2}\rfloor}\sum_{k=0}^{j}d_{2k+1}\omega_{2j+2-2k}(1;\tau)\frac{z^{n-1-2j}}{(n-1-2j)!}
\end{align}
and the sums $\E_n(t,1,q)$ for $n\geq 2$ can entirely be expressed by means of
the elliptic polylogarithms on the torus
\begin{align}\label{sec:connecting_eMPL:EnandGamma}
\E_n(t,1,q)
&=(2\pi i)^{n-1}\left( \Gt(\underbrace{\begin{smallmatrix}0 &\dots &0&1\\ 0&\dots&0& 0\end{smallmatrix}}_{n}; z,\tau)+\sum_{j=0}^{\lfloor\frac{n-1}{2}\rfloor}\sum_{k=0}^{j}d_{2k+1}\omega_{2j+2-2k}(1;\tau)\frac{z^{n-1-2j}}{(n-1-2j)!}\right)\,.
\end{align}
\medskip

Employing similar calculations, it is possible to relate iterated integrals
$\Gt$ of weight higher than one to the $\ELi$-functions.
The $q$-expansions~\eqref{sec:toolbox_eMPL:qExpansiong2n} and
\eqref{sec:toolbox_eMPL:qExpansiong2n1} of $g^{(m+1)}$ for $m>0$ lead to
\begin{align}\label{Ell:EBW:E0m}
\E_{0,-m}(t,1,q)&= \frac{m!}{(2\pi i)^{m+1}}\left(g^{(m+1)}(z,\tau)+(1+(-1)^{m+1})\zeta_{m+1} \right)
\end{align}
and therefore, since $\E_{n,m}$ satisfies the same partial differential equation as $\E_n$,
\begin{align}\label{sec:connecting_pdeEnm}
 \frac{\partial}{\partial z}\E_{n,m}(z,1,\tau)&=2\pi i \E_{n-1,m}(z,1,\tau)\,,
\end{align}
the following relations can be identified: for $n=1, m>0$
\begin{align}\label{Ell:EBW:E1mGamma}
\E_{1,-m}(t,1,q)&= \int_0^zdz' \frac{\partial}{\partial z'}\E_{1,-m}(z',1,\tau)+\E_{1,-m}(1,1,q)\nonumber\\
&= \frac{m!}{(2\pi i)^{m}}  \Gt(\begin{smallmatrix}m+1\\  0\end{smallmatrix}; z,\tau)+\frac{m!}{(2\pi i)^{m}} (1+(-1)^{m+1})\zeta_{m+1}z +\E_{1,-m}(1,1,q)\,,
\end{align}
for $n=2,m>0$
\begin{align}\label{Ell:EBW:E2mGamma}
\E_{2,-m}(t,1,q)&= \int_0^zdz' \frac{\partial}{\partial z'}\E_{2,-m}(z',1,\tau)+\E_{2,-m}(1,1,q)\nonumber\\
&= \frac{m!}{(2\pi i)^{m-1}}   \Gt(\begin{smallmatrix}0&m+1\\ 0& 0\end{smallmatrix}; z,\tau)+\frac{m!}{(2\pi i)^{m-1}}  (1+(-1)^{m+1})\zeta_{m+1}\frac{z^2}{2} \nonumber\\
&\phantom{=}+2\pi i \E_{1,-m}(1,1,q)z +\E_{2,-m}(1,1,q)\,.
\end{align}
A recursion leads to the general formula for $n>0,m>0$
\begin{align}\label{sec:connecting_eMPL:EnmandGammaPnm}
 \E_{n,-m}(t,1,q)&=m!(2\pi i)^{n-m-1} \Gt(\underbrace{\begin{smallmatrix}0&\dots&0&m+1\\ 0&\dots&0& 0\end{smallmatrix}}_{n}; z,\tau)+P_{n,m}(z,q)\,,
\end{align}
where 
\begin{align}\label{sec:connecting_eMPL:Pnm}
 P_{n,m}(z,q)&=m!(2\pi i)^{n-m-1}(1+(-1)^{m+1})\zeta_{m+1} \frac{z^n}{n!}+\sum_{j=1}^{n}(2\pi i)^{n-j} \E_{j,-m}(1,1,q)\frac{z^{n-j}}{(n-j)!}\,.
\end{align}
As in the case $m=0$, evaluation of $\E_{n,-m}(t,1,q)$ given in
\eqn{sec:connecting_eMPL:EnmandGamma} and the fact that $\E_{n,-m}(1,1,q)$
vanishes for $n+m$ even leads to an expression of the integration constants
$\E_{n,-m}(1,1,q)$ in terms of elliptic zeta values. The calculation is shown
in \appref{app:eMZV} and the result is given in \eqn{app:intConst:EnmExplicit},
i.e.\
\begin{align}\label{sec:connecting_eMPL:EnmandGamma}
\E_{n,-m}(1,1,q)&=\begin{cases}m!(2\pi i)^{n-1-m}\sum_{k=0}^{\lfloor \frac{n}{2} \rfloor}d_{2k+1}\omega_{n+1-2k}(m+1;\tau)& n+m\text{ odd}\\
0&n+m\text{ even,}
\end{cases}
\end{align}
such that
\begin{small}
\begin{align}
&\E_{n,-m}(t,1,q)\nonumber\\
&=\begin{cases}m!(2\pi i)^{n-1-m}\left(\Gt(\underbrace{\begin{smallmatrix}0&\dots&0&m+1\\ 0&\dots&0& 0\end{smallmatrix}}_{n}; z,\tau)+\sum_{j=0}^{\lfloor \frac{n}{2} \rfloor} \sum_{k=0}^{ j }d_{2k+1}\omega_{2j+1-2k}(m+1;\tau)\frac{z^{n-2j}}{(n-2j)!}\right)& m\text{ odd}\\
m!(2\pi i)^{n-1-m}\left(\Gt(\underbrace{\begin{smallmatrix}0&\dots&0&m+1\\ 0&\dots&0& 0\end{smallmatrix}}_{n}; z,\tau)+\sum_{j=0}^{\lfloor \frac{n-1}{2} \rfloor} \sum_{k=0}^{ j }d_{2k+1}\omega_{2j+2-2k}(m+1;\tau)\frac{z^{n-1-2j}}{(n-1-2j)!}\right)&m\text{ even.}
\end{cases}
\end{align}
\end{small}
For example and latter purposes, we find in particular the relations 
\begin{align}\label{Ell:EBW:E10Gamma}
\E_{1,0}(t,1,q)
&=\Gt(\begin{smallmatrix}1\\0\end{smallmatrix}; z,\tau)-\omega_2(1;\tau)+\frac{1}{2}\left(\Li_1(t)+\Li_1(t^{-1})\right),
\end{align}
\begin{align}\label{Ell:EBW:E20Gamma}
\E_{2,0}(t,1,q)
&=2\pi i \left(\Gt(\begin{smallmatrix}0&1\\0&0\end{smallmatrix}; z,\tau)-\omega_2(1;\tau)z\right) +\frac{1}{2}\left(\Li_2(t)-\Li_2(t^{-1}) \right)
\end{align}
and 
\begin{align}\label{Ell:EBW:E11Gamma}
\E_{1,-1}(t,1,q)&=\frac{1}{2\pi i}\left(\Gt(\begin{smallmatrix}2\\  0\end{smallmatrix}; z,\tau)-\omega_1(2;\tau) z\right)\nonumber\\
&=\frac{1}{2\pi i}\Gt(\begin{smallmatrix}2\\  0\end{smallmatrix}; z,\tau)+\frac{1}{\pi i}\zeta_2 z\,.
\end{align}
Thus, we have established a direct connection between the functions $\E_{n,-m}$
on the Tate curve and the iterated integrals of the form
$\Gt(\underbrace{\begin{smallmatrix}0&\dots&m\\ 0&\dots&
0\end{smallmatrix}}_{n}; z,\tau)$ for $n,m>0$, which are defined on the torus.

On the other hand, the elliptic
Bloch-Wigner function $\DD^{\E}$ can be rewritten in terms
of the above examples $\E_{1,0}$, $\E_{2,0}$ and $\E_{1,-1}$. This involves the identities
\begin{align}\label{sec:connection_eMPL:sumLi2ToE}
\sum_{l>0}\left(\Li_2(tq^l)-\Li_2(t^{-1}q^l)\right)&=-\E_{2,0}(t,1,q)
\end{align}
and
\begin{align}\label{sec:connection_eMPL:sumLogLi1ToE}
\sum_{l>0}\log(|tq^l|)\Li_{1}(tq^l)-\sum_{l>0}\log(|t^{-1}q^l|)\Li_{1}(t^{-1}q^l)&=-\log(|t|)\E_{1,0}(t,1,q)-\log(|q|)\E_{1,1}(t,1,q)\,,
\end{align}
which follow straightforwardly from the definition~\eqref{Ell:EBW:Enm} of
$\E_{n,m}(t,s,q)$. Therefore, the value of $\DD^{\E}$ can be expressed in terms of
the iterated integrals $\Gt$ on the torus as follows
\begin{align}\label{sec:connection_eMPL:DEandGamma}
\DD^{\E}(t,q)&=\sum_{l>0}\Im\left(\Li_2(tq^l)-\Li_2(t^{-1}q^l)\right)-\sum_{l>0}\Im\left(\log(|tq^l|)\Li_1(tq^l)-\log(|t^{-1}q^l|)\Li_1(t^{-1}q^l)\right)\nonumber\\
&\phantom{=}+\DD(t)\nonumber\\
&=-\Im\left(\E_{2}(t,1,q)\right)+\log(|t|) \Im\left(\E_{1}(t,1,q)\right)+\log(|q|) \Im\left(\E_{1,-1}(t,1,q)\right)\nonumber\\
&=\Im(\tau) \Re\left(\Gt(\begin{smallmatrix}2\\  0\end{smallmatrix}; z,\tau)\right)+2\pi \Re\left( \Gt(\begin{smallmatrix}1&0\\0&0\end{smallmatrix}; z,\tau)\right)-2\pi \Re(z) \Re\left(\Gt(\begin{smallmatrix}1\\0\end{smallmatrix}; z,\tau)\right)\nonumber\\
&\phantom{=}+2\Re(z)\left(\pi\Re\left(\omega_2(1;\tau)\right)+\zeta_2\Im(\tau)\right),
\end{align}
where the $q$-independent term $\DD(t)$ is absorbed in the second equality by
going from $\E_{n,m}(t,1,q)$ to $\E_{n}(t,1,q)$ according to
\eqn{Ell:EBW:BarEnDef}. The logarithmic factors with the absolute values of $t$
and $q$, respectively, yield contributions of the imaginary parts of $z$ and
$\tau$, respectively. The final expression explicitly involving the real part
of $z$ is obtained by using
equations~\eqref{Ell:EBW:E10Gamma}-\eqref{Ell:EBW:E11Gamma} and the identity
$\Re(z_1 z_2)+\Im(z_1)\Im(z_2)=\Re(z_1)\Re(z_2)$, where $z_1,z_2\in\ZC$, for
the last equality above.
The translation of the elliptic Bloch-Wigner function $\DE$ from the torus, as
given by \eqn{sec:connection_eMPL:DEandGamma}, to the projective elliptic curve
is based on the results in \rcite{Broedel:2017kkb}.  The iterated integrals
$\Gt$ on the torus can be expressed via the isomorphism $\xi_{\tau\ZZ+\ZZ,E}$
in terms of some iterated integrals on the projective elliptic curve, which are
defined as follows 
\begin{equation}\label{sec:connecting_eMPL:E3}
\E_3(\begin{smallmatrix}n_1&\dots&n_k\\c_1&\dots&c_k\end{smallmatrix}; x,\vec{e})=\int_0^xdx'\text{ }\varphi_{n_1}(c_1;x',\vec{a}) \E_3(\begin{smallmatrix}n_2&\dots&n_k\\c_2&\dots&c_k\end{smallmatrix}; x',\vec{e})\,, \qquad \E_3(; x,\vec{e})=1\,,
\end{equation}
with $c_i\in \ZC\cup\{\infty\}$, $\vec{e}=(e_1,e_2,e_3)$ is the vector of the
roots\footnote{Note that we use slightly different conventions than in
\rcite{Broedel:2017kkb}, where the defining cubic equation of the projective
curve is written in standard form $y^2=(x-a_1)(x-a_2)(x-a_3)$ in contrast to
our notation which only involves the Weierstrass form.} of the Weierstrass
equation and the integration kernels $\varphi_{n}(c;x,\vec{e})$ are defined
according to the construction of \rcite{Broedel:2017kkb}. For example, the
differential $\varphi_0(0,x,\vec{e})\,dx$ is simply the holomorphic differential
$dx/y$ which itself is the differential $dz$ on the torus
\begin{equation}\label{sec:connecting_eMPL:phi0}
\varphi_0(0,x,\vec{e})\,dx= \frac{dx}{y}=\frac{d\wp(z)}{\wp'(z)}=dz\,.
\end{equation}
The integration kernels $\varphi_n(\infty;x,\vec{e})$ for $n\geq 1$ are defined
as follows: first, define the integral of $x/y$ with an additional term as
follows
\begin{align}
Z_3(x_P,\vec{e})&=-\int_{e_1}^{x_P}dx\left(\frac{x}{ y}+2\frac{\eta_1}{y}\right).
\end{align}
This defines the kernel for $n=1$
\begin{align}
\varphi_1(\infty;x,\vec{e})&=\frac{1}{y}Z_3(x,\vec{e})\,.
\end{align}
The kernels for higher $n$ are defined by some polynomials $Z_3^{(n)}$, which
are of degree $n$ in $Z_3(x)$ with the coefficients being polynomials in $x$
and $y$ and that do not have any poles in $x$. For example in the case of
$n=2$, the integration kernel is defined as
\begin{equation}
\varphi_2(\infty;x,\vec{e})=\frac{1}{y}Z_3^{(2)}(x,\vec{e})=\frac{1}{y}\left(\frac{1}{8}Z_3(x,\vec{e})^2-\frac{x}{2}\right).
\end{equation}
The (explicit) construction of $Z_3^{(n)}$ is exactly the same as the
construction of $g^{(n)}(z,\tau)$ as a polynomial in $g^{(1)}(z,\tau)$ with
polynomial coefficients in $\wp(z)$ and $\wp'(z)$, see \rcite{Broedel:2017kkb}.
This leads to a very close relation between the kernels
$\varphi_n(\infty;x,\vec{e})$ and $g^{(n)}(z,\tau)$. For $n=0$, we first
rewrite\footnote{Note that for this calculation, we choose the sign of $y=\pm
\sqrt{4x^3-g_2x-g_3}$ in Abel's map
\eqref{sec:toolbox_ellipticfunctions:Abelsmap} such that we indeed obtain
\eqn{sec:connecting_eMPL:Z3g1} and not the negative of the right-hand side,
i.e.\ $Z_3(x,\vec{e})=-g^{(1)}(z,\tau)$.}
\begin{equation}\label{sec:connecting_eMPL:Z3g1}
Z_3(x,\vec{e})=\zeta(z)-2\eta_1 z=g^{(1)}(z,\tau)
\end{equation}
using \eqn{sec:toolbox_eMPL:g1Zeta}, such that
\begin{align}\label{sec:connecting_eMPL:phi1}
\varphi_1(\infty;x,\vec{e})\,dx&=g^{(1)}(z,\tau)\,dz\,.
\end{align}
Thus, the construction of $Z_3^{(n)}$ ensures that the same result holds for $n\geq 1$
\begin{align}\label{sec:connecting_eMPL:phin}
\varphi_n(\infty;x,\vec{e})\,dx&=g^{(n)}(z,\tau)\,dz\,,
\end{align}
which is all that is needed to rewrite $\DE$. With $z_0$ being a zero of $\wp$
such that $\wp'(z_0)>0$, the identification $x=\wp(z)$ and from the equations
\eqref{sec:connecting_eMPL:phi0}, \eqref{sec:connecting_eMPL:phi1} and
\eqref{sec:connecting_eMPL:phin} for the differentials, the iterated integrals
in \eqn{sec:connection_eMPL:DEandGamma} can be expressed as follows on the
projective elliptic curve
\begin{align}\label{sec:connection_eMPL:Gamma1000E3}
\Gt(\begin{smallmatrix}1&0\\0&0\end{smallmatrix}; z,\tau)&=\E_3(\begin{smallmatrix}1&0\\\infty&0\end{smallmatrix}; x,\vec{e})+\Gt(\begin{smallmatrix}1&0\\0&0\end{smallmatrix}; z_0,\tau)\,,
\end{align}
\begin{align}\label{sec:connection_eMPL:Gamma20E3}
\Gt(\begin{smallmatrix}2\\0\end{smallmatrix}; z,\tau)&=\E_3(\begin{smallmatrix}2\\\infty\end{smallmatrix}; x,\vec{e})+\Gt(\begin{smallmatrix}2\\0\end{smallmatrix}; z_0,\tau)\,,
\end{align}
\begin{align}\label{sec:connection_eMPL:Gamma10E3}
\Gt(\begin{smallmatrix}1\\0\end{smallmatrix}; z,\tau)&=\E_3(\begin{smallmatrix}1\\\infty\end{smallmatrix}; x,\vec{e})+\Gt(\begin{smallmatrix}1\\0\end{smallmatrix}; z_0,\tau)\,,
\end{align}
as well as
\begin{align}\label{sec:connection_eMPL:zE3}
z&=\E_3(\begin{smallmatrix}0\\0\end{smallmatrix}; x,\vec{e})+z_0\,.
\end{align}
Therefore, the elliptic Bloch-Wigner function takes the following form on the
projective elliptic curve 
\begin{align}\label{sec:connection_eMPL:DEandE3}
\DD^{\E}(t,q)
&=\Im(\tau) \Re\left(\Gt(\begin{smallmatrix}2\\  0\end{smallmatrix}; z,\tau)\right)+2\pi \Re\left( \Gt(\begin{smallmatrix}1&0\\0&0\end{smallmatrix}; z,\tau)\right)-2\pi \Re(z) \Re\left(\Gt(\begin{smallmatrix}1\\0\end{smallmatrix}; z,\tau)\right)\nonumber\\
&\phantom{=}+2\Re(z)\left(\pi\Re\left(\omega_2(1;\tau)\right)+\zeta_2\Im(\tau)\right)\nonumber\\
&=\Im(\tau) \Re\left(\E_3(\begin{smallmatrix}2\\\infty\end{smallmatrix}; x,\vec{e})+\Gt(\begin{smallmatrix}2\\0\end{smallmatrix}; z_0,\tau)\right)+2\pi \Re\left(\E_3(\begin{smallmatrix}1&0\\\infty&0\end{smallmatrix}; x,\vec{e})+\Gt(\begin{smallmatrix}1&0\\0&0\end{smallmatrix}; z_0,\tau)\right)\nonumber\\
&\phantom{=}-2\pi \Re\left(\E_3(\begin{smallmatrix}0\\0\end{smallmatrix}; x,\vec{e})+z_0\right) \Re\left(\E_3(\begin{smallmatrix}1\\\infty\end{smallmatrix}; x,\vec{e})+\Gt(\begin{smallmatrix}1\\0\end{smallmatrix}; z_0,\tau)\right)\nonumber\\
&\phantom{=}+2\Re\left(\E_3(\begin{smallmatrix}0\\0\end{smallmatrix}; x,\vec{e})+z_0\right)\left(\pi\Re\left(\omega_2(1;\tau)\right)+\zeta_2\Im(\tau)\right)\,.
\end{align}
The constant terms involving the iterated integrals on the torus evaluated at
$z_0$ and 1, respectively, drop out once the elliptic Bloch relation
\eqref{sec:toolbox_ellBloch:ellBlochRelation} is formed.

\subsection{Higher elliptic polylogarithms}\label{sec:connection:HigherEllPolylogs}
The translation procedure from the Tate curve to the torus described in the
previous section is applicable to elliptic generalisations of higher
polylogarithms. In this subsection we present two such families, both of which
include the elliptic Bloch-Wigner function, and show how they can be expressed
in terms of the elliptic integrals $\Gt$ on the torus. These families of
functions are not independent, the first one is actually a subclass of the
second.

The first construction of higher elliptic polylogarithms is based on the
averaging process over the Tate curve which was used to define the elliptic
Bloch-Wigner function in \eqn{sec:toolbox_Bloch:DE}. The single-valued
polylogarithms that are to be averaged were first described by Ramakrishnan
\cite{Ramakrishnan86} and generalise the Bloch-Wigner function to higher
orders. They are defined by
\begin{align}
\mathcal{L}_n(t)&=\mathcal{R}_n\left(\sum_{k=0}^{n-1}\frac{2^k B_k}{k!}\log^k(|t|)\Li_{n-k}(t)\right),
\end{align}
where $\mathcal{R}_n$ denotes the imaginary or real part if $n$ is even or odd,
respectively, and $B_k$ the $k$-th Bernoulli number. The Bloch-Wigner function
$\DD$ is obtained for $n=2$, and these functions also satisfy a
similar inversion relation as $\DD$, namely
\begin{align}
\mathcal{L}_n(t^{-1})&=(-1)^{n-1}\mathcal{L}_n(t)\,.
\end{align}
The elliptic generalisation used in \rcite{ZagierGangl} and proposed in
\rcite{Zagier}, as linear combinations of the more general class described
below, is
\begin{align}
\mathcal{L}_n^{\E}(t,q)&=\sum_{l\in\mathbb{Z}}\mathcal{L}_n(tq^l)\nonumber\\
&=\sum_{k=0}^{n-1}\frac{2^k B_k}{k!}\mathcal{R}_n\left(\sum_{l>0}\log^k(|tq^l|)\Li_{n-k}(tq^l)+(-1)^{n-1}\sum_{l>0}\log^k(|t^{-1}q^l|)\Li_{n-k}(t^{-1}q^l)\right)\nonumber\\
&\phantom{=}+\mathcal{L}_n(t)\,,
\end{align}
such that in particular $\mathcal{L}_2^{\E}=\DE$. By a similar calculation as for
$\DE$, which e.g.\ involves the identity 
\begin{align}\label{sec:connection:sumLogToE}
&\sum_{l>0}\log^k(|tq^l|)\Li_{n-k}(tq^l)+(-1)^{n-1}\sum_{l>0}\log^k(|t^{-1}q^l|)\Li_{n-k}(t^{-1}q^l)\nonumber\\
&=-\sum_{m=0}^k \binom{k}{m}\log^{k-m}(|t|)\log^{m}(|q|)\E_{n-k,-m}
\end{align}
generalising equations~\eqref{sec:connection_eMPL:sumLi2ToE}
and~\eqref{sec:connection_eMPL:sumLogLi1ToE}, the elliptic polylogarithms
$\mathcal{L}_n^{\E}$ turn out to be related to the functions $\E_{n,-m}$ according
to
\begin{align}\label{sec:connection:LEm}
\mathcal{L}_n^{\E}(t,q)&=-\sum_{m=0}^k \binom{k}{m}\frac{2^k B_k}{k!}\log^{k-m}(|t|)\log^{m}(|q|)\mathcal{R}_n\left(\E_{n-k,-m}\right)+\mathcal{L}_n(t)\,.
\end{align}
Just like in the dilogarithmic case of $n=2$, this result can immediately be
expressed in terms of the iterated integrals on the torus and the projective
curve using the results of the previous section. 
\medskip

The more general class of single-valued elliptic polylogarithms,
introduced in \rcite{Zagier} and used in \rcite{DHoker:2015wxz} in the context
of modular graph functions for one-loop closed string amplitudes, can be
constructed from the single-valued sum
\begin{align}\label{sec:toolbox_eMPL:Dab}
\DD_{a,b}(t)&=(-1)^{a-1}\sum_{n=a}^{a+b-1}\binom{n-1}{a-1}\frac{(-2\log(|t|))^{a+b-1-n}}{(a+b-1-n)!}\Li_n(t)\nonumber\\
&\phantom{=}+(-1)^{b-1}\sum_{n=b}^{a+b-1}\binom{n-1}{b-1}\frac{(-2\log(|t|))^{a+b-1-n}}{(a+b-1-n)!}\overline{\Li_n(t)},
\end{align}
which satisfies $\overline{\DD_{a,b}(t)}=\DD_{b,a}(t)$, where the overline
denotes complex conjugation. The functions $\mathcal{L}_n$ above are linear
combinations of $\DD_{a,b}$ and hence, a subclass of the latter \cite{Zagier}.
For example, 
\begin{align}\label{sec:connecting_eMPL:D12}
\DD_{1,2}(t)=2i\DD(t)+2\log(|t|)\log(|1-t|)\,,
\end{align}
such that the Bloch-Wigner function can be written as
$\DD(t)=\frac{1}{4i}\left(\DD_{1,2}(t)-\DD_{2,1}(t)\right)$.  The elliptic
generalisation is similar to the previous average over the Tate curve and given
by \cite{Zagier}
\begin{align}\label{sec:toolbox_eMPL:DEab}
\DD^{\E}_{a,b}(t,q)&=\sum_{l\geq 0}\DD_{a,b}(tq^l)+(-1)^{a+b}\sum_{l>0}\DD_{a,b}(t^{-1}q^l)+\frac{(4\pi\Im(\tau))^{a+b-1}}{(a+b)!}B_{a+b}(u)\,,
\end{align}
where $B_{n}$ is the $n$-th Bernoulli polynomial and $z=u\tau+v$ with $u,v\in [0,1]$. 
For example, the elliptic Bloch-Wigner function can be expressed as
\begin{align}\label{sec:toolbox_eMPL:DEandDEab}
\DE(t,q)&=-\frac{1}{2}\Im(\DD^{\E}_{2,1}(t,q))\,.
\end{align}
In order to express the functions $\DD^{\E}_{a,b}$ in terms of $\E_{n,m}$, the
relevant prefactor in $\DD^{\E}_{a,b}(t,q)$ for the translation has to be determined.
This is the factor obtained by plugging the right-hand side of the definition
\eqref{sec:toolbox_eMPL:Dab} of $\DD_{a,b}$ into \eqn{sec:toolbox_eMPL:DEab} and
pushing the sum over $l$ to the logarithmic functions depending on this
summation index, i.e.\
\begin{align}\label{sec:connecting_eMPL:DabErelevantFactor}
&\sum_{l> 0}\left(\log(|tq^l|)^{a+b-1-n}\Li_n(tq^l)+(-1)^{a+b}\log(|t^{-1}q^l|)^{a+b-1-n}\Li_n(t^{-1}q^l)\right)\nonumber\\
&=-\sum_{m=0}^{a+b-1-n}\binom{a+b-1-n}{m}\log(|t|)^{a+b-1-n-m}\log(|q|)^m\E_{n,-m}(t,1,q)\,
\end{align}
where we used \eqn{sec:connection:sumLogToE}. This leads to an expression of
$\DD^{\E}_{a,b}(t,q)$ as a linear combination of terms of the form $\E_{n,-m}$ and
complex conjugates thereof, such that, according to the previous section, it is
indeed a linear combination of (powers of) the iterated integrals $\Gt$ and
their complex conjugates. The explicit result is rather lengthy and can be
found in \appref{app:DEab}. In particular, it matches the result for $\DE$
given in \eqn{sec:connection_eMPL:DEandGamma}. 
\smallskip

Let us make a comment about the $K$-theoretic use of the elliptic Bloch-Wigner
function $\DE$ in the construction of a regulator map $\RR:
K_2(E)\rightarrow\ZC$ in equation $(8.1.1)$ of \rcite{BlochRelation}, where
$K_2(E)$ is the second $K$-group associated to an elliptic curve $E$ over $\ZC$.
The non-elliptic version of the map $\RR$ generalised to higher $K$-groups is of
particular interest in the formulation of the conjectures of \rcite{ZagierGangl},
which relate the Dedekind zeta function $\zeta_F(m)$ of a number field $F$ to
special values of the single-valued polylogarithms $\mathcal{L}_m$, and which
are also used in the description of the $m$-th Bloch group. The elliptic
version $\RR$ can be used in the construction of the second elliptic Bloch
group, see e.g.\ \rcite{ZagierGangl}, and its imaginary part is the elliptic
Bloch-Wigner function $\DE$. In order to describe its real part, let 
\begin{align}\label{sec:connecting_eMPL:J}
\JJ(t)&=\log(|t|)\log(|1-t|)\,,
\end{align}
such that the real part of the regulator map $\RR$ is given by 
\begin{align}\label{sec:connecting_eMPL:originalJE}
\JJ^{\E}(t,q)&=\sum_{l\geq 0}\JJ(t q^l)-\sum_{l >0}\JJ(t^{-1} q^l)\,.
\end{align}
Comparing equations~\eqref{sec:connecting_eMPL:D12} and
\eqref{sec:connecting_eMPL:J} as well as the definitions of their elliptic
generalisations~\eqref{sec:toolbox_eMPL:DEab} and
\eqref{sec:connecting_eMPL:originalJE}, leads to the conclusion that
\begin{align}\label{sec:connecting_eMPL:JE}
\JJ^{\E}(t,q)&=\frac{1}{2}\Re(\DD^{\E}_{1,2}(t,q))+\frac{(4\pi\Im(\tau))^{2}}{6}B_{3}(u)\,.
\end{align}
Therefore, according to \eqn{sec:toolbox_eMPL:DEandDEab}, the regulator map
$\RR$ equals one half of $\DD^{\E}_{1,2}$ up to the last term in
\eqn{sec:connecting_eMPL:JE}, such that, as for its imaginary part, i.e.\ the
elliptic Bloch-Wigner function, the whole regulator map $\RR$ can immediately
be translated to the iterated integrals on the torus and the projective elliptic curve,
as described above. 

\subsection{The elliptic Bloch relation on the torus}\label{sec:BlochRelationTorus}

The connections between the different notions of elliptic (multiple)
polylogarithms found in the previous subsections \ref{sec:connection:TatetoTorus}
and \ref{sec:connection:HigherEllPolylogs} can be exploited to translate and to
compare various concepts and structures among them. In this section we show how the
elliptic Bloch relation~\eqref{sec:toolbox_ellBloch:ellBlochRelation}
translates to the torus, discover more general relations thereon and hence,
provide an alternative proof of the elliptic Bloch relation. In doing so, we
will show, how the Bloch relation can be interpreted in terms of differentials of 
iterated integrals or, more generally, in terms of the elliptic symbol calculus
introduced in \rcite{Broedel:2018iwv}.

Let $F$ be an elliptic function on the Tate curve with the following divisor
\begin{equation}
\Div(F)=\sum_i d_i (a_i)\,,\qquad \sum_i d_i =0\,,\qquad \prod_i a_i^{d_i}=1\,.
\end{equation}
Formulated on the torus the above equation translates into, 
\begin{equation}
\Div(F)=\sum_i d_i (A_i)\,,\qquad\sum_i d_i =0\,,\qquad \sum_i d_iA_i=0\,.
\end{equation}
where $a_i=e^{2\pi i A_i}$.  Using
\eqn{sec:toolbox_ellipticfunctions:FandSigma}, one can express $F$ in terms of
a product of Weierstrass $\sigma$ functions
\begin{equation}
  \label{eqn452}
F(z)=s_A\prod_i\sigma(z-A_i)^{d_i}=s_A\exp\left(\sum_i d_i \int_{0}^{z-A_i}dz' \zeta(z')\right)
\end{equation}
for some scaling $s_A\in\ZC^{\ast}$ of $F$. Similarly,
for a given $\ka\in\ZC^{\ast}$, $\kappa-F$ can be represented by\footnote{Note
that here, the $e_j$ do not denote the roots of a Weierstrass equation, but the
orders of the zeros and poles of the elliptic function $\kappa-F$.}
\begin{equation}
  \label{eqn453}
\ka-F(z)=s_B\prod_j\sigma(z-B_j)^{e_j}=s_B\exp\left(\sum_j e_j \int_{0}^{z-B_j}dz' \zeta(z')\right),
\end{equation}
where $s_B\in \ZC^{\ast}$. For notational convenience, let us split the set of
zeros and poles of $F$ and $\ka-F$, denoted by $I$ and $J$, respectively, into
the zeros of $F$, $I'=\{A_i|d_i>0\}$, the zeros of $\ka-F$, $J'=\{B_j|e_j>0\}$,
and the common set of poles $K=\{A_i|d_i<0\}=\{B_j|e_j<0\}$. 
Using these
conventions, the elliptic Bloch relation
\eqref{sec:toolbox_ellBloch:ellBlochRelation} can be rewritten by means of
\eqn{sec:connection_eMPL:DEandGamma} as 
\begin{subequations}
\label{sec:connection_Bloch:BlochRelationTorus}
\begin{align}
0&=\sum_{i,j}d_i e_j \DE\left(\frac{a_i}{b_j},q\right)\nonumber\\
 &=-2\pi \sum_{i,j}d_i e_j \Bigg(\Re\left( \Gt(\begin{smallmatrix}1&0\\0&0\end{smallmatrix}; A_i-B_j,\tau)\right)\label{eqn:EBRtorusa}\\
 &\qquad\qquad\qquad\quad + \Re\Big(\frac{\tau}{2\pi i}\Big) \Re\left(\Gt(\begin{smallmatrix}2\\  0\end{smallmatrix}; A_i-B_j,\tau)\right)\label{eqn:EBRtorusb}\\
&\qquad\qquad\qquad\quad-\Re(A_i-B_j) \Re\left(\Gt(\begin{smallmatrix}1\\0\end{smallmatrix}; A_i-B_j,\tau)\right) \Bigg),\label{eqn:EBRtorusc}
\end{align}
\end{subequations}
where $B_j$ is given by $b_j=e^{2\pi i B_j}$ and the summation indices $(i,j)$
run over $I\times J$, unless mentioned otherwise. 

We give an alternative proof of equation
\eqref{sec:connection_Bloch:BlochRelationTorus}, which we refer to as the
elliptic Bloch relation on the torus, in the following paragraphs by showing
that the sums over the single iterated integrals $\Gt$ occurring in the above
formula vanish separately (and for the first two also their imaginary parts,
yielding two holomorphic analogues of the elliptic Bloch relation). Note that
since we are interested in generating functional equations we consider the
zeros and poles $A_i$ and $B_j$ as well as the scaling factors $s_A$ and $s_B$
to be (not independent) variables, e.g.\ depending on variable coefficients of
the rational function on the elliptic curve that determine $F$, cf. the
examples in \subsecref{sec:toolbox:ellBloch}. 

Let us start with the first term of the elliptic Bloch relation on the torus,
\eqn{eqn:EBRtorusa}: naturally, the zeros and poles satisfy the constraints
$\sum_i d_i A_i=0$ and $\sum_j e_j B_j=0$ as functional identities. Hence, the
functional identity
\begin{equation}
\ka=\ka-F(A_i)=s_B\prod_{j}\sigma(A_i-B_j)^{e_j}
\end{equation}
holds for $i\in I'$, such that taking the total differential of both sides and
using \eqn{sec:toolbox_eMPL:g1Zeta}, i.e.\ $\zeta(z)=g^{(1)}(z,\tau)+2\eta_1
z$, as well as the representations \eqref{eqn452} and~\eqref{eqn453} the
differential equation
\begin{align}\label{sec:connection_Bloch:IprimeDeqSumJ}
\sum_j e_j  g^{(1)}(A_i-B_j)d(A_i-B_j)&= -d\log(s_B)-c_1\sum_j e_j B_j dB_j
\end{align}
can be obtained. For $k\in K$, a functional identity involving the residue instead
of the infinite value $\ka-F(A_k)$ can be used for a similar calculation: since
by convention $\sigma'(0)=1$, the residue of $\ka-F$ at $A_k$ is
\begin{align}
\Res_{A_k}(\ka-F)&=s_B\prod_{j\neq k}\sigma(A_k-B_j)^{e_j},
\end{align}
which implies that
\begin{align}\label{sec:connection_Bloch:KDeqSumJ}
\sum_{j\neq k}e_j g^{(1)}(A_k-B_j)d(A_k-B_j)&=d\log\left(\Res_{A_k}(\ka-F)\right)-d\log(s_B)-c_1\sum_{j}e_jB_jdB_j.
\end{align}
Two similar differential equations for sums over $I$ can be found, the first
one starting from $\ka=F(B_j)$, where $j\in J'$,
\begin{align}\label{sec:connection_Bloch:JprimeDeqSumI}
\sum_i d_i  g^{(1)}(A_i-B_j)d(A_i-B_j)&= -d\log(s_A)-c_1\sum_i d_i A_i dA_i.
\end{align}
With $k\in K$ and using that $\Res_{A_k}(F)=-\Res_{A_k}(\ka-F)$, the last such
differential equation turns out to be
\begin{align}\label{sec:connection_Bloch:KDeqSumI}
\sum_{i\neq k}d_i g^{(1)}(A_k-A_i)d(A_k-A_i)&=d\log\left(\Res_{A_k}(\ka-F)\right)-d\log(s_A)-c_1\sum_{i}d_iA_idA_i.
\end{align}
Going through an elaborate calculation, whose details we have outsourced to
\appref{app:vanish}, the four differential
equations~\eqref{sec:connection_Bloch:IprimeDeqSumJ},
\eqref{sec:connection_Bloch:KDeqSumJ},
\eqref{sec:connection_Bloch:JprimeDeqSumI} and
\eqref{sec:connection_Bloch:KDeqSumI} can be combined into the differential
equation 
\begin{align}\label{sec:connection_Bloch:vanishingSumzg1}
\sum_{i,j}d_i e_j (A_i-B_j)g^{(1)}(A_i-B_j,\tau)d(A_i-B_j)&=0\,.
\end{align}
For integration paths with $d\tau=0$, the differential of the iterated
integral $\Gt(\begin{smallmatrix}1&0\\0&0\end{smallmatrix}; z,\tau)$ is given
by
\begin{align}
d\Gt(\begin{smallmatrix}1&0\\0&0\end{smallmatrix}; z,\tau)&=zg^{1}(z,\tau)dz\,.
\end{align}
Accordingly, \eqn{sec:connection_Bloch:vanishingSumzg1} implies that
\begin{align}
\sum_{i,j}d_i e_j \Gt(\begin{smallmatrix}1&0\\0&0\end{smallmatrix}; A_i-B_j,\tau)&=c_2
\end{align}
for some constant $c_2\in \ZC$. In general, the zeros and poles of $F$ are only
constrained by $\sum_i d_i A_i=0=\sum_i d_i$, thus, it may be assumed that they
can be split in a way such that the divisor of $F$ consists of triplets with
two of them being unconstrained and the third one being given by
$A_3=-A_1-A_2$. An alternative way of saying this is that divisors of the form
$(A_1)+(A_2)-(0)-(A_1+A_2)$ span the set of principal divisors, which was
encountered in \subsecref{sec:toolbox:divisorFunction}, cf.\
\eqn{sec:toolbox_ellipticfunctions:spanningFlambda}. Thus, by continuity, the
above equation can be evaluated at the point where all $A_i=0$ to determine
\begin{equation}
c_2=\sum_{j} e_j \Gt(\begin{smallmatrix}1&0\\0&0\end{smallmatrix}; -B_j,\tau)\sum_i d_i=0.
\end{equation}
Therefore, we find a holomorphic analogue of the elliptic Bloch relation
\begin{align}\label{sec:connection_Bloch:BlochRelationGamma1000}
\sum_{i,j}d_i e_j \Gt(\begin{smallmatrix}1&0\\0&0\end{smallmatrix}; A_i-B_j,\tau)&=0.
\end{align}

Similar arguments apply for the term \eqref{eqn:EBRtorusc} involving the
iterated integral $z\Gt(\begin{smallmatrix}1\\0\end{smallmatrix}; z,\tau)$ in
  the elliptic Bloch relation on the torus
  \eqref{sec:connection_Bloch:BlochRelationTorus}. Let $i\in I'$ and write
\begin{equation}
  \ka=\ka-F(A_i)=s_B\exp\left(\sum_j e_j \int_{0}^{A_i-B_j}dz g^{(1)}(z,\tau)+\frac{c_1}{2}\sum_j e_j B_j^2\right),
\end{equation}
such that
\begin{align}\label{sec:connection_Bloch:IprimezGamma10SumJ}
\sum_j e_j \Gt(\begin{smallmatrix}1\\0\end{smallmatrix}; A_i-B_j,\tau)&=\log(\ka)-\log(s_B)-\frac{c_1}{2}\sum_j e_j B_j^2-2\pi i m_1,
\end{align}
for some $m_1\in\ZZ$, which holds for
$\Gt(\begin{smallmatrix}1\\0\end{smallmatrix}; A_i-B_j,\tau)$ being
  the regularised or unregularised iterated integral, because the factor
  $\sum_j e_j=0$ cancels the logarithmic singularity. For $k\in K$ and with
  $\sigma(z)=s_C \exp\left(\int_{z_0}^z dz' \zeta(z')\right)$ such that
  $\sigma'(0)=1$, the same calculation as before leads to
\begin{align}
\Res_{A_k}(\ka-F)&=s_Bs_C\exp\left(\sum_{j\neq k} e_j \int_{z_0}^{A_k-B_j}dz \zeta(z) \right)\nonumber\\
&=s_Bs_C\exp\left(\sum_{j\neq k} e_j \int_{0}^{A_k-B_j}dz g^{(1)}(z,\tau)+\frac{c_1}{2} \sum_{j}e_jB_j^2 +\int_0^{z_0}dz \zeta(z) \right)
\end{align}
which implies that
\begin{align}\label{sec:connection_Bloch:KzGamma10SumJ}
\sum_{j\neq k} e_j \int_{0}^{A_k-B_j}dz g^{(1)}(z,\tau)&=\log(\Res_{A_k}(\ka-F))-\log(s_C)-\log(s_B)-\frac{c_1}{2}\sum_j e_j B_j^2\nonumber\\
&\phantom{=}-\int_0^{z_0}dz \zeta(z)-2\pi i m_2,
\end{align}
where $m_2\in \ZZ$, and analogously for the sum over $I\setminus\{k\}$
\begin{align}\label{sec:connection_Bloch:KzGamma10SumI}
\sum_{i\neq k} d_i \int_{0}^{A_k-A_i}dz g^{(1)}(z,\tau)&=\log(\Res_{A_k}(F))-\log(s_C)-\log(s_A)-\frac{c_1}{2}\sum_i d_i A_i^2\nonumber\\
&\phantom{=}-\int_0^{z_0}dz \zeta(z)-2\pi i m_3,
\end{align}
for $m_3\in\ZZ$. A similar result holds for $j\in J'$,
\begin{align}\label{sec:connection_Bloch:JprimezGamma10SumI}
\sum_i d_i \Gt(\begin{smallmatrix}1\\0\end{smallmatrix}; A_i-B_j,\tau)&=\log(\ka)-\log(s_A)-\frac{c_1}{2}\sum_i d_i A_i^2-2\pi i m_4,
\end{align}
where $m_4\in\ZZ$. Since $\log(\Res_{A_k}(F))=\log(\Res_{A_k}(1-F))+i\pi$,
equations~\eqref{sec:connection_Bloch:KzGamma10SumJ} and
\eqref{sec:connection_Bloch:KzGamma10SumI} lead to
\begin{align}
&\sum_j e_j \Gt(\begin{smallmatrix}1\\0\end{smallmatrix}; A_k-B_j,\tau)-\sum_i d_i \Gt(\begin{smallmatrix}1\\0\end{smallmatrix}; A_k-A_i,\tau)\nonumber\\
&=-i\pi(1+2m_2-2m_3) + \log(s_A)+\frac{c_1}{2}\sum_i d_i A_i^2-\log(s_B)-\frac{c_1}{2}\sum_j e_j B_j^2.
\end{align}
Finally, using the equations~\eqref{sec:connection_Bloch:IprimezGamma10SumJ},
\eqref{sec:connection_Bloch:KzGamma10SumJ},
\eqref{sec:connection_Bloch:KzGamma10SumI} and
\eqref{sec:connection_Bloch:JprimezGamma10SumI} all together, the identities
\begin{align}\label{sec:connection_Bloch:BlochRelationRezReGamma10}
\sum_{i,j}d_i e_j \Re\left(A_i-B_j\right)\Re\left(\Gt(\begin{smallmatrix}1\\0\end{smallmatrix}; A_i-B_j,\tau)\right)&=0
\end{align}
and 
\begin{align}\label{sec:connection_Bloch:BlochRelationImzReGamma10}
\sum_{i,j}d_i e_j \Im\left(A_i-B_j\right)\Re\left(\Gt(\begin{smallmatrix}1\\0\end{smallmatrix}; A_i-B_j,\tau)\right)&=0.
\end{align}
can be obtained, see \appref{app:vanish} for the calculation. 

Now, we are left with the term \eqref{eqn:EBRtorusb} involving
$\Gt(\begin{smallmatrix}2\\0\end{smallmatrix}; z,\tau)$. Let us take the
  partial derivative of \eqn{sec:connection_Bloch:IprimezGamma10SumJ} with
  respect to $\tau$ and use the partial differential
  \eqn{sec:toolbox_eMPL:gnHeatEq} of the integration kernel, i.e.\ $2\pi i
  \frac{\partial}{\partial \tau}g^{(1)}(z,\tau)=\frac{\partial}{\partial
  z}g^{(2)}(z,\tau)$, to find
\begin{align}\label{sec:connection_Bloch:Iprimeg2SumJ}
\sum_j e_j g^{(2)}(A_i-B_j,\tau)&=-2\pi i\frac{\partial}{\partial \tau}\frac{c_1}{2}\sum_j e_j B_j^2-2\pi i\sum_j e_jg^{(1)}(A_i-B_j,\tau)\frac{\partial}{\partial \tau}(A_i-B_j),
\end{align}
valid for $i\in I'$. A similar result holds for $j\in J'$
\begin{align}\label{sec:connection_Bloch:Jprimeg2SumI}
\sum_i d_i g^{(2)}(A_i-B_j,\tau)&=-2\pi i \frac{\partial}{\partial \tau}\frac{c_1}{2}\sum_i d_i A_i^2-2\pi i\sum_i d_ig^{(1)}(A_i-B_j,\tau)\frac{\partial}{\partial \tau}(A_i-B_j)
\end{align}
and for $k\in K$ 
\begin{align}\label{sec:connection_Bloch:Kg2SumIPlusJ}
&\sum_j e_j g^{(2)}(A_k-B_j)-\sum_i d_i g^{(2)}(A_k-A_i)\nonumber\\
&=-2\pi i\frac{\partial}{\partial \tau}\frac{c_1}{2}\sum_i d_i A_i^2-2\pi i\frac{\partial}{\partial \tau}\frac{c_1}{2}\sum_j e_j B_j^2\nonumber\\
&\phantom{=}-2\pi i\sum_j e_j g^{(1)}(A_k-B_j)\frac{\partial}{\partial \tau}(A_k-B_j)+2\pi i\sum_i d_i g^{(1)}(A_k-A_i) \frac{\partial}{\partial \tau}(A_k-B_j).
\end{align}
The equations~\eqref{sec:connection_Bloch:Iprimeg2SumJ},
\eqref{sec:connection_Bloch:Jprimeg2SumI} and
\eqref{sec:connection_Bloch:Kg2SumIPlusJ} imply that for paths with $d\tau=0$
the differential equation
\begin{align}\label{sec:connection_Bloch:vanishingSumdGamma20}
d \sum_{ij}d_i e_j \Gt(\begin{smallmatrix}2\\0\end{smallmatrix}; A_i-B_j,\tau)
&=0
\end{align}
holds, the explicit calculation is shown in the \appref{app:vanish}. By
the same argument as for \eqn{sec:connection_Bloch:BlochRelationGamma1000}, we
therefore find another functional identity which can be interpreted as a
holomorphic analogue of the elliptic Bloch relation on the torus
\begin{align}\label{sec:connection_Bloch:BlochRelationGamma20}
\sum_{i,j}d_i e_j \Gt(\begin{smallmatrix}2\\0\end{smallmatrix}; A_i-B_j,\tau)=0.
\end{align}
To summarise, we managed to express the elliptic Bloch relation
\eqref{sec:connection_Bloch:BlochRelationTorus} in terms of iterated
integrals on the torus. 

Let us comment on the two holomorphic functional equations
\eqref{sec:connection_Bloch:BlochRelationGamma1000} and
\eqref{sec:connection_Bloch:BlochRelationGamma20} respectively, in terms of the
iterated integrals $\Gt$ on the torus which have the same structure as the
original elliptic Bloch relation: in the language of \rcite{BlochRelation}, it
turns out that the iterated integrals
$\Gt(\begin{smallmatrix}1&0\\0&0\end{smallmatrix}; z,\tau)$ and
$\Gt(\begin{smallmatrix}2\\0\end{smallmatrix}; z,\tau)$ are Steinberg
  functions. However, we have to be careful when using these functional
  identities: these iterated integrals are multi-valued and in order to
  reproduce
  \eqns{sec:connection_Bloch:BlochRelationGamma1000}{sec:connection_Bloch:BlochRelationGamma20}
  they have to be evaluated on the representatives of the zeros and poles of
  $F$ and $\ka-F$ which satisfy $\sum_i d_i A_i=0=\sum_j e_j B_j$, and not only
  such that these sums lie in the lattice $\Lambda$. These equations have been
  obtained by differential calculus of iterated integrals, which is simply the
  symbol calculus of an iterated integral with depth 1. Thus, together with
  \eqn{sec:connection_Bloch:BlochRelationRezReGamma10} we provide an
interpretation of the elliptic Bloch relation using the elliptic symbol
calculus of the iterated integrals $\Gt$ on the torus.

\subsection{The elliptic Bloch relation in the projective formulation}
By means of equations
\eqref{sec:connection_eMPL:Gamma1000E3}-\eqref{sec:connection_eMPL:zE3}, the
elliptic Bloch relation~\eqref{sec:connection_Bloch:BlochRelationTorus} can
also be expressed on the projective elliptic curve
\begin{align}\label{sec:connection_Bloch:BlochRelationProjCurve}
0&=\sum_{i,j}d_i e_j \DE\left(\frac{a_i}{b_j},q\right)\nonumber\\
&=-2\pi \sum_{i,j}d_i e_j \Big(\Re\left( \Gt(\begin{smallmatrix}1&0\\0&0\end{smallmatrix}; A_i-B_j,\tau)\right)+ \Re(\frac{\tau}{2\pi i}) \Re\left(\Gt(\begin{smallmatrix}2\\  0\end{smallmatrix}; A_i-B_j,\tau)\right)\nonumber\\
&\phantom{=}-\Re(A_i-B_j) \Re\left(\Gt(\begin{smallmatrix}1\\0\end{smallmatrix}; A_i-B_j,\tau)\right) \Big)\nonumber\\
&=-2\pi \sum_{i,j}d_i e_j \Big(\Re\left( \E_3(\begin{smallmatrix}1&0\\\infty&0\end{smallmatrix}; x_{ij},\vec{e})\right)+ \Re(\frac{\tau}{2\pi i}) \Re\left(E_3(\begin{smallmatrix}2\\\infty\end{smallmatrix}; x_{ij},\vec{e})\right)\nonumber\\
&\phantom{=}-\Re(\E_3(\begin{smallmatrix}0\\0\end{smallmatrix}; x_i,\vec{e})-\E_3(\begin{smallmatrix}0\\0\end{smallmatrix}; x_j,\vec{e})) \Re\left(\E_3(\begin{smallmatrix}1\\\infty\end{smallmatrix}; x_{ij},\vec{e})\right) \Big),
\end{align}
where $x_i=\wp(A_i)$, $x_j=\wp(B_j)$ and $x_{ij}=\wp\left(A_i-B_j\right)$. Similarly, the holomorphic functional relations~\eqref{sec:connection_Bloch:BlochRelationGamma1000} and \eqref{sec:connection_Bloch:BlochRelationGamma20} translate to 
\begin{align}\label{sec:connection_Bloch:holoRel10Curve}
\sum_{i,j}d_i e_j  \E_3(\begin{smallmatrix}1&0\\\infty&0\end{smallmatrix}; x_{ij},\vec{e})&=0
\end{align}
and
\begin{align}\label{sec:connection_Bloch:holoRel2Curve}
\sum_{i,j}d_i e_j  \E_3(\begin{smallmatrix}2\\\infty\end{smallmatrix}; x_{ij},\vec{e})&=0.
\end{align}

\section{Conclusions}\label{sec:conclusion}
In this article, we have investigated the elliptic Bloch-Wigner function $\DE$
in order to obtain functional relations of the iterated integrals
$\Gt$ on the torus and especially to formulate an elliptic analogue
of the five-term identity on the torus. This analysis led to several results:

\begin{itemize}
  \item The elliptic Bloch-Wigner function $\DE$, which is usually defined on
    the Tate curve, has been translated into the language of iterated elliptic
    integrals $\Gt$ on the torus. This was the precondition for the application
    of the elliptic symbol calculus.

  \item We have been extending the translation to the torus for two additional
    classes of functions:   the first class are the sums $\DD_{a,b}^{\E}$
    \cite{Zagier} on the Tate curve defined in \eqn{sec:toolbox_eMPL:DEab}.
    These functions play a crucial role in the calculation of modular graph
    functions \cite{DHoker:2015wxz}. The final formul\ae{} can be obtained by
    combining \eqn{app:eqDEab} with equations \eqref{Ell:EBW:BarE0g1},
    \eqref{sec:connecting_eMPL:EnandGamma}, \eqref{Ell:EBW:E0m} and
    \eqref{sec:connecting_eMPL:EnmandGamma}.  The representation of functions
    $\DD_{a,b}^{\E}$ in terms of $\Gt$'s on the torus allows for series expansions
    and therefore the investigation of relations between different modular
    graph functions. In particular, those representations might shed some light
    on the explicit construction of a representation of a single-valued map for
    genus-one string amplitudes as suggested in \rcite{Broedel:2018izr}. 

    The second class is a particular subclass of the former and is given by the functions $\mathcal{L}_n^{\E}$, defined
    in \eqn{sec:connection:LEm}, which are based on Ramakrishnan's single-valued
    polylogarithms: their representation on the torus follows from a combination
    of \eqn{sec:connection:LEm} with equations \eqref{Ell:EBW:BarE0g1},
    \eqref{sec:connecting_eMPL:EnandGamma}, \eqref{Ell:EBW:E0m} and
    \eqref{sec:connecting_eMPL:EnmandGamma}.

  \item Once representations on the torus do exist, it is just one further step
    to translate \cite{Broedel:2017kkb} those into representations in the
    projective formulation of the elliptic curve. In particular, we have chosen
    to express the elliptic Bloch-Wigner function in terms of iterated
    integrals $\E_3$ on the projective elliptic curve. For the two general classes of
    functions mentioned above this can be done in a straightforward manner as
    well.  
  
    The above expressions on the Tate curve and on the torus can be extended to
    the projective elliptic curve as well. In the case of the elliptic Bloch-Wigner
    function $\DE$, we have found \eqn{sec:connection_eMPL:DEandE3}, which
    expresses the value of $\DE$ on the Tate curve, the torus and the projective
    curve.
\end{itemize}
\begin{itemize}
  \item Employing the above translations, we have taken the elliptic Bloch
    relation \eqref{sec:toolbox_ellBloch:ellBlochRelation}, which is defined in
    terms of the elliptic Bloch-Wigner function $\DE$, from the Tate curve to
    the torus and the projective elliptic curve: the result is noted in
    \eqn{sec:connection_Bloch:BlochRelationProjCurve}. 

  \item The investigation of the elliptic Bloch relation on the torus led to
	  the holomorphic analogues of the elliptic Bloch relation given by
	  equations \eqref{sec:connection_Bloch:BlochRelationGamma1000} and
	  \eqref{sec:connection_Bloch:BlochRelationGamma20} as well as the
	  non-holomorphic equations
	  \eqref{sec:connection_Bloch:BlochRelationRezReGamma10} and
	  \eqref{sec:connection_Bloch:BlochRelationImzReGamma10}. Since
	  validity of those equations can be proved using the elliptic symbol
	  calculus on the torus, we hereby found an alternative proof and
	  interpretation of the elliptic Bloch relation.

  \item Translating the elliptic Bloch relation even one step further to the
    projective elliptic curve yields a possibility to write down functional equations in
    terms of algebraic arguments. However, this is possible only when
    restricting the parametrising rational function to lines.  Beyond lines,
    the complexity of the calculation of zeros of the parametrising rational
    function prevents the corresponding functional relation from being purely
    algebraic in the arguments. 

    In particular, this applies to the elliptic analogue of the five-term
    identity, i.e.\ the functional relation induced via the elliptic Bloch
    relation by the elliptic generalisation
    $\eqref{sec:toolbox_ellipticfunctions:Ffiveterm}$ of the rational function
    \eqref{sec:invitation:Bloch:5termf} parametrising the classical five-term
    identity.
    
    In general, due to the complexity of Abel's map, it can not be expected,
    that generic functional relations generated by the Bloch relation may be
    formulated explicitly in terms of algebraic arguments on the torus or the
    Tate curve, respectively. 

\end{itemize}
In the classical case, the five-term identity is conjectured to generate all
other functional identities between the dilogarithm.  It would be interesting
to investigate, whether a similar conjecture can be formulated for the elliptic
case. As described at the end of \subsecref{sec:toolbox:ellBloch}, Gangl and
Zagier state \cite{ZagierGangl} that the elliptic Bloch relation, the symmetry
and the duplication relation are expected to generate all the functional
relations of the elliptic Bloch group associated to the elliptic Bloch-Wigner
function. However, the construction of higher elliptic Bloch groups and in
particular the corresponding group of functional relations awaits further
investigation. 
 
In particular it would be nice, if this
conjecture came along with a geometric interpretation of the elliptic
analogue of the five-term identity, similar, but for sure more complicated,
to the classical case.  

While the elliptic Bloch relation is capable of providing a large class
of functional relations, it is not clear, whether there might exist other
structures or similar mechanisms generating further relations, for example
based on functions beyond the elliptic Bloch-Wigner function. Based on the
experience gained from the current implementation and the investigations in
this article, a straightforward generalisation seems unlikely. 

In the context of Feynman diagrams leading to elliptic polylogarithms, the main
recent focus was usually on evaluating the integrals. Despite the lack of a
dedicated application of the functional relations investigated in this article,
we hope that our results will facilitate simplifications and unravel new
structures in further elliptic Feynman calculations to come. 

%%%%%%%%%%%%%%%%%%%%%%%%%%%%%%%%%%%%%%%%%%%%%%%%%%%%%%%%%%%%%%%%%%%%%%%%%%%%%%
%%%%%%%%%%%%%%%%%%%%%%%%%%%%%%%%%%%%%%%%%%%%%%%%%%%%%%%%%%%%%%%%%%%%%%%%%%%%%%
%%%%%%%%%%%%%%%%%%%%%%%%%%%%%%%%%%%%%%%%%%%%%%%%%%%%%%%%%%%%%%%%%%%%%%%%%%%%%%

\subsection*{Acknowledgments}

We are grateful to Herbert Gangl and Nils Matthes for the very careful reading
and their helpful comments on the draft of this article. Furthermore, we would
like to thank Rob Klabbers for various interesting discussions. JB would like
to thank the Institute for Theoretical Studies at ETH Z\"urich for organisation
of the program "Modular forms, periods and scattering amplitudes", where part
of this work was done. AK would like to thank the IMPRS for Mathematical and
Physical Aspects of Gravitation, Cosmology and Quantum Field Theory, of which
he is a member and which renders his studies possible.  Furthermore, AK is
supported by the Swiss Studies Foundation, to which he would like to express
his gratitude. 

%%%%%%%%%%%%%%%%%%%%%%%%%%%%%%%%
%%%%%%%%%%%%%%%%%%%%%%%%%%%%%%%%

\newpage

\section*{Appendix}
\appendix

%%%%%%%%%%%%%%%%%%%%%%%%%%%%%%%%
%%%%%%%%%%%%%%%%%%%%%%%%%%%%%%%%

\section{Group addition on $E(\ZC)$}\label{app:chord-tangent}
The geometric picture of the addition on the elliptic curve is that two
distinct points $P_1=\pc{x_1}{y_1}$ and $P_2=\pc{x_2}{y_2}$ with $y_1\neq \pm
y_2$ form a line which intersects the elliptic curve $y^2=4x^3-g_2 x-g_3$ at a
third point $-P_3=\pc{x_3}{-y_3}$. The sum $P_3=P_1+P_2$ is defined as being
the projection of $-P_3=\pc{x_3}{-y_3}$ to the negative $y$-coordinate
$P_3=\pc{x_3}{y_3}$. Thus, two points with their $y$-coordinate being of the
opposite sign are indeed the inverse of each other with $\infty=[0\!:1\!:0]$
being the unit element since the line defined by $P_3$ and $-P_3$ intersects
the elliptic curve only at infinity. The algebraic description is the
following. For $P_1$ and $P_2$ as above, the line intersecting them is given by
$y=\lambda x+\mu$, where
\begin{equation}
\lambda=\frac{y_2-y_1}{x_2-x_1}\,,\qquad \mu=\frac{y_1x_2-y_2x_1}{x_2-x_1}\,.
\end{equation}
The x-coordinate of the third point $-P_3=\pc{x_3}{-y_3}$ intersecting the line
and the elliptic curve is the third solution (besides $x_1$ and $x_2$) of the
cubic equation $(\lambda x+\mu)^2=4x^3-g_2x-g_3$, which is in terms of $x_1$
and $x_2$ 
\begin{align}
x_3=-x_1-x_2+\frac{\lambda^2}{4}\,.
\end{align}
The $y$ coordinate of $P_3$ is then simply the negative of the $y$ coordinate determined by the line and $x_3$,
\begin{align}
y_3=-\lambda x_3-\mu\,.
\end{align}
The last case we need to consider is if the points $P_1$ and $P_2$ are
identical and not the unit element, i.e.\ $P_1=P_2=P=\pc{x_P}{y_P}$. For
$y_P\neq 0$, the above description of taking the line intersecting $P_1$ and
$P_2$ degenerates to taking the tangent on the elliptic curve at $P$. The sum
$2P=P+P=\pc{x_{2P}}{y_{2P}}$ is then again the projection of the second point
lying on this tangent and the elliptic curve with respect to the
$x$-coordinate. Algebraically, this corresponds to 
\begin{equation}
\lambda=\frac{12 x_P^2-g_2}{2 y_P}\,,\qquad \mu = y_P-\lambda x_P
\end{equation}
and 
\begin{equation}
x_{2P}=-2x_P+\frac{\lambda^2}{4}\,,\qquad y_{2P}=-\lambda x_{2P}-\mu
\end{equation}
as before. In the case of $y_P=0$, the point $P$ is inverse to itself, such
that $P+P=P-P=\infty$. These addition rules exactly agree with the well-known
addition formula of the Weierstrass $\wp$ function 
\begin{align}
\wp(x_1+x_2)&=-\wp(x_1)-\wp(x_2)+\frac{1}{4}\left(\frac{\wp'(x_2)-\wp'(x_1)}{\wp(x_2)-\wp(x_1)}\right)^2
\end{align}
for $x_1\neq x_2$ and similar for its derivative. This ensures that
$\xi_{\Lambda,E}$ defined in \eqn{sec:toolbox_ellipticfunctions:xiLE} is indeed
a homomorphism.
\section{$q$-expansion of integration kernels and elliptic polylogarithms}\label{app:qExpansion}
Starting from the $q$-expansion of the Jacobi $\theta$ function, the
$q$-expansion of the integration kernels are obtained via the generating
Eisenstein-Kronecker series \cite{Broedel:2014vla} and are given by
\begin{align}\label{sec:toolbox_eMPL:qExpansiong0}
g^{(0)}(z,\tau) &=1\,,
\end{align}
\begin{align}\label{sec:toolbox_eMPL:qExpansiong1}
g^{(1)}(z,\tau)&=\pi \cot(\pi z)+ 4\pi \sum_{k,l>0}\sin(2\pi k z)q^{kl}
\end{align}
and for $m>0$ by
\begin{align}\label{sec:toolbox_eMPL:qExpansiong2n}
g^{(2m)}(z,\tau)&=-2\zeta_{2m}-2\frac{(2\pi i)^{2m}}{(2m-1)!} \sum_{k,l>0}\cos(2\pi k z) l^{2m-1}q^{kl}\,,
\end{align}
as well as by
\begin{align}\label{sec:toolbox_eMPL:qExpansiong2n1}
g^{(2m+1)}(z,\tau)&=-2 i \frac{(2\pi i)^{2m+1}}{(2m)!} \sum_{k,l>0}\sin(2\pi k z) l^{2m}q^{kl}\,.
\end{align}
The $(n-1)$-fold integration of the regularised integral \eqref{sec:toolbox_eMPL:qExpG1}, i.e.\
\begin{align}
\Gt_{\text{reg}}(\begin{smallmatrix}1\\0 \end{smallmatrix}; z,\tau)
&= \log(1-e^{2\pi i z})-\pi i z+4\pi \sum_{k,l>0}\frac{1}{2\pi k}\left(1-\cos(2\pi k z)\right)q^{kl}\,,
\end{align}
and the $n$-fold integration of the above integration kernels $g^{(m)}(z,\tau)$
for $m>1$ can be determined analytically. This yields the following efficient
method to write down their $q$-expansion and, hence, for their numerical
evaluation. The central observation is that for $n\geq 0$ the $2n$-fold
integration of $\sin(2 \pi k z)$ with $k\in \ZZ$ is given by 
\begin{align}
\int_0^zdz_1\int_0^{z_1}dz_2\dots\int_0^{z_{2n-1}}dz_{2n}\sin(2\pi k z_{2n})&=\frac{(-1)^n}{(2 \pi k)^{2n}}\sin(2 \pi k z)\nonumber\\
&\phantom{=}+\sum_{j=1}^n \frac{(-1)^{n-j}}{(2 \pi k)^{2n+1-2j}}\frac{z^{2j-1}}{(2j-1)!}
\end{align}
and the $(2n+1)$-fold integration by
\begin{align}
\int_0^zdz_1\int_0^{z_1}dz_2\dots\int_0^{z_{2n}}dz_{2n+1}\sin(2\pi k z_{2n+1})&=\frac{(-1)^{n+1}}{(2 \pi k)^{2n+1}}\cos(2 \pi k z)\nonumber\\
&\phantom{=}+\sum_{j=0}^n \frac{(-1)^{n-j}}{(2 \pi k)^{2n+1-2j}}\frac{z^{2j}}{(2j)!}\,.
\end{align}
A similar result holds for the iterative integration of $\cos(2 \pi k z)$,
\begin{align}
\int_0^zdz_1\int_0^{z_1}dz_2\dots\int_0^{z_{2n-1}}dz_{2n}\cos(2\pi k z_{2n})&=\frac{(-1)^n}{(2 \pi k)^{2n}}\cos(2 \pi k z)\nonumber\\
&\phantom{=}+\sum_{j=0}^{n-1} \frac{(-1)^{n+1-j}}{(2 \pi k)^{2n-2j}}\frac{z^{2j}}{(2j)!}
\end{align}
and
\begin{align}
\int_0^zdz_1\int_0^{z_1}dz_2\dots\int_0^{z_{2n}}dz_{2n+1}\cos(2\pi k z_{2n+1})&=\frac{(-1)^n}{(2 \pi k)^{2n+1}}\sin(2 \pi k z)\nonumber\\
&\phantom{=}+\sum_{j=1}^n \frac{(-1)^{n-j}}{(2 \pi k)^{2n+2-2j}}\frac{z^{2j-1}}{(2j-1)!}\,.
\end{align}
Combining the above results yields the following $q$-expansions of the elliptic
polylogarithms of the form $\Gt(\underbrace{\begin{smallmatrix}0 &\dots &0&m\\
0&\dots &0&0 \end{smallmatrix}}_{n}; z,\tau)$ for $n\geq 1$:
\begin{small}
\begin{align}\label{app:qExp:G12n}
\Gt(\underbrace{\begin{smallmatrix}0 &\dots &0&1\\ 0&\dots &0&0 \end{smallmatrix}}_{2n}; z,\tau)&=-\frac{1}{(2\pi i)^{2n-1}}\Li_{2n}(e^{2\pi i z})+\sum_{j=1}^{2n-1}\frac{\zeta_{j+1}}{(2 \pi i)^{j}}\frac{z^{2n-1-j}}{(2n-1-j)!}-\pi i \frac{z^{2n}}{(2n)!}\nonumber\\
&\phantom{=}+(-1)^n 4\pi\sum_{k,l>0}\frac{1}{(2 \pi k)^{2n}}\left(\sin(2 \pi k z)+\sum_{j=1}^n\frac{(-1)^{j}}{(2 \pi k)^{1-2j}}\frac{z^{2j-1}}{(2j-1)!} \right)q^{kl}\,,
\end{align}
\begin{align}\label{app:qExp:G12n1}
\Gt(\underbrace{\begin{smallmatrix}0 &\dots &0&1\\ 0&\dots &0&0 \end{smallmatrix}}_{2n+1}; z,\tau)&=-\frac{1}{(2\pi i)^{2n}}\Li_{2n+1}(e^{2\pi i z})+\sum_{j=1}^{2n}\frac{\zeta_{j+1}}{(2 \pi i)^{j}}\frac{z^{2n-j}}{(2n-j)!}-\pi i \frac{z^{2n+1}}{(2n+1)!}\nonumber\\
&\phantom{=}+(-1)^{n+1} 4\pi\sum_{k,l>0}\frac{1}{(2 \pi k)^{2n+1}}\left(\cos(2 \pi k z)+\sum_{j=0}^{n} \frac{(-1)^{j+1}}{(2 \pi k)^{-2j}}\frac{z^{2j}}{(2j)!}\right)q^{kl}
\end{align}
\end{small}
and for $m>1$ and $n\geq 0$
\begin{small}
\begin{align}\label{app:qExp:G2m2n}
\Gt(\underbrace{\begin{smallmatrix}0 &\dots &0&2m\\ 0&\dots &0&0 \end{smallmatrix}}_{2n}; z,\tau)&=-2\zeta_{2m}\frac{z^{2n}}{(2n)!}\nonumber\\
&\phantom{=}+(-1)^{n+1}2\frac{(2\pi i)^{2m}}{(2m-1)!} \sum_{k,l>0}\frac{1}{(2 \pi k)^{2n}}\left(\cos(2 \pi k z)+\sum_{j=0}^{n-1} \frac{(-1)^{1+j}}{(2 \pi k)^{-2j}}\frac{z^{2j}}{(2j)!}\right) l^{2m-1}q^{kl}\,,
\end{align}
\begin{align}\label{app:qExp:G2m2n1}
\Gt(\underbrace{\begin{smallmatrix}0 &\dots &0&2m\\ 0&\dots &0&0 \end{smallmatrix}}_{2n+1}; z,\tau)&=-2\zeta_{2m}\frac{z^{2n+1}}{(2n+1)!}\nonumber\\
&\phantom{=}+(-1)^{n+1}2\frac{(2\pi i)^{2m}}{(2m-1)!} \sum_{k,l>0}\frac{1}{(2\pi k)^{2n+1}}\left(\sin(2 \pi k z)+\sum_{j=1}^n \frac{(-1)^{j}}{(2 \pi k)^{1-2j}}\frac{z^{2j-1}}{(2j-1)!}\right) l^{2m-1}q^{kl}\,,
\end{align}
\end{small}
as well as
\begin{small}
\begin{align}\label{app:qExp:G2m12n}
\Gt(\underbrace{\begin{smallmatrix}0 &\dots &0&2m+1\\ 0&\dots &0&0 \end{smallmatrix}}_{2n}; z,\tau)&=(-1)^{n+1}2 i \frac{(2\pi i)^{2m+1}}{(2m)!} \sum_{k,l>0}\frac{1}{(2\pi k)^{2n}}\left(\sin(2 \pi k z)+\sum_{j=1}^n \frac{(-1)^{j}}{(2 \pi k)^{1-2j}}\frac{z^{2j-1}}{(2j-1)!}\right) l^{2m}q^{kl}\,,
\end{align}
\begin{align}\label{app:qExp:G2m12n1}
\Gt(\underbrace{\begin{smallmatrix}0 &\dots &0&2m+1\\ 0&\dots &0&0 \end{smallmatrix}}_{2n+1}; z,\tau)&=(-1)^n2 i \frac{(2\pi i)^{2m+1}}{(2m)!} \sum_{k,l>0}\frac{1}{(2\pi k)^{2n+1}}\left(\cos(2 \pi k z)+\sum_{j=0}^n \frac{(-1)^{1+j}}{(2 \pi k)^{-2j}}\frac{z^{2j}}{(2j)!}\right) l^{2m}q^{kl}\,
\end{align}
\end{small}
where, in the above formula, we denote the integration kernels by
$\Gt(\underbrace{\begin{smallmatrix}0 &\dots &0&m\\ 0&\dots &0&0
\end{smallmatrix}}_{0}; z,\tau)=g^{(m)}(z,\tau)$.

%%%%%%%%%%%%%%%%%%%%%%%%%%%%%%%%%%%%
%%%%%%%%%%%%%%%%%%%%%%%%%%%%%%%%%%%%

\section{Integration constants as elliptic zeta values}\label{app:eMZV}
This section is dedicated to the calculation of the integration constants from
\subsecref{sec:connection:TatetoTorus}, i.e.\

\begin{align}\label{Ell:EBW:Enm2}
\E_{n,-m}(1,1,q)&=-(1-(-1)^{n+m})\ELi_{n,-m}(1,1,q)
\end{align}
for $n\geq 1$ and $m\geq 0$, where
\begin{align}
\ELi_{n,m}(1,1,q)&= \sum_{k,l>0}\frac{1}{k^n}\frac{1}{l^m}q^{kl}\,,
\end{align}
in terms of the elliptic zeta values 
\begin{align}
\omega_n(m;\tau)&=\Gt(\underbrace{\begin{smallmatrix}0 &\dots &0&m\\ 0&\dots &0&0 \end{smallmatrix}}_{n}; 1,\tau)
\end{align}
defined in \eqn{sec:toolbox:ezv}.

Let us begin with the case $m=0$, where we consider for $n\geq 2$
\begin{equation}
\E_n(1,1,q)=
-\frac{1}{2}\left(1-(-1)^n\right)\zeta_n+\E_{n,0}(1,1,q)=\begin{cases}
0&n\text{ even}\\ -\zeta_n-2\ELi_{n,0}(1,1,q)&n\text{ odd}
\end{cases}
\end{equation}
as defined in \eqn{Ell:EBW:BarEnDef}. While $\E_n(1,1,q)$ vanishes for even
$n$, for odd $n\geq 3$ the integration constants $\E_n(1,1,q)$ will turn out to
be linear combinations of
$\omega_1(1;\tau),\omega_3(1;\tau),\dots,\omega_n(1;\tau)$, which we derive
similarly as the result for $n=1$ given by \eqn{sec:connecting_eMPL:ELi1}
\begin{align}
2  \ELi_{1,0}(1,1,q)&=\omega_2(1;\tau)+\frac{\pi i}{2}\,.
\end{align}
In order to do so, let $n\geq 4$ be even. In this case, the recursion given by
\eqn{sec:connecting_eMPL:EnandGamma} evaluated at one, which is based on the
partial differential equation \eqref{sec:connecting_eMPL:deqEn(z)},
\begin{align}
\frac{\partial}{\partial z}\E_n(z,1,\tau)&= 2\pi i \E_{n-1}(z,1,\tau)\,,
\end{align}
takes the explicit form
\begin{align}
0&=\E_n(1,1,q)\nonumber\\
&=2\pi i\int_0^1 dz_0 E_{n-1}(z_0,1,\tau)\nonumber\\
&=(2\pi i)^2\int_0^1 dz_0\int_0^{z_0}dz_1 E_{n-2}(z_1,1,\tau)+2\pi i \E_{n-1}(1,1,q)\nonumber\\
&=(2\pi i)^4\int_0^1 dz_0\int_0^{z_0}dz_1\int_0^{z_1}dz_2\int_0^{z_2}dz_3 E_{n-4}(z_3,1,\tau)\nonumber\\
&\phantom{=}+\frac{(2\pi i)^3}{3!}\E_{n-3}(1,1,q)+2\pi i \E_{n-1}(1,1,q)\nonumber\\
&=(2\pi i)^{n-2}\int_0^1 dz_0\int_0^{z_0}dz_1\dots\int_0^{z_{n-4}}dz_{n-3} E_{2}(z_{n-3},1,\tau)\nonumber\\
&\phantom{=}+\frac{(2\pi i)^{n-3}}{(n-3)!}\E_{3}(1,1,q)+\dots+2\pi i \E_{n-1}(1,1,q)\,.
\end{align}
Plugging in
\begin{align}
\E_2(t,1, q)&=2\pi i \left(\Gt(\begin{smallmatrix}0&1\\0&0\end{smallmatrix}; z,\tau)-\omega_2(1;\tau)z\right)
\end{align}
from \eqn{Ell:EBW:E2Gamma}, and solving for $\E_{n-1}(1,1,q)$ leads to the recursive formula 
\begin{align}\label{app:intConst:EnRecursive}
\E_{n-1}(1,1,q)&=(2\pi i)^{n-2}\left(\frac{\omega_2(1;\tau)}{(n-1)!}-\omega_n(1;\tau)\right)-\frac{(2\pi i)^{n-4}}{(n-3)!}\E_{3}(1,1,q)-\dots-\frac{(2\pi i)^2}{3!} \E_{n-3}(1,1,q)\,.
\end{align}
The first examples are $n=4$
\begin{align}\label{app:intConst:E3}
\E_3(1,1,q)&=(2\pi i)^{2}\left(\frac{\omega_2(1;\tau)}{3!}-\omega_4(1;\tau)\right)\,,
\end{align}
$n=6$
\begin{align}\label{app:intConst:E5}
\E_5(1,1,q)&=(2\pi i)^{4}\left(\frac{\omega_2(1;\tau)}{5!}-\omega_6(1;\tau)\right)-\frac{(2\pi i)^{2}}{3!}\E_{3}(1,1,q)\nonumber\\
&=(2\pi i)^{4}\left(\frac{\omega_2(1;\tau)}{5!}-\omega_6(1;\tau)\right)-\frac{(2\pi i)^{2}}{3!}(2\pi i)^{2}\left(\frac{\omega_2(1;\tau)}{3!}-\omega_4(1;\tau)\right)\nonumber\\
&=(2\pi i)^4\left(\left(\frac{1}{5!}-\frac{1}{3!3!}\right)\omega_2(1;\tau)+\frac{1}{3!}\omega_4(1;\tau)-\omega_6(1;\tau)\right)
\end{align}
and $n=8$
\begin{align}\label{app:intConst:E7}
\E_7(1,1,q)&=(2\pi i)^{6}\left(\frac{\omega_2(1;\tau)}{7!}-\omega_8(1;\tau)\right)-\frac{(2\pi i)^{4}}{5!}\E_{3}(1,1,q)-\frac{(2\pi i)^{2}}{3!}\E_{5}(1,1,q)\nonumber\\
&=(2\pi i)^{6}\left(\frac{\omega_2(1;\tau)}{7!}-\omega_8(1;\tau)\right)-\frac{(2\pi i)^{4}}{5!}(2\pi i)^{2}\left(\frac{\omega_2(1;\tau)}{3!}-\omega_4(1;\tau)\right)\nonumber\\
&\phantom{=}-\frac{(2\pi i)^{2}}{3!}(2\pi i)^4\left(\left(\frac{1}{5!}-\frac{1}{3!3!}\right)\omega_2(1;\tau)+\frac{1}{3!}\omega_4(1;\tau)-\omega_6(1;\tau)\right)\nonumber\\
&=(2\pi i)^{6}\Bigg(\left(\frac{1}{7!}-\frac{1}{5!3!}-\frac{1}{3!5!}+\frac{1}{3!3!3!}\right)\omega_2(1;\tau)\nonumber\\
&\phantom{=}+\left(\frac{1}{5!}-\frac{1}{3!3!}\right)\omega_4(1;\tau)+\frac{1}{3!}\omega_6(1;\tau)-\omega_8(1;\tau)\Bigg).
\end{align}
This recursive structure can be expressed explicitly in terms of the series
\begin{equation}\label{app:intConst:akDef}
d_k=\begin{cases}
-1&k=1\\
0&k\text{ even}\\
-\frac{d_1}{k!}-\frac{d_{3}}{(k-2)!}-\dots-\frac{d_{k-2}}{3!}&k\text{ odd}
\end{cases}
\end{equation}
with the final result being for any natural number $n\geq 1$
\begin{align}\label{app:intConst:EnExplicit}
E_{2n+1}(1,1,q)&=(2\pi i)^{2n}\sum_{k=0}^{n}d_{2k+1}\omega_{2n+2-2k}(1;\tau)\,, 
\end{align}
which can be checked inductively as follows: first, note that the series $d_k$ begins with
\begin{equation}
d_1=-1\,,\qquad d_3=-\frac{1}{3!} (-1)=\frac{1}{3!}\,,\qquad d_5=-\frac{1}{5!}(-1)-\frac{1}{3!}\frac{1}{3!}=\frac{1}{5!}-\frac{1}{3!3!}
\end{equation}
and
\begin{equation}
 d_7=\frac{1}{7!}-\frac{1}{5!}\frac{1}{3!}-\frac{1}{3!}\left(\frac{1}{5!}-\frac{1}{3!3!}\right)=\frac{1}{7!}-\frac{1}{5!3!}-\frac{1}{3!5!}+\frac{1}{3!3!3!}\,,
\end{equation}
such that for $n=1,2,3$ the explicit formula \eqref{app:intConst:EnExplicit} is
indeed in agreement with the first three examples
\eqref{app:intConst:E3}-\eqref{app:intConst:E7}. For the general case, let
$n>1$ and assume that the explicit formula \eqref{app:intConst:EnExplicit}
holds for $n-1$, such that the recursive formula
\eqref{app:intConst:EnRecursive} implies
\begin{align}
\E_{2n+1}(1,1,q)&=(2\pi i)^{2n}\left(\frac{\omega_2(1;\tau)}{(2n+1)!}-\omega_{2n+2}(1;\tau)\right)\nonumber\\
&\phantom{=}-\frac{(2\pi i)^{2n-3}}{(2n-1)!}\E_{3}(1,1,q)-\dots-\frac{(2\pi i)^2}{3!} \E_{2n-1}(1,1,q)\nonumber\\
&=(2\pi i)^{2n}\left(\frac{\omega_2(1;\tau)}{(2n+1)!}-\omega_{2n+2}(1;\tau)\right)\nonumber\\
&\phantom{=}-\frac{(2\pi i)^{2n-2}}{(2n-1)!}\left((2\pi i)^{2}\sum_{k=0}^{1}d_{2k+1}\omega_{4-2k}(1;\tau)\right)-\dots\nonumber\\
&\phantom{=}-\frac{(2\pi i)^2}{3!} \left((2\pi i)^{2n-2}\sum_{k=0}^{n-1}d_{2k+1}\omega_{2n-2k}(1;\tau)\right)\nonumber\\
&=(2\pi i)^{2n}
\left(\sum_{l=0}^{n-1}\sum_{k=0}^l\frac{(-d_{2k+1})}{(2n+1-2l)!}\omega_{2l+2-2k}(1;\tau)-\omega_{2n+2}(1;\tau)\right)\nonumber\\
&=(2\pi i)^{2n}
\left(\sum_{k=0}^{n-1}\sum_{l=k}^{n-1}\frac{(-d_{2k+1})}{(2n+1-2l)!}\omega_{2(l-k)+2}(1;\tau)-\omega_{2n+2}(1;\tau)\right)\nonumber\\
&=(2\pi i)^{2n}
\left(\sum_{k=0}^{n-1}\sum_{m=0}^{n-k-1}\frac{-d_{2k+1}}{(2(n-m)+1-2k)!}\omega_{2m+2}(1;\tau)-\omega_{2n+2}(1;\tau)\right)\nonumber\\
&=(2\pi i)^{2n}
\left(\sum_{m=0}^{n-1}\left(\sum_{k=0}^{n-m-1}\frac{-d_{2k+1}}{(2(n-m)+1-2k)!}\right)\omega_{2m+2}(1;\tau)-\omega_{2n+2}(1;\tau)\right)\nonumber\\
&=(2\pi i)^{2n}
\left(\sum_{m=0}^{n-1}d_{2(n-m)+1}\omega_{2m+2}(1;\tau)-\omega_{2n+2}(1;\tau)\right)\nonumber\\
&=(2\pi i)^{2n}
\sum_{m=0}^{n}d_{2(n-m)+1}\omega_{2m+2}(1;\tau)\nonumber\\
&=(2\pi i)^{2n}
\sum_{k=0}^{n}d_{2k+1}\omega_{2n+2-2k}(1;\tau)\,,
\end{align}
where we used the definition \eqref{app:intConst:akDef} of $d_{2n+1}$ for $n>1$, i.e.\
\begin{align}
d_{2n+1}&=\sum_{k=0}^{n-1}\frac{-d_{2k+1}}{(2n+1-2k)!}\,.
\end{align}
This calculation proves the explicit formula \eqref{app:intConst:EnExplicit}.
\smallskip

For $m\neq 0$, the two trivial cases, where the $\E_{n,-m}(1,1,q)$ vanish by
definition, are $n$ and $m$ both being either even or both being odd and have
to be distinguished. Starting with the former and using the partial
differential equation \eqref{sec:connecting_pdeEnm}
\begin{align}
\frac{\partial}{\partial z}\E_{n,m}(z,1,\tau)&=2\pi i \E_{n-1,m}(z,1,\tau)\,,
\end{align}
a similar recursion formula as above, which corresponds to the evaluation of
\eqn{sec:connecting_eMPL:EnmandGamma} at one, can be obtained for even $m\geq
1, n\geq 4$
\begin{align}
0&=\E_{n,-m}(1,1,q)\nonumber\\
&=2\pi i\int_0^1 dz_0 E_{n-1,-m}(z_0,1,\tau)\nonumber\\
&=(2\pi i)^2\int_0^1 dz_0\int_0^{z_0}dz_1 E_{n-2,-m}(z_1,1,\tau)+2\pi i \E_{n-1,-m}(1,1,q)\nonumber\\
&=(2\pi i)^4\int_0^1 dz_0\int_0^{z_0}dz_1\int_0^{z_1}dz_2\int_0^{z_2}dz_3 E_{n-4,-m}(z_3,1,\tau)\nonumber\\
&\phantom{=}+\frac{(2\pi i)^3}{3!}\E_{n-3,-m}(1,1,q)+2\pi i \E_{n-1,-m}(1,1,q)\nonumber\\
&=(2\pi i)^{n-2}\int_0^1 dz_0\int_0^{z_0}dz_1\dots\int_0^{z_{n-4}}dz_{n-3} E_{2,-m}(z_{n-3},1,\tau)\nonumber\\
&\phantom{=}+\frac{(2\pi i)^{n-3}}{(n-3)!}\E_{3,-m}(1,1,q)+\dots+2\pi i \E_{n-1,-m}(1,1,q)\,.
\end{align}
This can be solved for $\E_{n-1,-m}(1,1,q)$ using the result from equation \eqref{Ell:EBW:E2mGamma} for $m$ even, i.e.\
\begin{align}
\E_{2,-m}(t,1,q)
&= \frac{m!}{(2\pi i)^{m-1}}   \Gt(\begin{smallmatrix}0&m+1\\ 0& 0\end{smallmatrix}; z,\tau)+2\pi i \E_{1,-m}(1,1,q)z \,,
\end{align}
which leads to
\begin{align}
\E_{1,-m}(1,1,q)
&=- \frac{m!}{(2\pi i)^{m}}   \omega_{2}(m+1;\tau) \,
\end{align}
upon evaluation at one, such that
\begin{align}\label{app:intConst:EnmEvenRecursive}
\E_{n-1,-m}(1,1,q)&=m!(2\pi i)^{n-m-2}\left(\frac{\omega_{2}(m+1;\tau)}{(n-1)!}-\omega_{n}(m+1;\tau)\right)\nonumber\\
&\phantom{=}-\frac{(2\pi i)^{n-4}}{(n-3)!}\E_{3,-m}(1,1,q)-\dots-\frac{(2\pi i)^2}{3!} \E_{n-3,-m}(1,1,q)\,.
\end{align}
This evaluates e.g.\ for $n=4$ to 
\begin{align}\label{app:intConst:E3mEven}
\E_{3,-m}(1,1,q)&=m!(2\pi i)^{2-m}\left(\frac{\omega_{2}(m+1;\tau)}{3!}-\omega_{4}(m+1;\tau)\right)
\end{align}
and for $n=6$ to
\begin{small}
\begin{align}\label{app:intConst:E5mEven}
\E_{5,-m}(1,1,q)&=m!(2\pi i)^{4-m}\left(\frac{\omega_{2}(m+1;\tau)}{5!}-\omega_{6}(m+1;\tau)\right)-\frac{(2\pi i)^{2}}{3!}\E_{3,-m}(1,1,q)\nonumber\\
&=m!(2\pi i)^{4-m}\left(\frac{\omega_{2}(m+1;\tau)}{5!}-\omega_{6}(m+1;\tau)\right)\nonumber\\
&\phantom{=}-\frac{(2\pi i)^{2}}{3!}m!(2\pi i)^{2-m}\left(\frac{\omega_{2}(m+1;\tau)}{3!}-\omega_{4}(m+1;\tau)\right)\nonumber\\
&=m!(2\pi i)^{4-m}\left(\left(\frac{1}{5!}-\frac{1}{3!3!}\right)\omega_{2}(m+1;\tau)+\frac{1}{3!}\omega_{4}(m+1;\tau)-\omega_{6}(m+1;\tau)\right)\,.
\end{align}
\end{small}
Since this recursion is the same as the one for $m=0$ up to the factor $m!(2\pi
i)^{-m}$ and the higher elliptic zeta values, the explicit formula solving this
recursion corresponds to the previous formula given in
\eqn{app:intConst:EnExplicit} and can immediately be written down and proven as
before. The result is that for any natural number $n\geq 0$ and even $m\geq 1$ 
\begin{align}\label{app:intConst:EnmEvenExplicit}
E_{2n+1,-m}(1,1,q)&=m!(2\pi i)^{2n-m}\sum_{k=0}^{n}a_{2k+1}\omega_{2n+2-2k}(m+1;\tau)\,.
\end{align}

The remaining case is $m\geq 1$, $n\geq 3$ both odd. The recursive formula can
be obtained as before
\begin{align}
0&=\E_{n,-m}(1,1,q)\nonumber\\
&=2\pi i\int_0^1 dz_0 E_{n-1,-m}(z_0,1,\tau)\nonumber\\
&=(2\pi i)^2\int_0^1 dz_0\int_0^{z_0}dz_1 E_{n-2,-m}(z_1,1,\tau)+2\pi i \E_{n-1,-m}(1,1,q)\nonumber\\
&=(2\pi i)^{n-1}\int_0^1 dz_0\int_0^{z_0}dz_1\dots\int_0^{z_{n-3}}dz_{n-2} E_{1,-m}(z_{n-2},1,\tau)\nonumber\\
&\phantom{=}+\frac{(2\pi i)^{n-2}}{(n-2)!}\E_{2,-m}(1,1,q)+\dots+2\pi i \E_{n-1,-m}(1,1,q)\,.
\end{align}
As above, we can plug in $\E_{1,-m}(t,1,q)$ given by \eqn{Ell:EBW:E1mGamma} for $m$ odd, i.e.\
\begin{align}\label{app:intConst:E1mOddRecursive}
\E_{1,-m}(t,1,q)
&= \frac{m!}{(2\pi i)^{m}} \left( \Gt(\begin{smallmatrix}m+1\\  0\end{smallmatrix}; z,\tau)+2 \zeta_{m+1}z\right) \,,
\end{align}
and solve for $\E_{n-1,-m}(1,1,q)$, which yields
the recursive formula
\begin{align}\label{app:intConst:EnmOddRecursive}
\E_{n-1,-m}(1,1,q)&=m!(2\pi i)^{n-m-2}\left(-2\frac{\zeta_{m+1}}{n!}-\omega_{n}(m+1;\tau)\right)\nonumber\\
&\phantom{=}-\frac{(2\pi i)^{n-3}}{(n-2)!}\E_{2,-m}(1,1,q)-\dots-\frac{(2\pi i)^2}{3!} \E_{n-3,-m}(1,1,q)\,.
\end{align}
Evaluation of \eqn{app:intConst:E1mOddRecursive} at one, or considering the $q$-expansion \eqref{app:qExp:G2m2n1}, leads to the following connection between the even zeta values and the elliptic zeta values
\begin{align}
\omega_{1}(m+1;\tau)&=-2\zeta_{m+1}\,,
\end{align}
such that the above recursion can be expressed in the more familiar form 
\begin{align}\label{app:intConst:EnmOddRecursiveAlt}
\E_{n-1,-m}(1,1,q)&=m!(2\pi i)^{n-m-2}\left(\frac{\omega_{1}(m+1;\tau)}{n!}-\omega_{n}(m+1;\tau)\right)\nonumber\\
&\phantom{=}-\frac{(2\pi i)^{n-3}}{(n-2)!}\E_{2,-m}(1,1,q)-\dots-\frac{(2\pi i)^2}{3!} \E_{n-3,-m}(1,1,q)\,.
\end{align}
This yields for $n=3$
\begin{align}\label{app:intConst:E2mOdd}
\E_{2,-m}(1,1,q)&=m!(2\pi i)^{1-m}\left(\frac{\omega_{1}(m+1;\tau)}{3!}-\omega_{3}(m+1;\tau)\right)
\end{align}
and for $n=5$
\begin{small}
\begin{align}\label{app:intConst:E4mOdd}
\E_4(1,1,q)&=m!(2\pi i)^{3-m}\left(\frac{\omega_{1}(m+1;\tau)}{5!}-\omega_{5}(m+1;\tau)\right)-\frac{(2\pi i)^2}{3!}\E_{2,-m}(1,1,q)\nonumber\\
&=m!(2\pi i)^{3-m}\left(\frac{\omega_{1}(m+1;\tau)}{5!}-\omega_{5}(m+1;\tau)\right)\nonumber\\
&\phantom{=}-\frac{(2\pi i)^2}{3!}m!(2\pi i)^{1-m}\left(\frac{\omega_{1}(m+1;\tau)}{3!}-\omega_{3}(m+1;\tau)\right)\nonumber\\
&=m!(2\pi i)^{3-m}\left(\left(\frac{1}{5!}-\frac{1}{3!3!}\right)\omega_{1}(m+1;\tau)+\frac{1}{3!}\omega_{3}(m+1;\tau)-\omega_{5}(m+1;\tau)\right)\,.
\end{align}
\end{small}
Thus, the explicit solution can combinatorially be deduced as the ones above,
which leads for $n\geq 1$ a natural number and $m\geq 1$ odd to 
\begin{align}\label{app:intConst:EnmOddExplicit}
\E_{2n,-m}(1,1,q)&=m!(2\pi i)^{2n-1-m}\sum_{k=0}^{n}d_{2k+1}\omega_{2n+1-2k}(m+1;\tau)\,.
\end{align}

The above results \eqref{app:intConst:EnmEvenExplicit} and
\eqref{app:intConst:EnmOddExplicit} for $m\neq 0$ can conveniently be
summarised in one single formula: the values $\E_{n,-m}(1,1,q)$ for $n,m\geq 1$
can be expressed as the following linear combinations of elliptic zeta values
\begin{align}\label{app:intConst:EnmExplicit}
\E_{n,-m}(1,1,q)&=\begin{cases}m!(2\pi i)^{n-1-m}\sum_{k=0}^{\lfloor \frac{n}{2} \rfloor}d_{2k+1}\omega_{n+1-2k}(m+1;\tau)& n+m\text{ odd}\\
0&n+m\text{ even.}
\end{cases}
\end{align}

%%%%%%%%%%%%%%%%%%%%%%%%%%%%%%%%%%%%%%%
%%%%%%%%%%%%%%%%%%%%%%%%%%%%%%%%%%%%%%%
%
\section{$\DD_{a,b}^{\E}$ on the torus}\label{app:DEab}
In the following equation the functions $\DD^{\E}_{a,b}$, defined in
\eqn{sec:toolbox_eMPL:DEab}, are related to the sums $\E_{n,-m}$ as introduced
in \eqn{Ell:EBW:Enm}:
\begin{align}\label{app:eqDEab}
	&\DD^{\E}_{a,b}(t,q)\nonumber\\
	&=\sum_{l\geq 0}\DD_{a,b}(tq^l)+(-1)^{a+b}\sum_{l>0}\DD_{a,b}(t^{-1}q^l)+\frac{(4\pi\Im(\tau))^{a+b-1}}{(a+b)!}B_{a+b}(u)\nonumber\\
	&=(-1)^{a-1}\sum_{n=a}^{a+b-1}\binom{n-1}{a-1}\frac{(-2)^{a+b-1-n}}{(a+b-1-n)!}\nonumber\\
	&\phantom{=}\sum_{l> 0}\left(\log(|tq^l|)^{a+b-1-n}\Li_n(tq^l)+(-1)^{a+b}\log(|t^{-1}q^l|)^{a+b-1-n}\Li_n(t^{-1}q^l)\right)\nonumber\\
	&\phantom{=}+(-1)^{b-1}\sum_{n=b}^{a+b-1}\binom{n-1}{b-1}\frac{(-2)^{a+b-1-n}}{(a+b-1-n)!}\nonumber\\
	&\phantom{=}\sum_{l> 0}\left(\log(|tq^l|)^{a+b-1-n}\overline{\Li_n(tq^l)}+(-1)^{a+b}\log(|t^{-1}q^l|)^{a+b-1-n}\overline{\Li_n(t^{-1}q^l)}\right)\nonumber\\
	&\phantom{=}+\DD_{a,b}(t) +\frac{(4\pi\Im(\tau))^{a+b-1}}{(a+b)!}B_{a+b}(u)\nonumber\\
	&=(-1)^{a}\sum_{n=a}^{a+b-1}\binom{n-1}{a-1}\frac{(-2)^{a+b-1-n}}{(a+b-1-n)!}\nonumber\\
	&\phantom{=}\sum_{m=0}^{a+b-1-n}\binom{a+b-1-n}{m}\log(|t|)^{a+b-1-n-m}\log(|q|)^m\E_{n,-m}(t,1,q)\nonumber\\
	&\phantom{=}+(-1)^{b}\sum_{n=b}^{a+b-1}\binom{n-1}{b-1}\frac{(-2)^{a+b-1-n}}{(a+b-1-n)!}\nonumber\\
	&\phantom{=}\sum_{m=0}^{a+b-1-n}\binom{a+b-1-n}{m}\log(|t|)^{a+b-1-n-m}\log(|q|)^m\overline{\E_{n,-m}}(t,1,q)\nonumber\\
	&\phantom{=}+\DD_{a,b}(t) +\frac{(4\pi\Im(\tau))^{a+b-1}}{(a+b)!}B_{a+b}(u)\,.
\end{align}
The sums $\E_{n,-m}$ can be written in terms of the iterated integrals on the
torus as shown in \subsecref{sec:connection:TatetoTorus}. This provides an
explicit translation of $\DD_{a,b}^{\E}$ on the Tate curve to the elliptic
integrals $\Gt$ on the torus.
%
%%%%%%%%%%%%%%%%%%%%%%%%%%%%%%%%%%%%%%
%%%%%%%%%%%%%%%%%%%%%%%%%%%%%%%%%%%%%%
%
%
\section{Vanishing sums over integration kernels}
\label{app:vanish}
In this section we show some explicit calculations used in the main part of
this article. First, let us show how we can get from equations
\eqref{sec:connection_Bloch:IprimeDeqSumJ},
\eqref{sec:connection_Bloch:KDeqSumJ},
\eqref{sec:connection_Bloch:JprimeDeqSumI} and
\eqref{sec:connection_Bloch:KDeqSumI} to
\eqn{sec:connection_Bloch:vanishingSumzg1}, i.e.\
\begin{align}
\sum_{i,j}d_i e_j (A_i-B_j)g^{(1)}(A_i-B_j,\tau)d(A_i-B_j)&=0\,.
\end{align}
In order to apply the initial equations, the sum has to be split correctly, the
equations have to be plugged in, and the sum pulled together again, such that
$\sum_i d_i A_i=0=\sum_j e_j B_j$ can be used. Explicitly, this is the
following calculation
\begin{align}\label{app:vanish:1stCalculation}
&\sum_{i,j}d_i e_j (A_i-B_j)g^{(1)}(A_i-B_j,\tau)d(A_i-B_j)\nonumber\\
&=\sum_{i\in I'}d_iA_i\sum_{j\in J} e_jg^{(1)}(A_i-B_j,\tau)d(A_i-B_j)+\sum_{k\in K}d_kA_k\sum_{j\in J\setminus\{k\}} e_jg^{(1)}(A_k-B_j,\tau)d(A_k-B_j)\nonumber\\
&\phantom{=}-\sum_{j\in J'}e_jB_j\sum_{i\in I} d_ig^{(1)}(A_i-B_j,\tau)d(A_i-B_j)-\sum_{k\in K}d_kA_k\sum_{i\in I\setminus\{k\}} d_ig^{(1)}(A_k-A_i,\tau)d(A_k-A_i)\nonumber\\
&=\sum_{i\in I'}d_iA_i\sum_{j\in J} e_jg^{(1)}(A_i-B_j,\tau)d(A_i-B_j)-\sum_{j\in J'}e_jB_j\sum_{i\in I} d_ig^{(1)}(A_i-B_j,\tau)d(A_i-B_j)\nonumber\\
&\phantom{=}+\sum_{k\in K}d_kA_k\left(\sum_{j\in J\setminus\{k\}} e_jg^{(1)}(A_k-B_j,\tau)d(A_k-B_j)-\sum_{i\in I\setminus\{k\}} d_ig^{(1)}(A_k-A_i,\tau)d(A_k-A_i)\right)\nonumber\\
&=\sum_{i\in I'}d_iA_i\left(-d\log(s_B)-c_1\sum_{j\in J} e_j B_j dB_j\right)-\sum_{j\in J'}e_jB_j\left(\sum_{i\in I} -d\log(s_A)-c_1\sum_{i\in I} d_i A_i dA_i\right)\nonumber\\
&\phantom{=}+\sum_{k\in K}d_kA_k\left(d\log\left(s_A\right)-d\log\left(s_B\right)+c_1\sum_{i\in I}d_iA_idA_i-c_1\sum_{j\in J}e_jB_jdB_j\right)\nonumber\\
&=-\sum_{i\in I}d_iA_i\left(d\log(s_B)+c_1\sum_{j\in J} e_j B_j dB_j\right)+\sum_{j\in J}e_jB_j\left(\sum_{i\in I} d\log(s_A)+c_1\sum_{i\in I} d_i A_i dA_i\right)\nonumber\\
&=0\,.
\end{align}
A similar calculation leads from
equations~\eqref{sec:connection_Bloch:IprimezGamma10SumJ},
\eqref{sec:connection_Bloch:KzGamma10SumJ},
\eqref{sec:connection_Bloch:KzGamma10SumI} and
\eqref{sec:connection_Bloch:JprimezGamma10SumI} to the equation
\begin{align}\label{app:vanishingsums:fctrel2}
&\sum_{i,j}d_i e_j \Re\left(A_i-B_j\right)\Gt(\begin{smallmatrix}1\\0\end{smallmatrix}; A_k-B_j,\tau)
\nonumber\\
&=-i\pi\Re\left(2m_1\sum_{i\in I'}d_i A_i-2m_4\sum_{j\in J'}e_j  B_j+(1+2m_2-2m_3)\sum_{k\in K}d_k  A_k\right)\,,
\end{align}
which implies \eqn{sec:connection_Bloch:BlochRelationRezReGamma10} upon taking
the real part. \Eqn{sec:connection_Bloch:BlochRelationImzReGamma10} can be
obtained by the same calculation with $\Re(A_i-B_j)$ being replaced by
$\Im(A_i-B_j)$. Note that the following sum is valid for the regularised as
well as for the unregularised version of
$\Gt(\begin{smallmatrix}1\\0\end{smallmatrix}; z,\tau)$. With this in mind, let
  us calculate \eqref{app:vanishingsums:fctrel2} and split the sum as before to
  find
\begin{align}
&\sum_{i,j}d_i e_j \Re\left(A_i-B_j\right)\Gt(\begin{smallmatrix}1\\0\end{smallmatrix}; A_k-B_j,\tau)\nonumber\\
&=\sum_{i\in I'}d_i  \Re\left(A_i\right)\sum_{j\in J} e_j\Gt(\begin{smallmatrix}1\\0\end{smallmatrix}; A_i-B_j,\tau)+\sum_{k\in K}d_k  \Re\left(A_k\right)\sum_{j\in J\setminus \{k\}} e_j\Gt_\unreg(\begin{smallmatrix}1\\0\end{smallmatrix}; A_k-B_j,\tau)\nonumber\\
&\phantom{=}-\sum_{j\in J'}e_j  \Re\left(B_j\right)\sum_{i\in I} d_i\Gt(\begin{smallmatrix}1\\0\end{smallmatrix}; A_i-B_j,\tau)-\sum_{k\in K}d_k  \Re\left(A_k\right)\sum_{i\in I\setminus \{k\}} d_i\Gt_\unreg(\begin{smallmatrix}1\\0\end{smallmatrix}; A_k-A_i,\tau)\nonumber\\
&=\sum_{i\in I'}d_i  \Re\left(A_i\right)\left(-2\pi i m_1+\log(\ka)-\log(s_B)-\frac{c_1}{2}\sum_j e_j B_j^2\right)\nonumber\\
&\phantom{=}-\sum_{j\in J'}e_j  \Re\left(B_j\right)\left(-2\pi i m_4+\log(\ka)-\log(s_A)-\frac{c_1}{2}\sum_i d_i A_i^2\right)\nonumber\\
&\phantom{=}+\sum_{k\in K}d_k  \Re\left(A_k\right)\Bigg(-i\pi(1+2m_2-2m_3) -\log(\ka)+ \log(s_A)\nonumber\\
&\phantom{============}+\frac{c_1}{2}\sum_i d_i A_i^2+\log(\ka)-\log(s_B)-\frac{c_1}{2}\sum_j e_j B_j^2\Bigg)\nonumber\\
&= \Re\left(\sum_{i\in I}d_i A_i\right)\left(\log(\ka)-\log(s_B)-\frac{c_1}{2}\sum_{j\in J} e_j B_j^2\right)\nonumber\\
&\phantom{=}-e_j  \Re\left(\sum_{j\in J}B_j\right)\left(\log(\ka)-\log(s_A)-\frac{c_1}{2}\sum_{i\in I} d_i A_i^2\right)\nonumber\\
&\phantom{=}-i\pi\Re\left(2m_1\sum_{i\in I'}d_i A_i-2m_4\sum_{j\in J'}e_j  B_j+(1+2m_2-2m_3)\sum_{k\in K}d_k  A_k\right)\nonumber\\
&=-i\pi\Re\left(2m_1\sum_{i\in I'}d_i A_i-2m_4\sum_{j\in J'}e_j  B_j+(1+2m_2-2m_3)\sum_{k\in K}d_k  A_k\right)\,.
\end{align}
The last calculation of this kind is the step getting from
equations~\eqref{sec:connection_Bloch:Iprimeg2SumJ},
\eqref{sec:connection_Bloch:Jprimeg2SumI} and
\eqref{sec:connection_Bloch:Kg2SumIPlusJ} to
\eqn{sec:connection_Bloch:vanishingSumdGamma20}, i.e.\
\begin{align}
d \sum_{i,j}d_i e_j \Gt(\begin{smallmatrix}2\\0\end{smallmatrix}; A_i-B_j,\tau)
&=0\,.
\end{align}
Here, we have to apply the above splitting of the sum twice to obtain for $d\tau=0$
\begin{align}
&d \sum_{i,j}d_i e_j \Gt(\begin{smallmatrix}2\\0\end{smallmatrix}; A_i-B_j,\tau)\nonumber\\
&=\sum_{i,j}d_i e_j g^{(2)}(A_i-B_j,\tau)d(A_i-B_j)\nonumber\\
&=\sum_{i\in I'}d_i dA_i\sum_{j\in J} e_j g^{(2)}(A_i-B_j,\tau)-\sum_{j\in J'}e_j dB_j\sum_{i\in I} d_i g^{(2)}(A_i-B_j,\tau)\nonumber\\
&\phantom{=}+\sum_{k\in K}d_k dA_k\left(\sum_{j\in J} e_j g^{(2)}(A_k-B_j,\tau)-\sum_{i\in I} d_i g^{(2)}(A_k-A_i,\tau)\right)\nonumber\\
&=\sum_{i\in I'}d_i dA_i\left(-2\pi i\frac{\partial}{\partial \tau}\frac{c_1}{2}\sum_{j\in J} e_j B_j^2-2\pi i\sum_{j\in J} e_jg^{(1)}(A_i-B_j,\tau)\frac{\partial}{\partial \tau}(A_i-B_j)\right)\nonumber\\
&\phantom{=}-\sum_{j\in J'}e_j dB_j\left(-2\pi i \frac{\partial}{\partial \tau}\frac{c_1}{2}\sum_{i\in I} d_i A_i^2-2\pi i\sum_{i\in I} d_ig^{(1)}(A_i-B_j,\tau)\frac{\partial}{\partial \tau}(A_i-B_j)\right)\nonumber\\
&\phantom{=}+\sum_{k\in K}d_k dA_k\Big(-2\pi i\frac{\partial}{\partial \tau}\frac{c_1}{2}\sum_{i\in I} d_i A_i^2-2\pi i\frac{\partial}{\partial \tau}\frac{c_1}{2}\sum_{j\in J} e_j B_j^2\nonumber\\
&\phantom{=}-2\pi i\sum_{j\in J} e_j g^{(1)}(A_k-B_j)\frac{\partial}{\partial \tau}(A_k-B_j)+2\pi i\sum_{i\in I} d_i g^{(1)}(A_k-A_i) \frac{\partial}{\partial \tau}(A_k-B_j)\big)\nonumber\\
&=\sum_{ij}d_i e_j \frac{\partial}{\partial \tau}(A_i-B_j)g^{(1)}(A_i-B_j,\tau)d(A_i-B_j)\nonumber\\
&=0\,,
\end{align}
where for the last equality, we split the sum once again and proceed as in the
calculation of \eqn{app:vanish:1stCalculation}.
\bibliography{fiveterm}

\end{document}